\documentclass[preprint]{elsarticle}
\usepackage{graphicx,color}

\usepackage{amssymb,amsmath,amsthm,graphicx,amscd}
\usepackage[mathscr]{eucal}
\usepackage{enumerate,color,verbatim,multirow,comment}
\usepackage{bm,ulem}

\journal{Annals of Physics}

\begin{document}

\baselineskip=18pt
\numberwithin{equation}{section}
\numberwithin{figure}{section}

\allowdisplaybreaks

\begin{frontmatter}

\title{{\bf Nonequilibrium Steady State \\ in Open Quantum Systems}:  Influence Action, \\ Stochastic Equation and Power Balance}

\author{J.-T. Hsiang}
\ead{cosmology@gmail.com}
\address{Department of Physics, National Dong Hwa University, Taiwan and \\ Center for Theoretical Physics, Fudan University, Shanghai, China}

\author{B.~L.~Hu}
\ead{blhu@umd.edu}
\address{Joint Quantum Institute and Maryland Center for Fundamental Physics, \\ University of Maryland, College Park, Maryland 20742, USA}

\date{May 28, 2014}


\begin{abstract}
The existence and uniqueness of a steady state for nonequilibrium systems (NESS) is a fundamental subject and a main theme of research in statistical mechanics for decades. For Gaussian systems, such as a chain of harmonic oscillators connected at each end to a heat bath, and for anharmonic oscillators under specified conditions, definitive answers exist in the form of proven theorems. Answering this question for quantum many-body systems poses a challenge for the present. In this work we address this issue by deriving the stochastic equations for the reduced system with self-consistent backaction from the two baths, calculating the energy flow from one bath to the chain to the other bath, and exhibiting a power balance relation in the total (chain + baths) system which testifies to the existence of a NESS in this system at late times. Its insensitivity to the initial conditions of the chain corroborates to its uniqueness. The functional method we adopt here entails the use of the influence functional, the coarse-grained and stochastic effective actions, from which one can derive the stochastic equations and calculate the average values of physical variables in open quantum systems. This involves both taking the expectation values of quantum operators of the system and the distributional averages of stochastic variables stemming from the coarse-grained environment. This method though formal in appearance is compact and complete. It can also easily accommodate perturbative techniques and diagrammatic methods from field theory. Taken all together it provides a solid platform for carrying out systematic investigations into the nonequilibrium dynamics of open quantum systems and quantum thermodynamics.
\end{abstract}

\begin{keyword}
Nonequilibrium steady state  \sep Open quantum systems \sep Influence functional formalism \sep stochastic density matrix \sep Langevin equation \sep  noise and fluctuations \sep Energy flow and power balance relations \sep Quantum transport \sep quantum thermodynamics
\end{keyword}

\end{frontmatter}



\tableofcontents

\vskip 1 cm
\section{Introduction}

\textit{Nonequilibrium stationary states} (NESS)  play a uniquely important role in many-body systems in contact with two or more heat baths at different temperatures, similar in importance to the \textit{equilibrium state} of a system in contact with one heat bath which is the arena for the conceptualization and utilization of the canonical ensemble in statistical thermodynamics. The statistical mechanics \cite{NESSSM} and thermodynamics \cite{NESSTD} of open systems \footnote{Defined in a broader sense (A) an open system is one where some of its information is difficult or impossible to obtain or retrieve,  or is coarse-grained away by design or by necessity, both in theoretical and practical terms, the latter referring to the limited capability of the measuring agent or the precision level of instrumentation. The more specific sense (B) used in nonequilibrium statistical mechanics \cite{Zwangzig} emphasizing the influence of a system's environment on its dynamics goes as follows: Start with a closed system comprising of two subsystems $S_{1}$ and $S_{2}$ with some interaction between the two, one can express the dynamics of $S_{1}$ including that of $S_{2}$ in terms of an integral differential equation. If one subsystem $S_{2}$ contains an overwhelmingly large number of degrees of freedom than the other, we call $S_{2}$ an environment $E$ of $S_{1}$. The influence of $E$ on $S_{1}$ is called the backaction.  When the environment can be characterized by thermodynamic parameters it is called a heat bath at temperature $T$ or a matter reservoir with chemical potential $\mu$, etc. When a great deal of microscopic information of $S_{2}$ is discarded or coarse-grained, as is the case when it is described only by a few macroscopic parameters, the effect of the environment can be characterized by noise and fluctuations \cite{Gardiner}, and their backaction on the system show up in the ``reduced" system's dynamics as dissipation \cite{Weiss}, diffusion (quantum diffusion is responsible for the decoherence \cite{Guilini} of quantum phase information). An open system thus carries the influence or the backaction of its environment. Oftentimes the main task in the treatment of open systems is to find the influence of the environment on the subsystem. Defined in this sense (B) it is synonymous with "reduced" system -- with the burden of explanation now shifted to what ``reduced" entails operationally.} in NESS have been the focus of investigation into the important features of nonequilibrium processes of both theoretical interests, such as providing the context for the celebrated classical and quantum fluctuation theorems, and acting as the fountainhead of a new field known as quantum thermodynamics \cite{QThermo,Mahler}, and a wide range of practical applications, extending from physics and chemistry to biology.

For classical many body systems the existence and uniqueness of NESS is a fundamental subject and a main theme of research by mathematical physicists in statistical mechanics for decades. For Gaussian systems (such as a chain of harmonic oscillators with two heat baths at the two ends of the chain) \cite{NESSclHOC} and anharmonic oscillators under general conditions \cite{NESSclAHO} there are definitive answers in the form of proven theorems. Answering this question for quantum many body systems is not so straightforward and poses a major challenge for the present. For quantum many body systems a new direction of research is asking whether \textit{closed} quantum systems can come to equilibrium and thermalize \cite{EqilQCS}. Equilibration of open quantum systems \cite{BP} with strong coupling to a heat bath also shows interesting new features \cite{SFTH}. Transport phenomena in \textit{open} spin systems has also seen a spur of recent activities \cite{SpinQOS}. Noteworthy in the mathematical properties is the role played by symmetry in the nonequilibrium dynamics of these systems \cite{Symmetry}.

\subsection{Issues}

Our current research program on the nonequilibrium dynamics of quantum open systems attempts to address four sets of issues with shared common basis pertaining to NESS:

\paragraph A \textit{The approach to NESS}. Instead of seeking mathematical proofs for these basic issues which are of great importance but not easy to come by it is helpful to see how these systems evolve in time and find out under what conditions one or more NESS may exist. For this we seek to derive the quantum stochastic equations (master, Langevin, Fokker-Planck) for prototypical quantum open systems (e.g., for two oscillators in contact with two heat baths and extension to chains and networks) so one can follow their dynamics explicitly, to examine whether NESS exist at late times, by checking if energy fluxes reach a steady state and whether under these conditions a energy flow (power) balance relation exist. This is probably the most explicit demonstration of the NESS possible. In addition, the stochastic equations can be used to calculate the evolution of key thermodynamic and quantum quantities such as entropy for equilibration / thermalization considerations and quantum entanglement for quantum information inquires.

\paragraph B \textit{Quantum transport}: Since the seminal paper of \cite{LLP}, the role of nonlinearity and nonintegrability in the violation of Fourier law \cite{Fourier} has been explored in a wide variety of representative classical systems with different nonlinear interactions, such as the Fermi-Pasta-Ulam (FPU) models \cite{FPU} or the Frenkel-Kontorova (FK) model \cite{BamLi} and baths of different natures \cite{Fourier,LLPrep,DharAP}.  For the original papers and current status we refer to two nice reviews \cite{LLPrep,DharAP}. For applications of heat conduction to phononics, see \cite{LiRMP}. (Note also the recent work on anomalous heat diffusion \cite{LiuPRL13}).  Numerical results are a lot more difficult to come by for quantum many body systems, thus analytic results, even perturbative,  for weak nonlinearity,  are valuable.  Finding solutions to the quantum stochastic equations have been attempted for simple systems like a quantum anharmonic oscillator chain coupled to two heat baths at the ends or  harmonic oscillators coupled nonlinearly, each with its own heat bath (namely, with or without pinning potentials). The related problem of \textit{equilibration }of open quantum systems with nonlinearity remains an open issue. Even at the classical level this is not a straightforward issue. The existence of breather modes  \cite{Dhar07} and `strange' behavior \cite{EckZab} have been noted. Nonlinearity in quantum system adds a new dimension bearing some similarity or maybe sharing same origins with the issue of how to decipher scars of classical chaos in corresponding quantum systems.

\paragraph C {\it Fluctuation Relations}: Entropy Production in nonequilibrium system and the role of large deviations in currents; Fluctuation Theorems both in the Gallavoti-Cohen vein \cite{GalCoh,Evans,Kurchan,LebSpoFluc,Seifert} and the Jarzynski-Crook relations \cite{Jar97,Crook,FlucThmRev,JarRev}.  Much work in this field is formulated in the context of nonequilibrium thermodynamics. The use of microphysics models such as quantum Brownian motion and open quantum systems techniques, including even decoherence history concepts (for the definition of trajectories), such as used in \cite{SubHu} (see references therein) can provide some new perspective into these powerful relations.

\paragraph D  \textit{Quantum entanglement at finite temperature} \cite{VedralRMP,WVB08} It is generally believed that at high temperatures thermal fluctuations will overshadow quantum entanglement. This problem was explored by Audenaert et al \cite{AudWer} who work out exact solutions  for a bisected closed harmonic chain at ground and thermal states, by Anders \cite{Anders} for a harmonic lattice in 1-3 dimensions and derived a critical temperature above which the quantum system becomes separable. Anders and Winters \cite{AndWin} further provided proof of theorems  and a phase diagram on this issue. Entanglement of a two particle Gaussian state interacting with a single heat bath is investigated recently in \cite{GheDor}. What makes this issue interesting is the suggestion \cite{VedralNature} that quantum entanglement can persist at high temperature in NESS. Recently \cite{GSP} showed by a coupled oscillator model that no thermal entanglement is found in the high temperature limit. However, the existence of quantum entanglement in NESS for \textit{driven systems} as claimed by the experiment of Galve et al \cite{Galve}, and the calculations for spin systems \cite{Znidaric}, remains an open issue.  We want to settle this issue theoretically, at least for harmonic oscillator systems, with the help of the formalism set up here.

Our first batch of papers will focus on Issues A and B, which we describe below. A parallel batch will address Issues C and D in later expositions.

\subsection{Models and Methodology}

The generic quantum open system we study is a simple 1-dimensional quantum oscillator chain, with the two end-oscillators interacting with its own heat bath, each described by a scalar field. The two baths combined make up the environment. We begin our analysis with two oscillators linearly coupled and explore whether a NESS exists for this open system at late times. We do this  by solving for the  stochastic effective action and the Langevin equations, which is possible for a Gaussian system. From this we can derive the expressions for the energy flow from one bath to another through the system.   This is the reason why we begin our study with this model, since in addition to its generic character and  versatility, it provides a nice platform for explaining the methodology we adopt. For the sake of clarity we will work out everything explicitly, so as to facilitate easier comparison with other approaches. We name two papers which are closest to ours, either in the model used or in the concerns expressed: the paper by Dhar, Saito and Hanggi \cite{DSH} uses the reduced density matrix approach to treat quantum transport, while that of Ghsquiere, Sinayskiy and Petruccione \cite{GSP} uses master equations to treat entropy and entanglement dynamics.

Similar in spirit is an earlier  paper by Chen, Lebowitz and Liverani \cite{Chen} which use the Keldysh techniques in a path integral formalism to consider the dissipative dynamics of an anharmonic oscillator in a bosonic heat bath, and recent papers of Zoli \cite{Zoli}, Aron et al \cite{Aron} for instance. The main tools in nonequilibrium quantum many-body dynamics such as the closed time path (CTP, in-in, or Schwinger-Keldysh) \cite{ctp} effective action, the two-particle irreducible (2PI) representation, the large $N$ expansion were introduced for the establishment of quantum kinetic field theory a quarter of centuries ago \cite{CalHu88} and perfected along the way \cite{Berges,Aart,Bodet}. Applications to problems in atomic-optical \cite{Rey}, condensed matter \cite{Zhang}, nuclear-particle \cite{{Bodeker}} and gravitation-cosmology \cite{LRR} have been on the rise in the last decade. A description of quantum field theoretic methods applied to nonequilibrium  processes in a relativistic setting can be found in \cite{CalHu08}. By contrast, there is far less applications of these  well-developed (powerful albeit admittedly  heavy-duty) methodology for the study of nonequilibrium steady state in open quantum systems in contact with two or more baths.   We make such an attempt here for the exploration of fundamental issues of nonequilibrium statistical mechanics for quantum many-body systems and to provide a solid micro-physics foundation for the treatment of problems in quantum thermodynamics which we  see will span an increasingly broader range of applications in physics, chemistry and biology.   Below we explain our methodology and indicate its advantage when appropriate,  while leaving the details of how it is related to other approaches in the sections proper.

\textit{The mathematical framework} of our methodology is the path-integral influence functional formalism \cite{FeyVer,CalLeg,GIS88}, under which the influence action \cite{HPZ1}, the coarse-grained effective action \cite{cgea} and the stochastic effective action \cite{JH1,GalHu} are defined. The stochastic equations such as the master equation (see, e.g., \cite{HPZ1}) and the Langevin equations (see, e.g., \cite{CRV}) can be obtained from taking the functional variations of these effective actions.

There are two main steps in this approach we devised:

\paragraph 1 {\it The derivation of the influence action $S_{IF}$ and coarse-grained effective actions $S_{CG}$} for the reduced system 
(composed of two linearly interacting oscillators, then extended to a harmonic chain) obtained by coarse-graining or integrating over the environmental variables (composed of two baths, coupled to the two end oscillators of a harmonic chain). The baths are here represented by two scalar fields \cite{UnrZur,HuBanff}. Noise  does not appear until the second stage.  This material is contained in Sec.~\ref{S:erbdjh}.

\paragraph 2  For Gaussian systems  the imaginary part of the influence action can be identified via the Feynman-Vernon integral identity with a classical stochastic force (see, e.g., \cite{HPZ2,RHA}). Expressing the exponential of the coarse-grained effective action $S_{CG}$ in the form of a functional integral over the noise distribution, the stochastic effective action $S_{SE}$ is identified as the exponent of the integrand.  Taking the functional variation of $S_{SE}$ yields a set of {\it Langevin equations for the reduced system}. Alternatively one can construct the stochastic reduced density matrix. {\it The averages of dynamical variables in a quantum open system} includes taking the expectation values of the canonical variables as quantum operators and the distributional averages of stochastic variables as classical noises.  We illustrate how to calculate these quantities with both methods in Sec.~\ref{S:reeih3} and~\ref{S:reeih4}. \\

Our methodology includes as subcomponents the so-called reduced density matrix approach (e.g., \cite{DSH,TWH}), the nonequilibrium Green function (NEGF) \cite{Wang,DasDhar,Ness}),  the quantum master equation and quantum Langevin equation approaches.  It is intimately related to the closed-time-path, Schwinger-Keldysh or in-in effective action method, where one can tap into the many useful field theoretical and diagrammatic methods developed.  The stochastic equations of motion \footnote{This could be any of the three kinds mentioned earlier: see e.g., \cite{HPZ1,HalYu,CRV} respectively for derivations of the master, the Fokker-Planck (Wigner) and the Langevin equations.} obtained from taking the functional variation of the stochastic effective action enjoy the desirable features that a) they are real and causal, which guarantee the positivity of the reduced density matrix, and  b) the backaction of the environment on the system is incorporated in a self-consistent way.  These conditions are crucial for the study of nonequilibrium quantum processes including the properties of NESS.

\textit{The physical question we ask} is whether a NESS exists at late times. Since we have the evolutionary equations and their solutions for this system we can follow the quantum dynamics (with dissipation and decoherence) of physical quantities under the influence of the environment (in the form of two noise sources). We describe the behavior of the energy flux and derive the balance relations in Sec.~\ref{S:reeih4}.

Paper II \cite{II} will treat the same system but allow for nonlinear interaction between the two oscillators. For this we shall develop a functional perturbation theory for treating weak nonlinearity.  Entanglement at high temperatures in quantum systems in NESS and equilibration in a quantum system with weak nonlinearity are the themes of planned Papers III, IV respectively \cite{III,IV}.

\subsection{Main Features and Findings}

\subsubsection{Approach}

For the description of the dynamics of an open quantum system  obtaining the time development of the reduced density operator pretty much captures its essence and evolution. We derive the reduced density operator with the influence functional and closed-time-path formalisms (for a `no-thrill' introduction, see, e.g., Chapters 5, 6 of \cite{CalHu08}).

With this reduced density operator one can compute the time evolution of the expectation values of the operators corresponding to physical variables in the reduced system \footnote{In the sense described in Footnote 1, a reduced system is an open system whose dynamics includes the backaction of its coarse-grained environment.} Here we are interested in the energy flux (heat current) flowing between a chain of $n$ identical coupled harmonic oscillators which together represent the system ($\mathfrak{S}=\sum_{k=1}^{n}O_{k}$). Let's call $B_{1}$ the bath which $O_{1}$ interacts with, and $B_{2}$ the bath oscillator $O_{2}$ interacts with. Thus $B_{1}$, $B_{2}$ are affectionately named our oscillators' `private' baths.

Writing the reduced density operator in terms of the stochastic effective action + the probability functional, one can compute the energy current between each oscillator and its private bath in the framework of the reduced density operator. This  functional method provides a useful platform for the construction of a perturbation theory, which we shall show in the next paper, in treating weakly nonlinear cases.

Alternatively, from the influence action one can derive the Langevin equation describing the dynamics of the reduced system under the influence of a noise obtained from the influence functional. This is probably a more intuitive and transparent pathway in visualizing the energy flow between the system and the two baths.

\subsubsection{Features}

{\it The fundamental solutions} which together determine the evolutionary operator of the reduced density operator all have an exponentially decaying factor. This has the consequences that

		(1)	the dependence on the system's initial conditions will quickly become insignificant as the system evolves in time. Because of the exponential decay, only during a short transient period  are the effects of initial conditions observable.  At late time, the behavior of the system is governed by the baths. In other words, for Gaussian initial states, the time evolution of the system is always attracted to the behavior controlled by the bath, independent of the initial conditions of the system.

		(2) the physical variables of interest here tend to relax to --  becoming  exponentially close to -- a fixed value in time. For example, the velocity variance will asymptotically go to a constant on a time scale longer than the inverse of the decay constant. In addition all oscillators $O_{k}$ along the chain have the same relaxation time scale.

{\it The energy currents} between $B_{1}$--$O_{1}$, or $O_{k}$--$O_{k+1}$, or $O_{n}$--$B_{n}$ in general all evolve with time, and will depend on the initial conditions. However, after the motion of the oscillators along the chain is fully relaxed, the energy currents between components approach time-independent values, with the same magnitude.

This time-independence establishes the \textit{existence of an equilibrium steady state}.   Its insensitivity to the initial conditions of the chain testifies to \textit{its uniqueness}, the same magnitude ensures there is no energy buildup in any component of the open system: Heat flows from one bath to another via the intermediary of the subsystems. To our knowledge, unlike for classical harmonic oscillators where mathematical proofs of the existence and uniqueness of the NESS have been provided, there is no such proofs for quantum harmonic systems. It is perhaps tempting to make such an assumption drawing the close correspondence between quantum and classical Gaussian systems this is what most authors tacitly assume (e.g. \cite{DSH}). We have not provided a mathematical proof of the existence and uniqueness of a NESS for this generic system under study. What we have is an explicit demonstration,  drawing our conclusions from solving the dynamics of this system under very general conditions -- the full time evolution of the nonequilibrium open system is perhaps more useful for solving physical problems.

\subsubsection{Results}

\begin{enumerate}

\item  We have obtained the full nonequilibrium time evolution of the reduced system (in particular, energy flow along a harmonic chain between $B_{1}$--$O_{1}$, or $O_{k}$--$O_{k+1}$, or $O_{n}$--$B_{n}$ in a harmonic chain) at all temperatures and couplings  with arbitrary strength \footnote{For comparison, \cite{GSP} made a weak coupling assumption when working at low temperatures. For Gaussian systems one can solve the full dynamics at least formally in the strong coupling regime -- this is well known, see, e.g., \cite{Adesso}, and is assumed so in \cite{DSH}. However,  when explicit results are desired,  one often has to make some compromised assumptions, such as weak coupling between the oscillators and their baths, as done in the last section of \cite{DSH}.}.

    The formal mathematical expressions of the energy current are given in
    \begin{itemize}
    	\item Eqs.~\eqref{E:mar}, \eqref{E:bnxiryt} and \eqref{E:msar} as well as \eqref{E:hgeywowa}, \eqref{E:uryebd} for a two-oscillator chain, and
    	\item Eqs.~\eqref{E:tvrsmd}, \eqref{E:hmcnvw} and \eqref{E:uvrsmd} for an $n$-oscillator chain,
	\end{itemize}
	from which we can obtain a profile of energy currents between the components.

\item We have established the steady state value of the energy flux in \eqref{E:tvrsmd}, \eqref{E:hmcnvw} and \eqref{E:uvrsmd}. Manifest equality and time-independence of these expressions implies stationarity.  There is no buildup or deficit of energy in any of the components.

\item We have demonstrate that the NESS current is independent of the initial (Gaussian) configurations of the chain after the transient period. It thus implies uniqueness.

\item We have obtained a Landauer-like formula in
		\begin{itemize}
			\item Eq.~\eqref{E:deukfhd} for a two-oscillator chain, and
			\item Eq.~\eqref{E:utrsmd} for an $n$-oscillator chain
		\end{itemize}

\item In particular for the case of two oscillators ($n=2$), we define heat conductance \eqref{E:bdhredw}, and have shown that
	\begin{itemize}
	\item in the high temperature limit,
		\begin{enumerate}
			\item The steady energy current is proportional to the temperature difference between the baths, in \eqref{E:cbvkds},

			\item The heat conductance is independent of the temperature of either bath, as is seen in \eqref{E:dfmsdrrer},

			\item The dependence of the conductance on two types of coupling constants is shown in \eqref{E:derjsfw} and in Fig.~\ref{Fi:conductance}.
		\end{enumerate}

	\item in the low temperature limit
		\begin{enumerate}
			\item the temperature dependence of the steady energy current, \eqref{E:nmeukfhd}, \eqref{E:dfberhe} and in Fig.~\ref{Fi:currents}, and

			\item the temperature dependence of the conductivity in \eqref{E:mnhruewa}.
		\end{enumerate}
	\end{itemize}
\item We also plot the general dependence of the NESS energy current on the length of the chain $n$ in Fig.~\eqref{Fi:scaling}, based on our analytical expressions \eqref{E:mcnvdh1}, \eqref{E:icmer} and \eqref{E:mcnvdh2}. It shows that
		\begin{itemize}
			\item for small $n$, the NESS current does depend on the length in a nontrivial way; however
			\item for sufficiently large $n$, the NESS current oscillates but converges to a constant independent of $n$.
		\end{itemize}
\end{enumerate}

\section{Coarse-Grained Effective Action for Open Quantum Systems}\label{S:erbdjh}

Consider a quantum system $\mathfrak{S}= S_1 + S_2$ made up of two subsystems $S_{1,2}$ each consisting of a harmonic oscillator $O_{1,2}$ interacting with its own bath $B_{1,2}$ at temperatures $T_{1,2}$ respectively (assume $T_1 > T_2$). The system by itself is closed while when brought in contact with heat baths becomes open, owing to the overwhelming degrees of freedom in the baths which are inaccessible or unaccountable for.  The situation of one oscillator interacting with one bath under the general theme of quantum Brownian motion (QBM) has been studied for decades and is pretty well-understood, extending to non-Markovian dynamics in a general environment.  Here we wish to extend this study to two such identical configurations, adding a coupling between $O_1$ and $O_2$ which is assumed to be linear in this paper and nonlinear in subsequent papers. Let's call $\mathfrak{S}$ the combined system of two coupled quantum Brownian oscillators each interacting with its private bath.  Assume that each oscillator is isolated from the other thermal bath, thus there is no direct contact between $O_1$ and $B_2$, but there is indirect influence through $O_1$'s coupling to $O_2$ and its interaction with $B_2$.  Assume also that initially the wave functions of the oscillators do not overlap and that the baths do not occupy the same spacetime region\footnote{This can be viewed as an idealization of finite-size bath in the limit that the bath degrees of freedom is sufficiently large and the size of the bath is much larger than the scales associated with the oscillator's motion.}.  The physics question we are interested in is whether a nonequilibrium steady state exists in $\mathfrak{S}$ and how it comes about, in terms of its time evolution. As a useful indicator we wish to describe the energy flow in the three segments: $B_1 \rightarrow O_1 \rightarrow O_2 \rightarrow B_2$. It is not clear \textit{a priori} why energy should flow in a fixed direction (indeed it does not, before each oscillator fully relaxes) and the flow is steady (time translational, namely, there is no energy localization or heat accumulation, especially when we extend the two oscillators to a chain). For this purpose we need to derive the evolution equations for the reduced density operator \cite{GIS88} of the system proper, $\mathfrak{S}$, after it is rendered open, as a result of tracing over the two baths they interact with and including their backaction which shows up as quantum dissipation and diffusion in the equations of motion for the reduced system.  We do this by way of functional formalisms operating at two levels: 1) at the influence or effective action level, familiar to those with experience of the Feynman-Vernon influence functional \cite{FeyVer} and the Schwinger-Keldysh (`in-in', or closed-time-path) \cite{ctp} methods;  2) at the equation of motion level,  obtained from the functional variation of the effective / influence action.  This includes the familiar stochastic equations -- the master equation (see, e.g., \cite{CalLeg,HPZ1} for derivations,   Fokker-Planck \cite{HalYu} or Langevin equations \cite{CRV}) which is probably more widely used.


Let each subsystem be a quantum oscillator following a prescribed trajectory $\mathbf{z}^{(i)}$ (see, e.g., \cite{RHA}), and its displacement is described by $\chi^{(i)}$. The baths are represented by a massless quantum scalar field $\phi^{(i)}$ (e.g., \cite{UnrZur}) at finite temperature. (This is what we refer to as a thermal field, it is a quantum, not a classical, field, although for Gaussian systems quantum and classical equations of motions have the same form.)  The action of the total system is given by
\begin{align}\label{E:cnmxrere}
	 S[\chi,\phi]&=\int_{0}^{t}\!ds\;\Bigl\{\sum_{i=1}^{2}\frac{m}{2}\Bigl[\dot{\chi}^{(i)2}(s)-\omega^{2}\chi^{(i)2}(s)\Bigr]-m\sigma\,\chi^{(1)}(s)\chi^{(2)}(s)\Bigr\}\notag\\
	 &\qquad\qquad+\sum_{i=1}^{2}\int_{0}^{t}\!d^{4}x_{i}\;e_{i}\chi^{(i)}(s)\delta^{3}(\mathbf{x}_{i}-\mathbf{z}^{(i)}(s))\phi^{(i)}(x_{i})\notag\\
	 &\qquad\qquad\qquad\qquad+\sum_{i=1}^{2}\int_{0}^{t}\!d^{4}x_{i}\;\frac{1}{2}\,\bigl[\partial_{\mu}\phi^{(i)}(x_{i})\bigr]\bigl[\partial^{\mu}\phi^{(i)}(x_{i})\bigr]\,,
\end{align}
among which we have the actions that describes the oscillators $S_{\chi}$, the bath fields $S_{\phi}$, the interaction between the two oscillators $S_{I}$  and between each oscillator and its bath $S_{II}$, respectively,
\begin{align*}
	S_{\chi}[\chi^{(i)}]&=\int_{0}^{t}\!ds\;\frac{m}{2}\Bigl[\dot{\chi}^{(i)2}(s)-\omega^{2}\chi^{(i)2}(s)\Bigr]\,,\\
	 S_{\phi}[\phi^{(i)}]&=\int_{0}^{t}\!d^{4}x_{i}\;\frac{1}{2}\,\bigl[\partial_{\mu}\phi^{(i)}(x_{i})\bigr]\bigl[\partial^{\mu}\phi^{(i)}(x_{i})\bigr]\,,\\
	S_{I}[\chi^{(1)},\chi^{(2)}]&=\int_{0}^{t}\!ds\;\Bigl[-m\sigma\chi^{(1)}(s)\chi^{(2)}(s)\Bigr]\,,\\
	 S_{I\!I}[\chi^{(i)},\phi^{(i)}]&=\int_{0}^{t}\!d^{4}x_{i}\;e_{i}\chi^{(i)}(s)\delta^{3}(\mathbf{x}_{i}-\mathbf{z}^{(i)}(s))\phi^{(i)}(x_{i})\,.
\end{align*}
Here we assume that each oscillator is linearly coupled to its own thermal bath with coupling strength\footnote{We assume that the coupling strength $e_{i}$ is not so strong as to displace the oscillators.} $e_{i}$, and the oscillators are coupled with each other in the forms of $(\chi^{(1)}-\chi^{(2)})^{2}$ or $\chi^{(1)}\chi^{(2)}$ (which are equivalent by a shift in the $\chi$ coordinate), with an interaction strength denoted by $\sigma$. For simplicity without loss of physical contents we let the two oscillators have the same mass $m$ and natural frequency $\omega$. We leave the prescribed trajectory $\mathbf{z}^{(i)}(s)$ general here because the position of the oscillator changes the configuration of the quantum field which in turn affects the other oscillator, aspects which need to be included in quantum entanglement considerations (see, e.g., \cite{LinHu08}) and in treating relativistic quantum information issues (e.g., \cite{HLL}). In a later section when we turn to calculating the energy flow we can safely assume that their external (centers of mass) variables are fixed in space, and only their internal  variables $\chi$ enter in the dynamics.

Now we assume that the initial state of the total system $\mathfrak{S}$ at time $t=0$ is in a factorizable form \footnote{For a discussion of the physical consequences of factorizable initial conditions and generalizations, see e.g., \cite{HPZ1,JH1,PazRom,FRH}.}.
\begin{equation}
	\rho(0)=\rho_{\chi}\otimes\rho_{\beta_{1}}\otimes\rho_{\beta_{2}}\,,
\end{equation}
where $\rho_{\chi}$ is the initial density operator for the system proper $\mathfrak{S}$, consisting of two oscillators,  with each oscillator described by a Gaussian wavefunction
\begin{equation}
	 \rho_{\chi}(\chi^{(i)}_{a},\chi'^{(i)}_{a};0)=\left(\frac{1}{\pi\varsigma^{2}}\right)^{1/2}\exp\left[-\frac{1}{2\varsigma^{2}}\bigl(\chi^{(i)2}_{a}+\chi'^{(i)2}_{a}\bigr)\right]\,.
\end{equation}
The parameter $\varsigma$ is the width of the wavepacket, and the parameters $\chi_{a}$, $\chi_{b}$ are the shorthand notations for $\chi$ evaluated at times $t=0$ and $t$ respectively, that is, $\chi_{a}=\chi(0)$ and $\chi_{b}=\chi(t)$. This subscript convention will also apply to other variables. Each bath is initially in its own thermal state at temperature $\beta_{i}^{-1}$, so the corresponding initial density matrix is
\begin{align}
	\rho_{\beta_{i}}(\phi^{(i)}_{a},\phi'^{(i)}_{a};0)&=\langle\phi^{(i)}_{a}\vert e^{-\beta_{i}H_{\phi}[\phi^{(i)}]}\vert\phi'^{(i)}_{a}\rangle
\end{align}
$H_{\phi}[\phi^{(i)}]$ is the free scalar field Hamiltonian associated with the action $S_{\phi}[\phi^{(i)}]$.

The density operator of the total system is evolved by the unitary evolution operator $U(t,0)$,
\begin{equation}\label{E:woejs}
	\rho(t)=\Bigl\{U(t,0)\,\rho(0)\,U^{-1}(t,0)\Bigr\}\,.
\end{equation}
In the path-integral representation the total density matrix at time $t$ is related to its values at an earlier moment $t=0$ by
\begin{align}
	&\qquad\rho(\chi_{b}^{(i)},\chi'^{(i)}_{b};\phi^{(i)}_{b},\phi'^{(i)}_{b};t)\notag\\
	 &=\left\{\prod_{i=1}^{2}\int_{-\infty}^{\infty}\!d\chi^{(i)}_{a}d\chi'^{(i)}_{a}\!\int_{-\infty}^{\infty}\!d\phi^{(i)}_{a}d\phi'^{(i)}_{a}\!\int_{\chi^{(i)}_{a}}^{\chi^{(i)}_{b}}\!\mathcal{D}\chi^{(i)}_{+}\!\int_{\chi'^{(i)}_{a}}^{\chi'^{(i)}_{b}}\!\mathcal{D}\chi^{(i)}_{-}\!\int_{\phi^{(i)}_{a}}^{\phi^{(i)}_{b}}\!\mathcal{D}\phi^{(i)}_{+}\!\int_{\phi'^{(i)}_{a}}^{\phi'^{(i)}_{b}}\!\mathcal{D}\phi^{(i)}_{-}\right\}\notag\\
	 &\qquad\qquad\quad\exp\Bigl(\sum_{i=1}^{2}i\,S_{\chi}[\chi^{(i)}_{+}]-i\,S_{\chi}[\chi^{(i)}_{-}]\Bigr)\times\exp\Bigl(i\,S_{I}[\chi^{(1)}_{+},\chi^{(2)}_{+}]-i\,S_{I}[\chi^{(1)}_{-},\chi^{(2)}_{-}]\Bigr)\notag\\
	 &\qquad\qquad\times\exp\Bigl(\sum_{i=1}^{2}i\,S_{\phi}[\phi^{(i)}_{+}]-i\,S_{\phi}[\phi^{(i)}_{-}]\Bigr)\times\exp\Bigl(\sum_{i=1}^{2}i\,S_{I\!I}[\chi^{(i)}_{+},\phi^{(i)}_{+}]-i\,S_{I\!I}[\chi^{(i)}_{-},\phi^{(i)}_{-}]\Bigr)\notag\\
	 &\qquad\qquad\qquad\qquad\times\rho_{\chi}(\chi^{(i)}_{a},\chi'^{(i)}_{a};0)\prod_{i=1}^{2}\rho_{\beta_{i}}(\phi^{(i)}_{a},\phi'^{(i)}_{a};0)\,,
\end{align}
The subscripts $+$, $-$ attached to each dynamical variable indicate that the variable is evaluated along the forward and backward time paths, respectively implied by $U$ and $U^{-1}$ in \eqref{E:woejs}.

\subsection{Reduced Density Operator and Green Functions}

When we focus on the dynamics of the oscillators $\mathfrak{S}$, accounting for only the gross influences of their environments but not the details, we work with the reduced density operator of $\mathfrak{S}$ obtained by tracing out the microscopic degrees of freedom of its environment, their two baths. We obtain
\begin{align}
	&\quad\rho_{\chi}(\chi^{(i)}_{b},\chi'^{(i)}_{b};t)
	 =\operatorname{Tr}_{\phi^{(1)}}\operatorname{Tr}_{\phi^{(2)}}\rho(\chi_{b}^{(i)},\chi'^{(i)}_{b};\phi^{(i)}_{b},\phi'^{(i)}_{b};t)\notag\\
	 &=\int_{-\infty}^{\infty}\!\left\{\prod_{i=1}^{2}d\chi^{(i)}_{a}d\chi'^{(i)}_{a}\right\}\;\rho_{\chi}(\chi^{(i)}_{a},\chi'^{(i)}_{a},t_{a})\left\{\prod_{i=1}^{2}\int_{\chi^{(i)}_{a}}^{\chi^{(i)}_{b}}\!\mathcal{D}\chi^{(i)}_{+}\!\int_{\chi'^{(i)}_{a}}^{\chi'^{(i)}_{b}}\!\mathcal{D}\chi^{(i)}_{-}\right\}\notag\\
	 &\qquad\times\exp\Bigl(\sum_{i=1}^{2}i\,S_{\chi}[\chi^{(i)}_{+}]-i\,S_{\chi}[\chi^{(i)}_{-}]\Bigr)\times\exp\Bigl(i\,S_{I}[\chi^{(1)}_{+},\chi^{(2)}_{+}]-i\,S_{I}[\chi^{(1)}_{-},\chi^{(2)}_{-}]\Bigr)\notag\\
	 &\qquad\times\prod_{i=1}^{2}\exp\biggl\{\frac{i}{2}\,e_{i}^{2}\int_{0}^{t}\!ds\,ds'\biggl(\Bigl[\chi^{(i)}_{+}(s)-\chi^{(i)}_{-}(s)\Bigr]G_{R,\,\beta_{i}}(s,s')\Bigl[\chi^{(i)}_{+}(s')+\chi^{(i)}_{-}(s')\Bigr]\biggr.\biggr.\notag\\
	 &\qquad\qquad\qquad+\biggl.\biggl.i\,\Bigl[\chi^{(i)}_{+}(s)-\chi^{(i)}_{-}(s)\Bigr]G_{H,\,\beta_{i}}(s,s')\Bigl[\chi^{(i)}_{+}(s')-\chi^{(i)}_{-}(s')\Bigr]\biggr)\biggr\}\,,\label{E:nwjdjaa}
\end{align}
where the \textbf{retarded Green's function} $G_{R,\,\beta_{i}}$ is defined by
\begin{align}
	 G_{R,\,\beta_{i}}(s,s')&=i\,\theta(s-s')\operatorname{Tr}\Bigl(\rho_{\beta_{i}}\Bigl[\phi^{(i)}(z^{(i)}(s),s),\phi^{(i)}(z^{(i)}(s'),s')\Bigr]\Bigr)\notag\\
		&=i\,\theta(s-s')\,\Bigl[\phi^{(i)}(z^{(i)}(s),s),\phi^{(i)}(z^{(i)}(s'),s')\Bigr]=G_{R}(s,s')\,,
\intertext{and the \textbf{Hadamard function} $G_{H,\,\beta_{i}}$ by}
	 G_{H,\,\beta_{i}}(s,s')&=\frac{1}{2}\operatorname{Tr}\Bigl(\rho_{\beta_{i}}\Bigl\{\phi^{(i)}(z^{(i)}(s),s),\phi^{(i)}(z^{(i)}(s'),s')\Bigr\}\Bigr)\,.
\end{align}
The Hadamard function is simply the expectation value of the anti-commutator of the quantum field $\phi^{(i)}$, and notice that the retarded Green's function does not have any temperature dependence. The exponential containing $G_{R,\,\beta_{i}}$ and $G_{H,\,\beta_{i}}$ in \eqref{E:nwjdjaa} is the Feynman-Vernon influence functional $\mathcal{F}$,
\begin{align}
	&\quad\mathcal{F}[\chi_{+},\chi_{-}]=e^{i\,S_{IF}[\chi_{+},\chi_{-}]}\notag\\
	 &=\prod_{i=1}^{2}\exp\biggl\{\frac{i}{2}\,e_{i}^{2}\int_{0}^{t}\!ds\,ds'\biggl(\Bigl[\chi^{(i)}_{+}(s)-\chi^{(i)}_{-}(s)\Bigr]G_{R,\,\beta_{i}}(s,s')\Bigl[\chi^{(i)}_{+}(s')+\chi^{(i)}_{-}(s')\Bigr]\biggr.\biggr.\notag\\
	 &\qquad\qquad+\biggl.\biggl.i\,\Bigl[\chi^{(i)}_{+}(s)-\chi^{(i)}_{-}(s)\Bigr]G_{H,\,\beta_{i}}(s,s')\Bigl[\chi^{(i)}_{+}(s')-\chi^{(i)}_{-}(s')\Bigr]\biggr)\biggr\}\,,
\end{align}
where $S_{IF}$ is called the influence action. It captures the influences of the environment on the system $\mathfrak{S}$.

\subsection{Coarse-Grained Effective Action}

The coarse-grained effective action (CG) $S_{CG}$ is made of the influence action $S_{IF}$ from the environment and the actions of the system by
\begin{align}
	&\quad S_{CG}[q^{(i)},r^{(i)}]\notag\\
	 &=\left\{\sum_{i=1}^{2}S_{\chi}[\chi^{(i)}_{+}]-S_{\chi}[\chi^{(i)}_{-}]\right\}+S_{I}[\chi^{(1)}_{+},\chi^{(2)}_{+}]-S_{I}[\chi^{(1)}_{-},\chi^{(2)}_{-}]+S_{IF}[\chi_{+},\chi_{-}]\notag\\
	 &=\int_{0}^{t}\!ds\;\biggl\{\sum_{i=1}^{2}\Bigl[m\dot{q}^{(i)}(s)\dot{r}^{(i)}(s)-m\omega^{2}q^{(i)}(s)r^{(i)}(s)\Bigr]\biggr.\label{E:oweioiana}\\
	 &\qquad\qquad\qquad\qquad\qquad\qquad-\biggl.m\sigma\Bigl[q^{(1)}(s)r^{(2)}(s)+q^{(2)}(s)r^{(1)}(s)\Bigr]\biggr\}\notag\\
	 &\qquad+\sum_{i=1}^{2}e_{i}^{2}\int_{0}^{t}\!ds\,ds'\biggl[q^{(i)}(s)G_{R}(s,s')r^{(i)}(s')+\frac{i}{2}\,q^{(i)}(s)G_{H,\,\beta_{i}}(s,s')q^{(i)}(s')\biggr]\,.
\end{align}
Here we have introduced the relative coordinate $q^{(i)}$ and the centroid coordinate $r^{(i)}$,
\begin{equation}
	q^{(i)}=\chi^{(i)}_{+}-\chi^{(i)}_{-}\,,\qquad\qquad\qquad r^{(i)}=\frac{1}{2}\bigl(\chi^{(i)}_{+}+\chi^{(i)}_{-}\bigr)\,.
\end{equation}
Anticipating the oscillator chain treated in a later section, it is convenient to introduce the vectorial notations by
\begin{align*}
	\mathbf{q}&=\begin{pmatrix} q^{(1)}\\[12pt]q^{(2)}\end{pmatrix}\,,\qquad\qquad\qquad\mathbf{r}=\begin{pmatrix} r^{(1)}\\[12pt]r^{(2)}\end{pmatrix}\,,\qquad\qquad\qquad\pmb{\Omega}^{2}=\begin{pmatrix}
				\omega^{2}	&\sigma\\[12pt]
				\sigma		&\omega^{2}
				\end{pmatrix}\,,\\\\
	\mathbf{G}_{R}(s,s')&=\begin{pmatrix}
				G_{R}(s,s')				&0\\[12pt]
				0								&G_{R}(s,s')
				\end{pmatrix}\,,\quad\mathbf{G}_{H}(s,s')=\begin{pmatrix}
				G_{H,\,\beta_{1}}(s,s')	&0\\[12pt]
				0								&G_{H,\,\beta_{2}}(s,s')
				\end{pmatrix}\,,
\end{align*}
and from now on assume the coupling strengths $e_{1}$ and $e_{2}$ are the same, that is, $e_{1}=e_{2}=e$. In so doing the coarse-grained effective action can be written into a more compact form
\begin{align}
	 S_{CG}&=\int_{0}^{t}\!ds\;\biggl\{m\,\dot{\mathbf{q}}^{T}(s)\cdot\dot{\mathbf{r}}(s)-m\,\mathbf{q}(s)\cdot\pmb{\Omega}^{2}\cdot\mathbf{r}(s)\biggr.\label{E:jiwaljndakl}\\
	 &\qquad\qquad\quad+\biggl.e^{2}\int_{0}^{s}\!ds'\biggl[\mathbf{q}^{T}(s)\cdot\mathbf{G}_{R}(s,s')\cdot\mathbf{r}(s')+i\,\mathbf{q}^{T}(s)\cdot\mathbf{G}_{H}(s,s')\cdot\mathbf{q}(s')\biggr]\,.\notag
\end{align}
Formally, \eqref{E:jiwaljndakl} is very general, and can be readily applied to the configuration that oscillators simultaneously interact with two different thermal baths.

Since the coarse-grain effective action $S_{CG}$ governors the dynamics of the system $\mathfrak{S}$ under the influence of the environments, the time evolution of the reduced density matrix can thus be constructed with $S_{CG}$. We write the reduced density matrix as
\begin{align}
	 \rho_{\chi}(\chi^{(i)}_{b},\chi'^{(i)}_{b};t)&=\int_{-\infty}^{\infty}\!\left\{\prod_{i=1}^{2}dq^{(i)}_{a}dr^{(i)}_{a}\right\}\;J(q^{(i)}_{b},r^{(i)}_{b},t;q^{(i)}_{a},r^{(i)}_{a},0)\;\rho_{\chi}(q^{(i)}_{a},r^{(i)}_{a};0)\,,\label{E:fnjdsa}
\end{align}
where
\begin{align}
	 J(q^{(i)}_{b},r^{(i)}_{b},t;q^{(i)}_{a},r^{(i)}_{a},0)&=\left\{\prod_{i=1}^{2}\int_{q^{(i)}_{a}}^{q^{(i)}_{b}}\!\mathcal{D}q^{(i)}\!\int_{r^{(i)}_{a}}^{r^{(i)}_{b}}\!\mathcal{D}r^{(i)}\right\}\exp\Bigl\{i\,S_{CG}[q^{(i)},r^{(i)}]\Bigr\}\,,
\end{align}
is the evolutionary operator for the reduced density matrix (from time $0$ to $t$). The path integral in the evolutionary operator $J$ can be evaluated exactly because the coarse-grained effective action \eqref{E:jiwaljndakl} is quadratic in $\mathbf{q}$ and $\mathbf{r}$. We won't pursuit this route in this paper but it will be used later for our study of nonlinear systems.

\section{Stochastic Effective Action and Langevin Equations}\label{S:reeih3}

We now proceed to derive the stochastic equations and find their solutions

\subsection{Stochastic Effective Action}

Using the Feynman-Vernon identity for Gaussian integrals we can express the imaginary part of the coarse-grained effective action $S_{CG}$ in \eqref{E:jiwaljndakl} in terms of the distributional integral of a Gaussian noise $\pmb{\xi}$,
\begin{align}
	 &\quad\exp\left[-\frac{e^{2}}{2}\int_{0}^{t}\!ds\!\int_{0}^{t}\!ds'\;\mathbf{q}^{T}(s)\cdot\mathbf{G}_{H}(s,s')\cdot\mathbf{q}(s')\right]\notag\\
	 &=\int\mathcal{D}\pmb{\xi}\;\mathcal{P}[\pmb{\xi}]\,\exp\left[i\int_{0}^{t}\!ds\;\mathbf{q}^{T}(s)\cdot\pmb{\xi}(s)\right]\,,
\end{align}
with the moments of the noise given by
\begin{equation}
	 \langle\pmb{\xi}(s)\rangle=0\,,\qquad\qquad\qquad\langle\pmb{\xi}(s)\cdot\pmb{\xi}^{T}(s')\rangle=e^{2}\mathbf{G}_{H}(s,s')\,.
\end{equation}
The angular brackets here denote the ensemble average over the probability distribution functional $\mathcal{P}[\pmb{\xi}]$. Thus we may write the exponential of the coarse-grained effective action $S_{CG}$ in a form of a distributional integral
\begin{align}
	 e^{i\,S_{CG}[\mathbf{q},\mathbf{r}]}&=\int\mathcal{D}\pmb{\xi}\;\mathcal{P}[\pmb{\xi}]\,\exp\left[i\int_{0}^{t}\!ds\;\biggl\{m\,\dot{\mathbf{q}}^{T}(s)\cdot\dot{\mathbf{r}}(s)-m\,\mathbf{q}(s)\cdot\pmb{\Omega}^{2}\cdot\mathbf{r}(s)\biggr.\right.\notag\\
	 &\qquad\qquad\qquad\qquad+\left.\biggl.\mathbf{q}^{T}(s)\cdot\pmb{\xi}(s)+\int_{0}^{s}\!ds'\;\mathbf{q}^{T}(s)\cdot\mathbf{G}_{R}(s,s')\cdot\mathbf{r}(s')\biggr\}\right]\notag\\
	&=\int\mathcal{D}\pmb{\xi}\;\mathcal{P}[\pmb{\xi}]\,e^{i\,S_{SE}[\mathbf{q},\mathbf{r},\pmb{\xi}]}\,,
\end{align}
where $S_{SE}$ is the \textit{stochastic effective action} \cite{JH1} given by
\begin{align}\label{E:hunxms}
	 S_{SE}[\mathbf{q},\mathbf{r},\pmb{\xi}]&=\int_{0}^{t}\!ds\;\biggl\{m\,\dot{\mathbf{q}}^{T}(s)\cdot\dot{\mathbf{r}}(s)-m\,\mathbf{q}(s)\cdot\pmb{\Omega}^{2}\cdot\mathbf{r}(s)+\mathbf{q}^{T}(s)\cdot\pmb{\xi}(s)\biggr.\notag\\
	 &\qquad\qquad\qquad\qquad+\biggl.\int_{0}^{s}\!ds'\;\mathbf{q}^{T}(s)\cdot\mathbf{G}_{R}(s,s')\cdot\mathbf{r}(s')\biggr\}\,.
\end{align}
At this point, we may use the stochastic effective action to either derive the Langevin equation, or to construct the stochastic reduced density matrix. We proceed with the former route below.

\subsection{Langevin Equations}

Taking the variation of $S_{SE}$ with respect to $\mathbf{q}$ and letting $\mathbf{q}=0$, we arrive at a set of Langevin equation,
\begin{align}\label{E:euwbnss}
	 m\,\ddot{\pmb{\chi}}(s)+m\,\pmb{\Omega}^{2}\cdot\pmb{\chi}(s)-\int_{0}^{s}\!ds'\;\mathbf{G}_{R}(s,s')\cdot\pmb{\chi}(s')&=\pmb{\xi}(s)\,.
\end{align}
Formally, this equation of motion describes the time evolution of the reduced system under the non-Markovian influence of the environment. The influence is manifested in the form of the local stochastic driving noise $\pmb{\xi}$ and the nonlocal dissipative force,
\begin{equation*}
	 \int_{0}^{s}\!ds'\;\mathbf{G}_{R}(s,s')\cdot\pmb{\chi}(s')\,.
\end{equation*}
In general, this nonlocal expression implies the evolution of the reduced system is history-dependent. However, in the current configuration, the retarded Green's functions matrix has a very simple form
\begin{equation}
	\mathbf{G}_{R}(s,s')=-\frac{e^{2}}{2\pi}\,\theta(s-s')\,\delta'(s-s')\begin{pmatrix}1&0\\[8pt]0&1\end{pmatrix}\,,
\end{equation}
so the Langevin equation reduces to a purely local form
\begin{align}
	 m\,\ddot{\pmb{\chi}}(s)+2m\gamma\,\dot{\pmb{\chi}}(s)+m\,\pmb{\Omega}_{R}^{2}\cdot\pmb{\chi}(s)&=\pmb{\xi}(s)\,,\label{E:knkdnfks}
\end{align}
where $\pmb{\Omega}_{R}^{2}$ is obtained by absorbing the divergence of $\mathbf{G}_{R}(s,s')$ into the diagonal elements of the original $\pmb{\Omega}^{2}$, and $\gamma=e^{2}/8\pi m>0$. We immediately see that eq.~\eqref{E:knkdnfks} in fact describes nothing but a bunch of coupled, driven, damped oscillators. Thus the Langevin equation has a very intuitive interpretation \footnote{Start with two noninteracting oscillators, each interacts with its own thermal bath.  The system is described by two decoupled yet almost identical Langevin equations, the only difference is in the noises of the two baths at different temperatures. Now turn on the interaction between the two oscillators, then each Langevin equation should acquire an extra force term associated with the other oscillator's variable.}.

The general solution to \eqref{E:euwbnss} or \eqref{E:knkdnfks} can be expanded in terms of fundamental solution matrices $\mathbf{D}_{1}$ and $\mathbf{D}_{2}$. They are simply the homogeneous solutions of the corresponding equation of motion but satisfy a particular set of initial conditions,
\begin{align}
	\mathbf{D}_{1}(0)&=1\,,&\dot{\mathbf{D}}_{1}(0)&=0\,,\\
	\mathbf{D}_{2}(0)&=0\,,&\dot{\mathbf{D}_{2}}(0)&=1\,.
\end{align}	
Thus the general solution is given by
\begin{equation}
	 \pmb{\chi}(s)=\mathbf{D}_{1}(s)\cdot\pmb{\chi}(0)+\mathbf{D}_{2}(s)\cdot\dot{\pmb{\chi}}(0)+\frac{1}{m}\int_{0}^{s}\!ds'\;\mathbf{D}_{2}(s-s')\cdot\pmb{\xi}(s')\,.
\end{equation}
This can be the starting point to compute the variance of physical observables, their correlation functions or the variance of the conjugated variables. For example, the symmetrized correlation functions of $\pmb{\chi}$ (where the curly brackets below represent the anti-commutator) are given by
\begin{align}
	 \frac{1}{2}\langle\{\pmb{\chi}(t)\cdot\pmb{\chi}^{T}(t')\}\rangle&=\mathbf{D}_{1}(t)\cdot\langle\pmb{\chi}(0)\cdot\pmb{\chi}^{T}(0)\rangle\cdot\mathbf{D}_{1}(t')+\mathbf{D}_{2}(t)\cdot\langle\dot{\pmb{\chi}}(0)\cdot\dot{\pmb{\chi}}^{T}(0)\rangle\cdot\mathbf{D}_{2}(t')\notag\\
	 &+\frac{e^{2}}{m^{2}}\int_{0}^{t}\!ds\!\int_{0}^{t'}\!ds'\;\mathbf{D}_{2}(t-s)\cdot\mathbf{G}_{H}(s,s')\cdot\mathbf{D}_{2}(t'-s')\,,\label{E:mnxeruea}
\end{align}
if initially $\pmb{\chi}(0)$ and $\dot{\pmb{\chi}}(0)$ are not correlated. Notice there that our choice of the parameters $m$, $\sigma$, $e$ and $\omega$ renders the fundamental solution matrices symmetric, so we do not explicit show the transposition superscript in the place it is needed.

This is a good point to comment on the Langevin equation \eqref{E:euwbnss} and the derived results such as \eqref{E:mnxeruea}. Compared to the equation of motion of a closed system the Langevin equation describing the dynamics of a reduced system has two additional features, a stochastic forcing term (noise) on the RHS and a dissipative term on the LHS.  The noise term is a representation of certain measure of coarse-graining of the environment and the backaction of the coarse-grained environment manifests as dissipative dynamics of the reduced system. In the influence functional framework the Langevin equation is obtained by taking the functional variation of the stochastic effective action $S_{SE}$ about the mean trajectory $q\to0$ in the evolution of the reduced system. One may wonder whether in this approach the derivation of the Langevin equation accounts only for the induced quantum effects from the environment but overlooks the intrinsic quantum nature of the reduced system. This is because the homogeneous solution of the Langevin equation has no explicit dependence on the stochastic variable $\pmb{\xi}$ and thus insensitive to taking the noise distributional average defined by the probability functional $\mathcal{P}[\pmb{\xi}]$. If one writes the initial conditions as quantum operators of the canonical variables, one may identify the homogeneous part of the complete solution as the quantum operators associated with the reduced system, whose dissipative behavior will in most cases diminish while the reduced system relaxes in time. To accommodate both the quantum and the stochastic aspects we only need to extend the meaning  of the angular brackets $\langle\cdots\rangle$ to that of both taking the expectation value and the distributional average we can properly incorporate the intrinsic quantum nature of the reduced system and the noise effects, as demonstrated in \eqref{E:mnxeruea}. This approach works very nicely for the quantities which take on symmetric ordering, and the result is consistent with that computed by the reduced density matrix~\cite{HZH}.  It may become problematic if the quantities of interest take on a different ordering from the symmetric one.

Likewise we can find the variances pertinent to the reduced system in the NESS configurations by the Langevin equation. For example, the elements of the covariance matrix are given by
\begin{align}
	 \langle\,\Delta^{2}\pmb{\chi}_{b}^{(l)}\rangle&=\frac{e^{2}}{m^{2}}\Bigl[\int^{t}_{0}\!ds\,ds'\;\mathbf{D}_{2}(s)\cdot\mathbf{G}_{H}(s-s')\cdot\mathbf{D}_{2}(s')\Bigr]_{ll}\,,\label{E:cnmer1}\\
	 \langle\,\Delta^{2}\mathbf{p}_{b}^{(l)}\rangle&=e^{2}\Bigl[\int^{t}_{0}\!ds\,ds'\;\dot{\mathbf{D}}_{2}(s)\cdot\mathbf{G}_{H}(s-s')\cdot\dot{\mathbf{D}}_{2}(s')\Bigr]_{ll}\,,\\
	 \frac{1}{2}\,\langle\,\{\Delta\pmb{\chi}_{b}^{(l)},\Delta\mathbf{p}_{b}^{(l)}\}\rangle&=\frac{e^{2}}{m}\Bigl[\int^{t}_{0}\!ds\,ds'\;\mathbf{D}_{2}(s)\cdot\mathbf{G}_{H}(s-s')\cdot\dot{\mathbf{D}}_{2}(s')\Bigr]_{ll}\,,\label{E:cnmer3}
\end{align}
at late time $t\gg\gamma^{-1}$. The contributions from the homogenous solutions are transient, exponentially decaying with time, so they almost vanish on the evolution time scale greater than $\gamma^{-1}$.

In the limit $t\to\infty$, eqs.~\eqref{E:cnmer1}--\eqref{E:cnmer3} become
\begin{align}
	 \langle\,\Delta^{2}\pmb{\chi}_{b}^{(l)}\rangle&=\frac{e^{2}}{m^{2}}\,\Bigl[\int^{\infty}_{-\infty}\!\frac{d\omega}{2\pi}\;\widetilde{\pmb{\mathfrak{D}}}_{2}(\omega)\cdot\widetilde{\mathbf{G}}_{H}(\omega)\cdot\widetilde{\pmb{\mathfrak{D}}}_{2}^{*}(\omega)\Bigr]_{ll}\,,\label{E:euhaja}\\
	 \langle\,\Delta^{2}\mathbf{p}_{b}^{(l)}\rangle&=e^{2}\Bigl[\int^{\infty}_{-\infty}\!\frac{d\omega}{2\pi}\;\omega^{2}\,\widetilde{\pmb{\mathfrak{D}}}_{2}(\omega)\cdot\widetilde{\mathbf{G}}_{H}(\omega)\cdot\widetilde{\pmb{\mathfrak{D}}}_{2}^{*}(\omega)\Bigr]_{ll}\,,\label{E:fuhaja}\\
	\frac{1}{2}\,\langle\,\{\Delta\pmb{\chi}_{b}^{(l)},\Delta\mathbf{p}_{b}^{(l)}\}\rangle&=0\,,
\end{align}
where $\widetilde{\pmb{\mathfrak{D}}}_{2}(\omega)$ is given by
\begin{equation}\label{E:hjmswz}
	 \widetilde{\pmb{\mathfrak{D}}}_{2}(\omega)=\Bigl[-\omega^{2}\,\mathbf{I}+\pmb{\Omega}_{R}^{2}-i\,2\gamma\omega\,\mathbf{I}\Bigr]^{-1}\,,\qquad\qquad\pmb{\Omega}_{R}^{2}=\begin{pmatrix}
				\omega_{R}^{2}	&\sigma\\[12pt]
				\sigma		&\omega_{R}^{2}
				\end{pmatrix}\,.
\end{equation}
We see that its inverse Fourier transformation $\pmb{\mathfrak{D}}_{2}(s)$ is the kernel to the Langevin equation, that is, it satisfies \eqref{E:knkdnfks} with an impulse force described by a delta function $\delta(s)$. Furthermore, $\pmb{\mathfrak{D}}_{2}(s)$ is related to the fundamental solution matrix $\mathbf{D}_{2}(s)$ by $\pmb{\mathfrak{D}}_{2}(s)=\theta(s)\,\mathbf{D}_{2}(s)$.

The results in \eqref{E:euhaja} and \eqref{E:fuhaja} cannot be further simplified due to the fact there are two thermal baths with different temperature. The \textit{fluctuation-dissipation relation} in this case becomes a matrix relation
\begin{align}
	\widetilde{\mathbf{G}}_{H}(\omega)&=\begin{pmatrix}\coth\dfrac{\beta_{1}\omega}{2}	 &0\\[8pt]0&\coth\dfrac{\beta_{2}\omega}{2}\end{pmatrix}\cdot\operatorname{Im}\widetilde{\mathbf{G}}_{R}(\omega)\,,\\
\intertext{or in this case}
	\begin{pmatrix}
		\widetilde{G}_{H,\,\beta_{1}}(\omega)		&0\\[20pt]
		0						&\widetilde{G}_{H,\,\beta_{2}}(\omega)
	\end{pmatrix}&=\begin{pmatrix}\coth\dfrac{\beta_{1}\omega}{2}	 &0\\[8pt]0&\coth\dfrac{\beta_{2}\omega}{2}\end{pmatrix}\cdot\begin{pmatrix}
		\operatorname{Im}\widetilde{G}_{R}(\omega)	&0\\[20pt]
		0						&\operatorname{Im}\widetilde{G}_{R}(\omega)
	\end{pmatrix}\,.\notag
\end{align}
Note that each subsystem with its private thermal bath still has its own fluctuation-dissipation relation
\begin{equation}
	 \widetilde{G}_{H,\,\beta_{i}}(\omega)=\coth\dfrac{\beta_{i}\omega}{2}\,\operatorname{Im}\widetilde{G}_{R}(\omega)\,.
\end{equation}
Although the fluctuation-dissipation relation is diagonal, the matrix $\widetilde{\pmb{\mathfrak{D}}}_{2}(\omega)$ in \eqref{E:euhaja} and \eqref{E:fuhaja} will blend together the effects of both thermal baths. For example, let us examine $\langle\,\Delta^{2}\pmb{\chi}_{b}^{(1)}\rangle$. In the asymptotic future, it is given by
\begin{equation}
	 \langle\,\Delta^{2}\pmb{\chi}_{b}^{(1)}\rangle=\frac{1}{m^{2}}\int_{-\infty}^{\infty}\!\frac{d\omega}{2\pi}\;\biggl\{\bigl|\,\widetilde{\pmb{\mathfrak{D}}}^{11}_{2}(\omega)\,\bigr|^{2}\widetilde{\mathbf{G}}^{11}_{H}(\omega)+\bigl|\,\widetilde{\pmb{\mathfrak{D}}}^{12}_{2}(\omega)\,\bigr|^{2}\widetilde{\mathbf{G}}^{22}_{H}(\omega)\biggr\}
\end{equation}
as seen from \eqref{E:euhaja}. Physically it is not surprising since each oscillator's dynamics needs to reckon with the other oscillator and its bath, albeit indirectly, and thus is determined by both baths.

Alternatively, this feature can be seen from the dynamics of the normal modes of the reduced system that diagonalize $\pmb{\Omega}_{R}^{2}$. Let $\mathbf{v}_{1}$ and $\mathbf{v}_{2}$ be the eigenvectors of $\pmb{\Omega}_{R}^{2}$, with eigenvalues $\omega_{+}^{2}=\omega^{2}_{R}+\sigma$ and $\omega_{-}^{2}=\omega_{R}^{2}-\sigma$, respectively.  The $2\times2$-matrix $\mathbf{U}=(\mathbf{v}_{1},\mathbf{v}_{2})$ can be used to diagonalize the $\pmb{\Omega}_{R}^{2}$ matrix into
\begin{equation}
	\pmb{\Lambda}^{2}=\mathbf{U}^{T}\cdot\pmb{\Omega}_{R}^{2}\cdot\mathbf{U}=\begin{pmatrix}\omega_{+}^{2} &0\\[12pt]0&\omega_{-}^{2}\end{pmatrix}\,.
\end{equation}
Correspondingly,  $\mathbf{q}$ and $\mathbf{r}$ will be rotated by $\mathbf{U}$ to
\begin{equation}
	\pmb{\mathfrak{u}}=\mathbf{U}^{T}\cdot\mathbf{q}\,,\qquad\qquad\pmb{\mathfrak{v}}=\mathbf{U}^{T}\cdot\mathbf{r}\,.
\end{equation}
If both oscillators are fixed in space, as is the case for our investigation of the NESS, then the Green's functions looks much simpler because they don't have spatial dependence. Thus the new Green's function matrix $\pmb{\mathfrak{G}}$, transformed by $\mathbf{U}$, is related to the original one $\mathbf{G}$ by
\begin{align}
	 \pmb{\mathfrak{G}}_{R}(s,s')&=\mathbf{U}^{T}\cdot\mathbf{G}_{R}(s,s')\cdot\mathbf{U}\,,&\pmb{\mathfrak{G}}_{H}(s,s')=\mathbf{U}^{T}\cdot\mathbf{G}_{H}(s,s')\cdot\mathbf{U}\,.
\end{align}
Writing out the matrix $\pmb{\mathfrak{G}}$ explicitly, e.g., taking $\pmb{\mathfrak{G}}_{H}$ as an example, yields
\begin{align}\label{E:nvtisddt}
	\pmb{\mathfrak{G}}_{H}(s,s')&=\mathbf{U}^{T}\cdot\mathbf{G}_{H}(s,s')\cdot\mathbf{U}\notag\\
	&=\frac{e^{2}}{2}\begin{pmatrix}
				1	&1\\[12pt]
				1	&-1
				\end{pmatrix}\cdot\begin{pmatrix}
				G_{H,\,\beta_{1}}(s,s')	&0\\[12pt]
				0								&G_{H,\,\beta_{2}}(s,s')
				\end{pmatrix}\cdot\begin{pmatrix}
				1	&1\\[12pt]
				1	&-1
				\end{pmatrix}\notag\\
	&=\frac{e^{2}}{2}\begin{pmatrix}
				G_{H,\,\beta_{1}}(s,s')+G_{H,\,\beta_{2}}(s,s')	 &G_{H,\,\beta_{1}}(s,s')-G_{H,\,\beta_{2}}(s,s')\\[12pt]
				G_{H,\,\beta_{1}}(s,s')-G_{H,\,\beta_{2}}(s,s')	&G_{H,\,\beta_{1}}(s,s')+G_{H,\,\beta_{2}}(s,s')
				\end{pmatrix}\,.
\end{align}
Again we see that when we decompose the degrees of the freedom of the oscillators into their normal modes, the effects from both thermal baths are superposed.

For later reference, we explicitly write down the elements of the matrix $\widetilde{\pmb{\mathfrak{D}}}_{2}(\omega)$ as,
\begin{equation}\label{E:zdkjres}
	 \widetilde{\pmb{\mathfrak{D}}}_{2}(\omega)=\Bigl[-\omega^{2}\,\mathbf{I}+\pmb{\Omega}_{R}^{2}-i\,2\gamma\omega\,\mathbf{I}\Bigr]^{-1}=\begin{pmatrix}
	\dfrac{-\omega^{2}+\omega_{R}^{2}-i\,2\gamma\omega}{\det\,\widetilde{\pmb{\mathfrak{D}}}^{-1}_{2}(\omega)}	 &-\dfrac{\sigma}{\det\,\widetilde{\pmb{\mathfrak{D}}}^{-1}_{2}(\omega)}\\[16pt]
	-\dfrac{\sigma}{\det\,\widetilde{\pmb{\mathfrak{D}}}^{-1}_{2}(\omega)}	 &\dfrac{-\omega^{2}+\omega_{R}^{2}-i\,2\gamma\omega}{\det\,\widetilde{\pmb{\mathfrak{D}}}^{-1}_{2}(\omega)}\end{pmatrix}\,,
\end{equation}
with $\det\,\widetilde{\pmb{\mathfrak{D}}}^{-1}_{2}(\omega)=\bigl(\omega^{2}-\omega_{R}^{2}+i\,2\gamma\omega\bigr)^{2}-\sigma^{2}$. Thus the  variance? of $\chi^{(1)}$ takes the form
\begin{align}\label{E:qoksjn}
	&\quad\langle\,\Delta^{2}\pmb{\chi}_{b}^{(1)}\rangle\\
	 &=\frac{1}{m^{2}}\int_{-\infty}^{\infty}\!\frac{d\omega}{2\pi}\;\left\{\frac{\left(\omega^{2}-\omega^{2}_{R}\right)^{2}+4\gamma^{2}\omega^{2}}{\left[\left(\omega^{2}-\omega^{2}_{R}+\sigma\right)^{2}+4\gamma^{2}\omega^{2}\right]\left[\left(\omega^{2}-\omega^{2}_{R}-\sigma\right)^{2}+4\gamma^{2}\omega^{2}\right]}\,\widetilde{\mathbf{G}}^{11}_{H}(\omega)\right.\notag\\
	 &\qquad\qquad\qquad+\left.\frac{\sigma^{2}}{\left[\left(\omega^{2}-\omega^{2}_{R}+\sigma\right)^{2}+4\gamma^{2}\omega^{2}\right]\left[\left(\omega^{2}-\omega^{2}_{R}-\sigma\right)^{2}+4\gamma^{2}\omega^{2}\right]}\,\widetilde{\mathbf{G}}^{22}_{H}(\omega)\right\}\,,\notag
\end{align}
and
\begin{equation}
	\widetilde{\mathbf{G}}^{ii}_{H}(\omega)=\frac{\omega}{4\pi}\,\coth\frac{\beta_{i}\omega}{2}\,.
\end{equation}
Apparently at late time $t\to\infty$, the displacement variance of $O_{1}$ in \eqref{E:qoksjn} approaches a time-independent constant.

\subsection{Stochastic Reduced Density Matrix}

In the context of open quantum systems we mention two ways to obtain the desired physical quantities associated with the  dynamics of the reduced system: one is by way of the Langevin equation, which is less formal, more flexible and physically intuitive. It is particularly convenient if the quantities at hand involve noise either from the environment or externally introduced. The other is by way of the reduced density operator approach, which is easy to account for  the intrinsic quantum dynamics of the reduced system and to enforce the operator ordering. The drawback is that it is less straightforward to use this method to compute the expectation values of operators corresponding to physical variables which contain the environmental noise because the influence functional does not have explicit dependence on the noise. Only after invoking the Feynman-Vernon Gaussian integral identity would the noise of the environment, now in the form of a classical stochastic forcing term, be made explicit.  We will show in this section a way to combine the advantages of these two approaches, by incorporating the noise from the environment in the reduced density matrix, whose dynamical equation is obtained by taking the functional variation of the stochastic effective action.

Let us rewrite the reduced density matrix \eqref{E:fnjdsa} in terms of the stochastic effective action $S_{SE}$ in \eqref{E:hunxms},
\begin{align}
	 \rho_{\chi}(\mathbf{q}_{b},\mathbf{r}_{b};t)&=\int_{-\infty}^{\infty}\!d\mathbf{q}_{a}d\mathbf{r}_{a}\;\rho_{\chi}(\mathbf{q}_{a},\mathbf{r}_{a};0)\int_{\mathbf{q}_{a}}^{\mathbf{q}_{b}}\!\mathcal{D}\mathbf{q}\!\int_{\mathbf{r}_{a}}^{\mathbf{r}_{b}}\!\mathcal{D}\mathbf{r}\;\exp\Bigl\{i\,S_{CG}[\mathbf{q},\mathbf{r}]\Bigr\}\notag\\
	 &=\int_{-\infty}^{\infty}\!d\mathbf{q}_{a}d\mathbf{r}_{a}\;\rho_{\chi}(\mathbf{q}_{a},\mathbf{r}_{a};0)\int_{\mathbf{q}_{a}}^{\mathbf{q}_{b}}\!\mathcal{D}\mathbf{q}\!\int_{\mathbf{r}_{a}}^{\mathbf{r}_{b}}\!\mathcal{D}\mathbf{r}\int\mathcal{D}\pmb{\xi}\;\mathcal{P}[\pmb{\xi}]\,e^{i\,S_{SE}[\mathbf{q},\mathbf{r},\pmb{\xi}]}\notag\\
	&=\int\mathcal{D}\pmb{\xi}\;\mathcal{P}[\pmb{\xi}]\,\rho_{\chi}(\mathbf{q}_{b},\mathbf{r}_{b},t_{};\pmb{\xi}]\,,
\end{align}
The term in the integrand $\rho_{\chi}(\mathbf{q}_{b},\mathbf{r}_{b},t_{a};\pmb{\xi}]$ is called the stochastic reduced density matrix which is seen to have explicit dependence on the noise $\pmb{\xi}$ of the environment:
\begin{equation}\label{E:eucbma}
	 \rho_{\chi}(\mathbf{q}_{b},\mathbf{r}_{b},t_{b};\pmb{\xi}]=\int_{-\infty}^{\infty}\!d\mathbf{q}_{a}d\mathbf{r}_{a}\;\rho_{\chi}(\mathbf{q}_{a},\mathbf{r}_{a},t_{a})\left\{\int_{\mathbf{q}_{a}}^{\mathbf{q}_{b}}\!\mathcal{D}\mathbf{q}\!\int_{\mathbf{r}_{a}}^{\mathbf{r}_{b}}\!\mathcal{D}\mathbf{r}\right\}\;\,e^{i\,S_{SE}[\mathbf{q},\mathbf{r},\pmb{\xi}]}\,.
\end{equation}
with the stochastic effective action given by \eqref{E:hunxms},
\begin{align}
	 S_{SE}[\mathbf{q},\mathbf{r},\pmb{\xi}]&=\int_{0}^{t}\!ds\;\biggl\{m\,\dot{\mathbf{q}}^{T}(s)\cdot\dot{\mathbf{r}}(s)-m\,\mathbf{q}(s)\cdot\pmb{\Omega}^{2}\cdot\mathbf{r}(s)+\mathbf{q}^{T}(s)\cdot\pmb{\xi}(s)\biggr.\notag\\
	 &\qquad\qquad\qquad\qquad+\biggl.\int_{0}^{s}\!ds'\;\mathbf{q}^{T}(s)\cdot\mathbf{G}_{R}(s,s')\cdot\mathbf{r}(s')\biggr\}\,.
\end{align}
In this rendition, the reduced system, now driven by a classical stochastic force of the environment (by virtue of the Feynman-Vernon transform) as part of the influence from the environment, is described by the stochastic density matrix. For each realization of the environmental noise, the reduced system evolves to a state described by the density matrix \eqref{E:eucbma}. Different realizations make the system end up at different final states with probability given by $\mathcal{P}[\pmb{\xi}]$.

The reduced system is ostensibly non-conservative with the presence of friction and noise terms, and rightly so, as these effects originate from the interaction between the system and its environment, which in the case understudy consists of two baths. These two processes are, however, constrained by the fluctuation-dissipation relation associated with each bath. This relation plays a fundamental role in the energy flow balance between the system and the bath: fluctuations in the environment show up as noise and its backaction on the system gives rise to dissipative dynamics. How this relation bears on the problem of equilibration for a quantum system interacting with one bath is pretty well known.  We will show explicitly below how this relation underscores the approach to steady state for systems in nonequilibrium.

To compute the quantum and stochastic average of a dynamical variable, say,  $f(\pmb{\chi};\pmb{\xi}]$ at time $t$, which contains both the stochastic variable $\pmb{\xi}$ and the canonical variables $\pmb{\chi}$ of the reduced system, we simply evaluate the trace associated with the system variables and the ensemble average associated with the environmental noise,
\begin{equation}\label{E:mcbers}
	\langle f(\pmb{\chi};\pmb{\xi}]\rangle=\int\mathcal{D}\pmb{\xi}\;\mathcal{P}[\pmb{\xi}]\,\operatorname{Tr}_{\chi}\rho_{\chi}(t;\pmb{\xi}]\,f(\pmb{\chi};\pmb{\xi}]\,.
\end{equation}
The procedure in \eqref{E:mcbers} is as follows: for each specific realization of the stochastic source, we first calculate the expectation value of the quantum operator $f(\pmb{\chi};\pmb{\xi}]$ for the state described by the reduced density operator $\rho_{\chi}(t;\pmb{\xi}]$. The obtained result, still dependent on the stochastic variable, will  then be averaged over according to the probability distribution $\mathcal{P}[\pmb{\xi}]$ of the noise.

As an example, we will compute the power $P_{\xi_{1}}$ delivered by the stochastic force (noise)  $\xi_{1}$ from Bath 1 to Oscillator 1. The power $P_{\xi_{1}}$ is defined by
\begin{equation}
	P_{\xi_{1}}(t)=\langle\,\xi_{1}(t)\,\dot{\chi}^{(1)}(t)\rangle\,,
\end{equation}	
and observe that $p^{(1)}=m\,\dot{\chi}^{(1)}$. Thus we have
\begin{align}
	P_{\xi_{1}}(t)&=\frac{1}{m}\,\langle\,\xi_{1}(t)\,p^{(1)}(t)\rangle\notag\\
			 &=-\frac{i}{m}\int\mathcal{D}\pmb{\xi}\;\mathcal{P}[\pmb{\xi}]\,\int_{-\infty}^{\infty}\!d\mathbf{q}_{b}d\mathbf{r}_{b}\;\delta(\mathbf{q}_{b})\,\xi_{1}(t)\,\frac{\partial}{\partial\chi^{(1)}_{b}}\,\rho_{\chi}(\mathbf{q}_{b},\mathbf{r}_{b},t;\pmb{\xi})\,,\label{E:tnbrex}
\end{align}
where the momentum $p^{(i)}$ canonical to the coordinate $\chi^{(i)}$ is given by
\begin{equation}
	p^{(i)}=-i\,\frac{\partial}{\partial\chi^{(i)}}\,,
\end{equation}
and the trace over the dynamical variables of the reduced system is defined as
\begin{equation}
	\operatorname{Tr}_{\chi}=\int_{-\infty}^{\infty}\!d\mathbf{q}_{b}d\mathbf{r}_{b}\;\delta(\mathbf{q}_{b})\,.
\end{equation}
Since the initial state of the reduced system is a Gaussian state and the stochastic effective action is quadratic in the system's variables, the final state will remain Gaussian and the corresponding reduced density operator thus can be evaluated exactly. To derive the explicit form of the reduced density matrix, we first evaluate the path integrals in \eqref{E:eucbma},
\begin{align}
	 &\quad\int_{\mathbf{q}_{a}}^{\mathbf{q}_{b}}\!\mathcal{D}\mathbf{q}\!\int_{\mathbf{r}_{a}}^{\mathbf{r}_{b}}\!\mathcal{D}\mathbf{r}\;\exp\left\{i\int_{0}^{t}\!ds\;\biggl[m\,\dot{\mathbf{q}}^{T}(s)\cdot\dot{\mathbf{r}}(s)-m\,\mathbf{q}(s)\cdot\pmb{\Omega}^{2}\cdot\mathbf{r}(s)+\mathbf{q}^{T}(s)\cdot\pmb{\xi}(s)\biggr.\right.\notag\\
	 &\qquad\qquad\qquad\qquad\qquad\qquad\qquad\qquad+\left.\biggl.\int_{0}^{s}\!ds'\;\mathbf{q}^{T}(s)\cdot\mathbf{G}_{R}(s,s')\cdot\mathbf{r}(s')\biggr]\right\}\notag\\
	 &=\mathcal{N}\,\exp\biggl[i\,m\,\mathbf{q}_{b}^{T}\cdot\dot{\overline{\mathbf{r}}}_{b}^{\vphantom{T}}-i\,m\,\mathbf{q}^{T}_{a}\cdot\dot{\overline{\mathbf{r}}}_{a}^{\vphantom{T}}\biggr]\,,
\end{align}
where $\mathcal{N}$ is the normalization constant, which can be determined by the unitarity requirement. It is given by
\begin{equation}
	\mathcal{N}=\left(\frac{m}{2\pi}\right)^{2}\det\dot{\pmb{\mu}}(0)\,.
\end{equation}
Note that the mean trajectories $\overline{\mathbf{q}}$, $\overline{\mathbf{r}}$ are solutions to the stochastic Langevin equation \eqref{E:knkdnfks} with the boundary conditions $\overline{\mathbf{q}}(t)=\mathbf{q}_{b}$, $\overline{\mathbf{q}}(0)=\mathbf{q}_{a}$ and $\overline{\mathbf{r}}(t)=\mathbf{r}_{b}$, $\overline{\mathbf{r}}(0)=\mathbf{r}_{a}$. Thus they and their time derivatives are functionals of the stochastic noise $\pmb{\xi}$. Explicitly, in terms of the boundary values, we can write $\overline{\mathbf{r}}(s)$ as
\begin{equation}
	 \overline{\mathbf{r}}(s)=\pmb{\nu}(s)\cdot\mathbf{r}_{a}+\pmb{\mu}(s)\cdot\mathbf{r}_{b}+\pmb{\mathcal{J}}_{r}(s)\,,
\end{equation}
for $0\leq s\leq t$. The functions $\pmb{\mu}(s)$, $\pmb{\nu}(s)$ are defined by
\begin{align}
	\pmb{\mu}(s)&=\mathbf{D}_{2}(s)\cdot\mathbf{D}^{-1}_{2}(t)\,,\label{E:zzz1}\\
	 \pmb{\nu}(s)&=\mathbf{D}_{1}(s)-\mathbf{D}_{2}(s)\cdot\mathbf{D}^{-1}_{2}(t)\cdot\mathbf{D}_{1}(t)\,,\label{E:zzz2}
\end{align}
and the current $\pmb{\mathcal{J}}_{r}(s)$ is given by
\begin{equation}
	 \pmb{\mathcal{J}}_{r}(s)=\frac{1}{m}\int_{0}^{s}\!ds'\;\mathbf{D}_{2}(s-s')\cdot\pmb{\xi}(s')-\frac{1}{m}\int_{0}^{t}\!ds'\;\mathbf{D}_{2}(s)\cdot\mathbf{D}^{-1}_{2}(t)\cdot\mathbf{D}_{2}(t-s')\cdot\pmb{\xi}(s')\,.
\end{equation}
Additionally, we can write the partial derivative $\partial/\partial\chi$ as
\begin{equation}
	\frac{\partial}{\partial\chi_{b}^{(1)}}=\frac{\partial}{\partial q_{b}^{(1)}}+\frac{1}{2}\frac{\partial}{\partial r_{b}^{(1)}}\,.
\end{equation}
Here as an example of the stochastic reduced density matrix approach, we will provide greater details in  the derivation of the power delivered by the stochastic force $\pmb{\xi}_{1}$ on $O_{1}$.  With these, \eqref{E:tnbrex} becomes
\begin{align}
	 P_{\xi_{1}}(t)&=-\frac{i}{m}\,\mathcal{N}\int\mathcal{D}\pmb{\xi}\;\mathcal{P}[\pmb{\xi}]\,\int_{-\infty}^{\infty}\!d\mathbf{q}_{b}\,d\mathbf{r}_{b}\;\delta(\mathbf{q}_{b})\int_{-\infty}^{\infty}\!d\mathbf{q}_{a}d\mathbf{r}_{a}\;\rho(\mathbf{q}_{a},\mathbf{r}_{a},0)\,\xi_{1}(t)\notag\\
	&\qquad\qquad\qquad\qquad\qquad\times\left[\frac{\partial}{\partial q_{b}^{(1)}}+\frac{1}{2}\frac{\partial}{\partial r_{b}^{(1)}}\right]\exp\biggl[i\,m\,\mathbf{q}_{b}^{T}\cdot\dot{\overline{\mathbf{r}}}_{b}^{\vphantom{T}}-i\,m\,\mathbf{q}^{T}_{a}\cdot\dot{\overline{\mathbf{r}}}_{a}^{\vphantom{T}}\biggr]\notag\\
	 &=-\frac{i}{m}\,\mathcal{N}\int\mathcal{D}\pmb{\xi}\;\mathcal{P}[\pmb{\xi}]\,\xi_{1}(t)\int_{-\infty}^{\infty}\!d\mathbf{q}_{b}\,d\mathbf{r}_{b}\;\delta(\mathbf{q}_{b})\int_{-\infty}^{\infty}\!d\mathbf{q}_{a}d\mathbf{r}_{a}\;\rho(\mathbf{q}_{a},\mathbf{r}_{a},0)\notag\\
	 &\times\biggl\{i\,m\,\dot{\overline{r}}_{b}^{(1)}+\frac{i\,m}{2}\biggl[\mathbf{q}_{b}^{T}\cdot\dot{\pmb{\mu}}(t)-\mathbf{q}_{a}^{T}\cdot\dot{\pmb{\mu}}(0)\biggr]_{11}\biggr\}\exp\biggl[i\,m\,\mathbf{q}_{b}^{T}\cdot\dot{\overline{\mathbf{r}}}_{b}^{\vphantom{T}}-i\,m\,\mathbf{q}^{T}_{a}\cdot\dot{\overline{\mathbf{r}}}_{a}^{\vphantom{T}}\biggr]\notag\\
	 &=\mathcal{N}\int\mathcal{D}\pmb{\xi}\;\mathcal{P}[\pmb{\xi}]\,\xi_{1}(t)\int_{-\infty}^{\infty}\!d\mathbf{r}_{b}\!\int_{-\infty}^{\infty}\!d\mathbf{q}_{a}d\mathbf{r}_{a}\;\rho(\mathbf{q}_{a},\mathbf{r}_{a},0)\notag\\
	 &\qquad\qquad\qquad\times\exp\biggl\{-i\,m\,\mathbf{q}^{T}_{a}\cdot\biggl[\dot{\pmb{\nu}}(0)\cdot\mathbf{r}_{a}+\dot{\pmb{\mu}}(0)\cdot\mathbf{r}_{b}+\dot{\pmb{\mathcal{J}}}_{r}(0)\biggr]\biggr\}\notag\\
	 &\qquad\qquad\qquad\qquad\qquad\times\biggl[\mathbf{r}_{a}^{T}\cdot\dot{\pmb{\nu}}^{T}(t)+\mathbf{r}_{b}^{T}\cdot\dot{\pmb{\mu}}^{T}(t)+\dot{\pmb{\mathcal{J}}}^{T}_{r}(s)-\frac{1}{2}\,\mathbf{q}_{a}^{T}\cdot\dot{\pmb{\mu}}(0)\biggr]_{11}\notag\\
	 &=\mathcal{N}\int\mathcal{D}\pmb{\xi}\;\mathcal{P}[\pmb{\xi}]\,\xi_{1}(t)\int_{-\infty}^{\infty}\!d\mathbf{r}_{b}\!\int_{-\infty}^{\infty}\!d\mathbf{q}_{a}d\mathbf{r}_{a}\;\rho(\mathbf{q}_{a},\mathbf{r}_{a},0)\notag\\
	 &\qquad\times\biggl[\dot{\pmb{\nu}}_{1m}(t)\,r_{a}^{(m)}+\dot{\mathcal{J}}_{r}^{(1)}(t)-\frac{1}{2}\,\dot{\pmb{\mu}}^{T}_{1m}(0)\,q_{a}^{(m)}+\frac{i}{m}\,\dot{\pmb{\mu}}_{1m}(t)\dot{\pmb{\mu}}^{-1}_{mn}(0)\,\frac{\partial}{\partial q_{a}^{(n)}}\biggr]\notag\\
	 &\qquad\qquad\qquad\qquad\qquad\times\exp\biggl\{-i\,m\,\mathbf{q}^{T}_{a}\cdot\biggl[\dot{\pmb{\nu}}(0)\cdot\mathbf{r}_{a}+\dot{\pmb{\mu}}(0)\cdot\mathbf{r}_{b}+\dot{\pmb{\mathcal{J}}}_{r}(0)\biggr]\biggr\}\notag\\
	 &=\mathcal{N}\left(\frac{2\pi}{m}\right)^{2}\left(\frac{1}{\pi\varsigma^{2}}\right)^{\frac{2}{2}}\int\mathcal{D}\pmb{\xi}\;\mathcal{P}[\pmb{\xi}]\,\xi_{1}(t)\int_{-\infty}^{\infty}\!d\mathbf{q}_{a}d\mathbf{r}_{a}\;\exp\biggl[-\frac{\mathbf{r}_{a}^{T}\cdot\mathbf{r}_{a}^{\vphantom{T}}}{\varsigma^{2}}-\frac{\mathbf{q}_{a}^{T}\cdot\mathbf{q}_{a}^{\vphantom{T}}}{4\varsigma^{2}}\biggr]\notag\\
	 &\qquad\qquad\qquad\qquad\qquad\qquad\times\exp\biggl\{-i\,m\,\mathbf{q}^{T}_{a}\cdot\biggl[\dot{\pmb{\nu}}(0)\cdot\mathbf{r}_{a}+\dot{\pmb{\mathcal{J}}}_{r}(0)\biggr]\biggr\}\notag\\
	 &\times\biggl[\dot{\pmb{\nu}}_{1m}(t)\,r_{a}^{(m)}+\dot{\mathcal{J}}_{r}^{(1)}(t)-\frac{1}{2}\,\dot{\pmb{\mu}}^{T}_{1m}(0)\,q_{a}^{(m)}+\frac{i}{m}\,\dot{\pmb{\mu}}_{1m}(t)\dot{\pmb{\mu}}^{-1}_{mn}(0)\,\frac{\partial}{\partial q_{a}^{(n)}}\biggr]\,\delta^{(2)}\bigl[\mathbf{q}^{T}_{a}\cdot\dot{\pmb{\mu}}(0)\bigr]\notag\\
	 &=\frac{\mathcal{N}}{\det\dot{\pmb{\mu}}(0)}\left(\frac{2\pi}{m}\right)^{2}\left(\frac{1}{\pi\varsigma^{2}}\right)^{\frac{2}{2}}\int\mathcal{D}\pmb{\xi}\;\mathcal{P}[\pmb{\xi}]\,\xi_{1}(t)\int_{-\infty}^{\infty}\!d\mathbf{q}_{a}d\mathbf{r}_{a}\;\delta^{(2)}\bigl(\mathbf{q}_{a}\bigr)\notag\\
	 &\qquad\exp\biggl[-\frac{\mathbf{r}_{a}^{T}\cdot\mathbf{r}_{a}^{\vphantom{T}}}{\varsigma^{2}}-\frac{\mathbf{q}_{a}^{T}\cdot\mathbf{q}_{a}^{\vphantom{T}}}{4\varsigma^{2}}\biggr]\times\exp\biggl\{-i\,m\,\mathbf{q}^{T}_{a}\cdot\biggl[\dot{\pmb{\nu}}(0)\cdot\mathbf{r}_{a}+\dot{\pmb{\mathcal{J}}}_{r}(0)\biggr]\biggr\}\notag\\
	 &\qquad\times\biggl\{\dot{\pmb{\nu}}(t)\cdot\mathbf{r}_{a}+\dot{\pmb{\mathcal{J}}}_{r}(t)-\frac{1}{2}\,\dot{\pmb{\mu}}^{T}(0)\cdot\mathbf{q}_{a}\biggr.\notag\\
	 &\qquad\qquad\quad-\biggl.\frac{i}{m}\,\dot{\pmb{\mu}}(t)\cdot\dot{\pmb{\mu}}^{-1}(0)\cdot\Bigl(-i\,m\Bigl[\dot{\pmb{\nu}}(0)\cdot\mathbf{r}_{a}+\dot{\pmb{\mathcal{J}}}_{r}(0)\Bigr]-\frac{1}{2\varsigma^{2}}\,\mathbf{q}_{a}\Bigr)\biggr\}_{11}\notag\\
	 &=\frac{\mathcal{N}}{\det\dot{\pmb{\mu}}(0)}\left(\frac{2\pi}{m}\right)^{2}\left(\frac{1}{\pi\varsigma^{2}}\right)^{\frac{2}{2}}\int\mathcal{D}\pmb{\xi}\;\mathcal{P}[\pmb{\xi}]\,\xi_{1}(t)\int_{-\infty}^{\infty}\!d\mathbf{r}_{a}\;\exp\biggl[-\frac{\mathbf{r}_{a}^{T}\cdot\mathbf{r}_{a}^{\vphantom{T}}}{\varsigma^{2}}\biggr]\notag\\
	 &\qquad\qquad\qquad\qquad\qquad\qquad\qquad\quad\;\times\Bigl[\dot{\pmb{\mathcal{J}}}_{r}(t)-\dot{\pmb{\mu}}(t)\cdot\dot{\pmb{\mu}}^{-1}(0)\cdot\dot{\pmb{\mathcal{J}}}_{r}(0)\Bigr]_{11}\notag\\
	 &=\frac{\mathcal{N}}{\det\dot{\pmb{\mu}}(0)}\left(\frac{2\pi}{m}\right)^{2}\int\mathcal{D}\pmb{\xi}\;\mathcal{P}[\pmb{\xi}]\;\xi_{1}(t)\Bigl[\dot{\pmb{\mathcal{J}}}_{r}(t)-\dot{\pmb{\mu}}(t)\cdot\dot{\pmb{\mu}}^{-1}(0)\cdot\dot{\pmb{\mathcal{J}}}_{r}(0)\Bigr]_{11}\,.
\end{align}
The expressions in the square brackets can be reduced to
\begin{align}
	 \dot{\pmb{\mathcal{J}}}_{r}(t)-\dot{\pmb{\mu}}(t)\cdot\dot{\pmb{\mu}}^{-1}(0)\cdot\dot{\pmb{\mathcal{J}}}_{r}(0)&=\frac{1}{m}\int_{0}^{t}\!ds'\;\dot{\mathbf{D}}_{2}(t-s')\cdot\pmb{\xi}(s')\,.
\end{align}
Thus the power delivered to Oscillator 1 from Bath 1 is equal to
\begin{align}
	 P_{\xi_{1}}(t)&=\frac{1}{m}\int\mathcal{D}\pmb{\xi}\;\mathcal{P}[\pmb{\xi}]\;\xi_{1}(t)\int_{0}^{t}\!ds'\;\dot{\mathbf{D}}_{2}^{1m}(t-s')\,\xi_{m}(s')\notag\\
	 &=\frac{1}{m}\int_{0}^{t}\!ds'\;\dot{\mathbf{D}}_{2}^{1m}(t-s')\int\mathcal{D}\pmb{\xi}\;\mathcal{P}[\pmb{\xi}]\;\xi_{1}(t)\xi_{m}(s')\notag\\
	&=\frac{e^{2}}{m}\int_{0}^{t}\!ds'\;\dot{\mathbf{D}}^{1m}_{2}(t-s')\,\mathbf{G}_{H}^{1m}(t-s')\,.\label{E:bvnorue}
\end{align}
We will see that this is exactly the same as \eqref{E:oeirncs} except that \eqref{E:oeirncs} is expressed in terms of the normal modes. Alternatively, we can compare \eqref{E:bvnorue} with \eqref{E:lernddf}. They are the same.



\section{Energy Transport, Power Balance and Stationarity Condition}\label{S:reeih4}

As an important application of the formalism developed so far we examine in this section how energy is transported in the combined system ${\bf S}$ of two oscillators with two baths, in the nature of heat flux, to see whether there is any build-up or localization of energy (the answer is no), or whether there is a balance in the energy flow which signifies the existence of a nonequilibrium steady state (the answer is yes, with several power balance relations).

\subsection{Energy Flow between Components}

To study the energy transport in the system, as we mentioned in the last section, it is physically more transparent to use the Langevin equations \eqref{E:knkdnfks} instead of the matrix form we obtained in the previous section.  In this section we will illustrate this approach by deriving the Langevin equations for each subsystem, then analyze the heat transfer and energy flux balance relations from them.
\begin{align}
	 m\,\ddot{\chi}^{(1)}(s)+2m\gamma\,\dot{\chi}^{(1)}(s)+m\omega_{R}^{2}\,\chi^{(1)}(s)+m\sigma\,\chi^{(2)}(s)&=\xi^{(1)}(s)\,,\label{E:hgwieywia}\\
	 m\,\ddot{\chi}^{(2)}(s)+2m\gamma\,\dot{\chi}^{(2)}(s)+m\omega_{R}^{2}\,\chi^{(2)}(s)+m\sigma\,\chi^{(1)}(s)&=\xi^{(2)}(s)\,.\label{E:igwieywia}
\end{align}
Let's look at the physics from these equations before delving into the calculations. The motion of any harmonic oscillator in the system, say,  $O_{1}$, is always affected by its neighboring oscillator(s), Oscillator 2 in the present two oscillator case, via their mutual coupling with strength $m\sigma$.  $O_{1}$ is also driven into random motion by a stochastic force $\xi^{(1)}$  associated with quantum and thermal fluctuations of Bath 1. Fluctuations in Bath 1, described here by a scalar-field, induces a dissipative force (in general, it is a reactive force)  on $O_{1}$. Thus the harmonic oscillator $O_{1}$, other than its own harmonic force, is simultaneously acted on by these three seemingly unrelated forces. Certain correlations will be established among them over time. We will examine the interplay among these forces and their roles in energy transport, the power they deliver and the possible existence of power balance relations, which provide the conditions for the establishment of a nonequilibrium steady state.

First we will find the normal modes of the coupled motion \eqref{E:hgwieywia}--\eqref{E:igwieywia}. By an appropriate linear combination of the original dynamical variables $\chi^{(1)}$, $\chi^{(2)}$
\begin{align}
	\chi_{+}&=[\chi^{(1)}+\chi^{(2)}]/2\,, &\chi_{-}&=\chi^{(1)}-\chi^{(2)}\,,
\end{align}
we decouple the motions of $O_{1}$ and $O_{2}$ and arrive at
\begin{align}
	 m\,\ddot{\chi}_{+}(s)+2m\gamma\,\dot{\chi}_{+}(s)+m\,\omega_{+}^{2}\chi_{+}(s)&=\frac{1}{2}\,\bigl[\xi^{(1)}(s)+\xi^{(2)}(s)\bigr]=\xi_{+}(s)\,,\label{E:keuskjfw}\\
	 m\,\ddot{\chi}_{-}(s)+2m\gamma\,\dot{\chi}_{-}(s)+m\,\omega_{-}^{2}\chi_{-}(s)&=\xi^{(1)}(s)-\xi^{(2)}(s)=\xi_{-}(s)\,\label{E:leuskjfw}
\end{align}
where $\omega_{\pm}^{2}=\omega_{R}^{2}\pm\sigma$. These are the normal modes of the coupled dynamics, and they act as two independent driven, damped oscillators. Here we require $\sigma<\omega_{R}^{2}$ to avoid any instability in the evolution of the normal modes. Assume the fundamental solutions to \eqref{E:keuskjfw} and \eqref{E:leuskjfw} are  given by $d_{i}^{(+)}(s)$, $d_{i}^{(-)}(s)$ respectively and satisfy the initial conditions,
\begin{align}
	d_{1}^{(+)}(0)&=1\,,&\dot{d}_{1}^{(+)}(0)&=0\,,&d_{2}^{(+)}(0)&=0\,,&\dot{d}_{2}^{(+)}(0)&=1\,,\\
	d_{1}^{(-)}(0)&=1\,,&\dot{d}_{1}^{(-)}(0)&=0\,,&d_{2}^{(-)}(0)&=0\,,&\dot{d}_{2}^{(-)}(0)&=1\,.
\end{align}
Thus the full solutions to the Langevin equations \eqref{E:keuskjfw} and \eqref{E:leuskjfw} are given by
\begin{align}
	 \chi_{+}(s)&=d_{1}^{(+)}(s)\chi_{+}(0)+d_{2}^{(+)}(s)\dot{\chi}_{+}(0)+\frac{1}{m}\int_{0}^{s}\!ds'\;d_{2}^{(+)}(s-s')\xi_{+}(s')\,,\\
	 \chi_{-}(s)&=d_{1}^{(-)}(s)\chi_{-}(0)+d_{2}^{(-)}(s)\dot{\chi}_{-}(0)+\frac{1}{m}\int_{0}^{s}\!ds'\;d_{2}^{(-)}(s-s')\xi_{-}(s')\,.
\end{align}
The corresponding solutions to $\chi^{(1)}(s)$ and $\chi^{(2)}(s)$ can be obtained by the superposition of the normal modes,
\begin{align}
	\chi^{(1)}(s)&=\chi_{+}(s)+\frac{1}{2}\,\chi_{-}(s)\,,\\
	\chi^{(2)}(s)&=\chi_{+}(s)-\frac{1}{2}\,\chi_{-}(s)\,,
\end{align}
such that
\begin{align}
	 \chi^{(1)}(s)&=\frac{1}{2}\Bigl[d_{1}^{(+)}(s)+d_{1}^{(-)}(s)\Bigr]\chi^{(1)}(0)+\frac{1}{2}\Bigl[d_{1}^{(+)}(s)-d_{1}^{(-)}(s)\Bigr]\chi^{(2)}(0)\notag\\
	 &\qquad\qquad+\frac{1}{2}\Bigl[d_{2}^{(+)}(s)+d_{2}^{(-)}(s)\Bigr]\dot{\chi}^{(1)}(0)+\frac{1}{2}\Bigl[d_{2}^{(+)}(s)-d_{2}^{(-)}(s)\Bigr]\dot{\chi}^{(2)}(0)\notag\\
	&\qquad\qquad+\frac{1}{2m}\int^{s}_{0}\!ds'\,\Bigl[d_{2}^{(+)}(s-s')+d_{2}^{(-)}(s-s')\Bigr]\xi_{1}(s')\notag\\
	&\qquad\qquad\qquad+\frac{1}{2m}\int^{s}_{0}\!ds'\,\Bigl[d_{2}^{(+)}(s-s')-d_{2}^{(-)}(s-s')\Bigr]\xi_{2}(s')\,,
\intertext{and likewise}
	 \chi^{(2)}(s)&=\frac{1}{2}\Bigl[d_{1}^{(+)}(s)-d_{1}^{(-)}(s)\Bigr]\chi^{(1)}(0)+\frac{1}{2}\Bigl[d_{1}^{(+)}(s)+d_{1}^{(-)}(s)\Bigr]\chi^{(2)}(0)\notag\\
	 &\qquad\qquad+\frac{1}{2}\Bigl[d_{2}^{(+)}(s)-d_{2}^{(-)}(s)\Bigr]\dot{\chi}^{(1)}(0)+\frac{1}{2}\Bigl[d_{2}^{(+)}(s)+d_{2}^{(-)}(s)\Bigr]\dot{\chi}^{(2)}(0)\notag\\
	&\qquad\qquad+\frac{1}{2m}\int^{s}_{0}\!ds'\,\Bigl[d_{2}^{(+)}(s-s')-d_{2}^{(-)}(s-s')\Bigr]\xi_{1}(s')\notag\\
	&\qquad\qquad\qquad+\frac{1}{2m}\int^{s}_{0}\!ds'\,\Bigl[d_{2}^{(+)}(s-s')+d_{2}^{(-)}(s-s')\Bigr]\xi_{2}(s')\,.
\end{align}
These are nothing but the superposition of normal modes in a tethered motion. Now we are ready to compute the power or energy flow or heat transfer between subsystems. As we have stressed before, we do not \textit{a priori assume} the existence of NESS in this system.  The energy flow between the neighboring components of the total system is not necessarily time-independent, let alone having the same magnitude. Rather, we seek to \textit{demonstrate the presence} of a steady energy flow from one bath to the other after the systems is fully relaxed on a time scale $t\gg\gamma^{-1}$.

\subsubsection{Energy Flow between $B_{1}$ and $S_{1}$}

The interactions between Subsystem 1 and Bath 1 are summarized in the stochastic force $\xi_{1}$ and the dissipative self-force of $-2m\gamma\,\dot{\chi}^{(1)}$, after we coarse-grained the degrees of freedom of $B_{1}$. These two forces mediate the energy flow between $S_{1}$ and $B_{1}$.

The average power delivered to Subsystem 1 by the stochastic force (noise) of Bath 1  is given by
\begin{align}\label{E:oeirncs}
	 P_{\xi_{1}}(t)=\langle\xi_{1}(t)\dot{\chi}^{(1)}(t)\rangle&=\frac{1}{2m^{2}}\int^{t}_{0}\!ds\,\Bigl[\dot{d}_{2}^{(+)}(t-s)+\dot{d}_{2}^{(-)}(t-s)\Bigr]\langle\,\xi_{1}(t)\xi_{1}(s)\,\rangle\notag\\
	 &=\frac{e^{2}}{2m}\int^{t}_{0}\!ds\,\Bigl[\dot{d}_{2}^{(+)}(t-s)+\dot{d}_{2}^{(-)}(t-s)\Bigr]\,G_{H}^{\beta_{1}}(t-s)\notag\\
	 y=t-s\quad&=\frac{e^{2}}{2m^{}}\int^{t}_{0}\!dy\,\Bigl[\dot{d}_{2}^{(+)}(y)+\dot{d}_{2}^{(-)}(y)\Bigr]\,G_{H}^{\beta_{1}}(y)\,,
\end{align}
with $\langle\,\xi_{1}(t)\xi_{1}(s)\,\rangle=e^{2}\,G_{H}^{\beta_{1}}(t-s)$. It tells us the rate at which the energy is transported to Subsystem 1 from Bath 1 by means of the stochastic noise.

Since we are particular interested in the existence of NESS, we will pay special attention to the late-time behavior of the energy transport. In the limit $t\to\infty$ when the motion of Subsystem 1 is fully relaxed and noting that the fundamental solutions $d_{i}(s)=0$ if $s<0$, we write this average power as
\begin{align}
	 P_{\xi_{1}}(\infty)&=\frac{e^{2}}{2m^{}}\int^{\infty}_{-\infty}\!dy\,\Bigl[\dot{d}_{2}^{(+)}(y)+\dot{d}_{2}^{(-)}(y)\Bigr]\,G_{H}^{\beta_{1}}(y)\notag\\
	 &=4\pi\gamma\int^{\infty}_{-\infty}\frac{d\omega}{2\pi}\;-i\,\omega\Bigl[\widetilde{d}_{2}^{(+)}(\omega)+\widetilde{d}_{2}^{(-)}(\omega)\Bigr]\,\widetilde{G}_{H}^{\beta_{1}}(\omega)\,,\label{E:hdswe}
\end{align}
where $\gamma=e^{2}/8\pi m$, and we have defined the Fourier transformation as
\begin{equation}
	f(t)=\int^{\infty}_{-\infty}\frac{d\omega}{2\pi}\;\tilde{f}(\omega)\,e^{-i\,\omega t}\,,\qquad\Leftrightarrow\qquad\widetilde{f}(\omega)=\int^{\infty}_{-\infty}\,dt\;f(t)\,e^{+i\,\omega t}\,,
\end{equation}
so that the convolution integrals are given by
\begin{align}
	 \int_{-\infty}^{\infty}\!dt\;f(t)g(t)&=\int_{-\infty}^{\infty}\frac{d\omega}{2\pi}\;\widetilde{f}(\omega)\widetilde{g}(-\omega)\,,\\
	 \int_{-\infty}^{\infty}\!dt\!\int_{-\infty}^{\infty}\!dt'\;f(t)g(t')h(t-t')&=\int_{-\infty}^{\infty}\frac{d\omega}{2\pi}\;\widetilde{f}(\omega)\widetilde{g}(-\omega)\widetilde{h}(\omega)\,.\label{E:ncerios}
\end{align}
The Fourier transforms $\widetilde{d}_{2}^{(+)}(\omega)$, $\widetilde{d}_{2}^{(+)}(\omega)$, and $\widetilde{G}_{H}^{\beta_{1}}(\omega)$ are
\begin{align}
	\widetilde{d}_{2}^{(\pm)}(\omega)&=\frac{1}{\omega_{\pm}^{2}-\omega^{2}-i\,2\gamma\omega}\,,\label{E:ddddd}\\
	 \widetilde{G}_{H}^{\beta}(\omega)&=\coth\frac{\beta\omega}{2}\operatorname{Im}\widetilde{G}_{R}(\omega)=\frac{\omega}{4\pi}\,\coth\frac{\beta\omega}{2}\,.
\end{align}
Here and henceforth, we will not make distinction between between $\pmb{\mathfrak{D}}_{i}(t)$ and $\mathbf{D}_{i}(t)$, as in \eqref{E:hjmswz}. They differs only by an unit-step function $\theta(t)$, that is, $\pmb{\mathfrak{D}}_{i}(t)=\theta(t)\,\mathbf{D}_{i}(t)$. Mathematically speaking they are totally different in nature; the formal is the inhomogeneous solution to the Langevin equation while the latter is the homogeneous solution with a special set of initial conditions. However, in practice, they serve the same purpose to the current case. Thus when we refer to the fundamental solution, we use the notations $\mathbf{D}_{i}(t)$ or $d_{i}(t)$ for both cases unless mentioned otherwise.

The power done by the \textit{dissipative force} $-2m\gamma\dot{\chi}^{(1)}$ of Subsystem  1 is
\begin{align}
	 P_{\gamma_{1}}(t)&=-2m\gamma\langle\,\dot{\chi}^{(1)\,2}(t)\,\rangle=-4\pi\gamma^{2}\int^{t}_{0}\!ds\!\int^{t}_{0}\!ds'\\
	 &\qquad\;\biggl\{\Bigl[\dot{d}_{2}^{(+)}(t-s)+\dot{d}_{2}^{(-)}(t-s)\Bigr]\Bigl[\dot{d}_{2}^{(+)}(t-s')+\dot{d}_{2}^{(-)}(t-s')\Bigr]\,G_{H}^{\beta_{1}}(s-s')\biggr.\notag\\
	 &\qquad+\biggl.\Bigl[\dot{d}_{2}^{(+)}(t-s)-\dot{d}_{2}^{(-)}(t-s)\Bigr]\Bigl[\dot{d}_{2}^{(+)}(t-s')-\dot{d}_{2}^{(-)}(t-s')\Bigr]\,G_{H}^{\beta_{2}}(s-s')\biggr\}\,.\notag
\end{align}
Here we have ignored the contributions independent of the baths, which will become exponentially negligible after the subsystems are fully relaxed. We also assumed that both baths are independent of each other and the initial states of both subsystems are not correlated with either bath. We see right away that the dissipation power already depends on both reservoirs; the connection of System 1 with Bath 2 is established through the coupling $m\sigma\chi^{(1)}\chi^{(2)}$ between the two subsystems. Thus in this case that one should not expect that at late times the power delivered by the stochastic force or noise be exactly in balance with the dissipative power as is the equilibrium case for a single Brownian oscillator in contact with one bath.

After the subsystem is fully relaxed, the power done by the self-force becomes
\begin{align}
	 P_{\gamma_{1}}(\infty)&=-4\pi\gamma^{2}\int^{\infty}_{-\infty}\frac{d\omega}{2\pi}\;\omega^{2}\biggl\{\Bigl[\tilde{d}_{2}^{(+)}(\omega)+\tilde{d}_{2}^{(-)}(\omega)\Bigr]\Bigl[\tilde{d}_{2}^{(+)}(\omega)+\tilde{d}_{2}^{(-)}(\omega)\Bigr]^{*}\,\tilde{G}_{H}^{\beta_{1}}(\omega)\biggr.\notag\\
	 &\qquad\qquad+\biggl.\Bigl[\tilde{d}_{2}^{(+)}(\omega)-\tilde{d}_{2}^{(-)}(\omega)\Bigr]\Bigl[\tilde{d}_{2}^{(+)}(\omega)-\tilde{d}_{2}^{(-)}(\omega)\Bigr]^{*}\,\tilde{G}_{H}^{\beta_{2}}(\omega)\biggr\}\,,\label{E:bvcwke}
\end{align}
where we have used the convolution integral \eqref{E:ncerios} and \eqref{E:ddddd}. It is instructive to take a closer look into the expressions in \eqref{E:bvcwke} that are associated with $\tilde{G}_{H}^{\beta_{1}}$ of Bath 1. We observe that
\begin{align}
	 P_{\gamma_{1}}^{(1)}(\infty)&=-4\pi\gamma^{2}\int^{\infty}_{-\infty}\frac{d\omega}{2\pi}\;\omega^{2}\Bigl[\tilde{d}_{2}^{(+)}(\omega)+\tilde{d}_{2}^{(-)}(\omega)\Bigr]\Bigl[\tilde{d}_{2}^{(+)}(\omega)+\tilde{d}_{2}^{(-)}(\omega)\Bigr]^{*}\,\tilde{G}_{H}^{\beta_{1}}(\omega)\notag\\
	 &=i\,\pi\gamma\int^{\infty}_{-\infty}\frac{d\omega}{2\pi}\;\omega\,\Bigl[\tilde{d}_{2}^{(+)}(\omega)+\tilde{d}_{2}^{(-)}(\omega)-\tilde{d}_{2}^{(+)}(-\omega)-\tilde{d}_{2}^{(-)}(-\omega)\Bigr]\,\tilde{G}_{H}^{\beta_{1}}(\omega)\notag\\
	 &\qquad\qquad+\int^{\infty}_{-\infty}\frac{d\omega}{2\pi}\;\omega^{2}\Bigl[\tilde{d}_{2}^{(-)}(\omega)\,\tilde{d}_{2}^{(+)*}(\omega)+\tilde{d}_{2}^{(+)}(\omega)\,\tilde{d}_{2}^{(-)*}(\omega)\Bigr]\,\tilde{G}_{H}^{\beta_{1}}(\omega)\notag\\
	 &=i\,2\pi\gamma\int^{\infty}_{-\infty}\frac{d\omega}{2\pi}\;\omega\,\Bigl[\tilde{d}_{2}^{(+)}(\omega)+\tilde{d}_{2}^{(-)}(\omega)\Bigr]\,\tilde{G}_{H}^{\beta_{1}}(\omega)\notag\\
	 &\qquad\qquad+\int^{\infty}_{-\infty}\frac{d\omega}{2\pi}\;\omega^{2}\Bigl[\tilde{d}_{2}^{(-)}(\omega)\,\tilde{d}_{2}^{(+)*}(\omega)+\tilde{d}_{2}^{(+)}(\omega)\,\tilde{d}_{2}^{(-)*}(\omega)\Bigr]\,\tilde{G}_{H}^{\beta_{1}}(\omega)\,,\notag
\end{align}
in which we have used
\begin{align}
	 \tilde{d}_{2}^{(\pm)}(\omega)\,\tilde{d}_{2}^{(\pm)*}(\omega)&=-\frac{i}{4\gamma\omega}\Bigl[\tilde{d}_{2}^{(\pm)}(\omega)-\tilde{d}_{2}^{(\pm)*}(\omega)\Bigr]=-\frac{i}{4\gamma\omega}\Bigl[\tilde{d}_{2}^{(\pm)}(\omega)-\tilde{d}_{2}^{(\pm)}(-\omega)\Bigr]\,,\\
	\widetilde{G}_{H}^{\beta}(\omega)&=\widetilde{G}_{H}^{\beta}(-\omega)\,.
\end{align}
Now if we combine this contribution $P_{\gamma_{1}}^{(1)}(\infty)$ in the total dissipative power $P_{\gamma_{1}}(\infty)$ with $P_{\xi_{1}}(\infty)$, we end up with
\begin{align}
	&\quad P_{\xi_{1}}(\infty)+P_{\gamma_{1}}^{(1)}(\infty)\notag\\
	 &=\left\{4\pi\gamma\int^{\infty}_{-\infty}\frac{d\omega}{2\pi}\;-i\,\omega\Bigl[\tilde{d}_{2}^{(+)}(\omega)+\tilde{d}_{2}^{(-)}(\omega)\Bigr]\,\tilde{G}_{H}^{\beta_{1}}(\omega)\right\}\notag\\
	 &\qquad\qquad+\left\{2\pi\gamma\int^{\infty}_{-\infty}\frac{d\omega}{2\pi}\;i\,\omega\Bigl[\tilde{d}_{2}^{(+)}(\omega)+\tilde{d}_{2}^{(-)}(\omega)\Bigr]\,\tilde{G}_{H}^{\beta_{1}}(\omega)\right.\notag\\
	 &\qquad\qquad\qquad\qquad+\left.\int^{\infty}_{-\infty}\frac{d\omega}{2\pi}\;\omega^{2}\Bigl[\tilde{d}_{2}^{(-)}(\omega)\,\tilde{d}_{2}^{(+)*}(\omega)+\tilde{d}_{2}^{(+)}(\omega)\,\tilde{d}_{2}^{(-)*}(\omega)\Bigr]\,\tilde{G}_{H}^{\beta_{1}}(\omega)\right\}\notag\\
	 &=-2\pi\gamma\int^{\infty}_{-\infty}\frac{d\omega}{2\pi}\;i\,\omega\Bigl[\tilde{d}_{2}^{(+)}(\omega)+\tilde{d}_{2}^{(-)}(\omega)\Bigr]\,\tilde{G}_{H}^{\beta_{1}}(\omega)\notag\\
	 &\qquad\qquad\qquad\qquad+\int^{\infty}_{-\infty}\frac{d\omega}{2\pi}\;\omega^{2}\Bigl[\tilde{d}_{2}^{(-)}(\omega)\,\tilde{d}_{2}^{(+)*}(\omega)+\tilde{d}_{2}^{(+)}(\omega)\,\tilde{d}_{2}^{(-)*}(\omega)\Bigr]\,\tilde{G}_{H}^{\beta_{1}}(\omega)\notag\\
	 &=4\pi\gamma^{2}\int^{\infty}_{-\infty}\frac{d\omega}{2\pi}\;\omega^{2}\Bigl[\tilde{d}_{2}^{(+)}(\omega)-\tilde{d}_{2}^{(-)}(\omega)\Bigr]\Bigl[\tilde{d}_{2}^{(+)}(\omega)-\tilde{d}_{2}^{(-)}(\omega)\Bigr]^{*}\,\tilde{G}_{H}^{\beta_{1}}(\omega)\,.
\end{align}
This implies that after the subsystems are fully relaxed, the net energy transport rate, or power input into Subsystem 1 from Bath 1 is given by
\begin{align}
	&\quad\;P_{\xi_{1}}(\infty)+P_{\gamma_{1}}(\infty)\notag\\
	&=P_{\xi_{1}}(\infty)+P_{\gamma_{1}}^{(1)}(\infty)+P_{\gamma_{1}}^{(2)}(\infty)\label{E:mar}\\
	 &=-4\pi\gamma^{2}\int^{\infty}_{-\infty}\frac{d\omega}{2\pi}\;\omega^{2}\Bigl[\tilde{d}_{2}^{(+)}(\omega)-\tilde{d}_{2}^{(-)}(\omega)\Bigr]\Bigl[\tilde{d}_{2}^{(+)}(\omega)-\tilde{d}_{2}^{(-)}(\omega)\Bigr]^{*}\Bigl[\tilde{G}_{H}^{\beta_{2}}(\omega)-\tilde{G}_{H}^{\beta_{1}}(\omega)\Bigr]\,.\notag
\end{align}
Note its dependence on $\tilde{G}_{H}^{\beta_{2}}(\omega)-\tilde{G}_{H}^{\beta_{1}}(\omega)$, which vanishes when there is no temperatures difference between the two baths, $\beta_{2}^{-1}=\beta_{1}^{-1}$.

\subsubsection{Energy Flow between $S_{1}$ and $S_{2}$}

Next we consider the energy flow between Subsystems 1 and 2. The power delivered by Subsystem 2 to Subsystem 1 is given by
\begin{align}
	P_{21}(t)&=-m\sigma\langle\,\chi^{(2)}(t)\dot{\chi}^{(1)}(t)\,\rangle\notag\\
	&=-\frac{\sigma}{4m}\int^{t}_{0}\!ds\!\int^{t}_{0}\!ds'\notag\\
	 &\quad\;\biggl\{\Bigl[d_{2}^{(+)}(t-s)-d_{2}^{(-)}(t-s)\Bigr]\Bigl[\dot{d}_{2}^{(+)}(t-s')+\dot{d}_{2}^{(-)}(t-s')\Bigr]\langle\,\xi_{1}(s)\xi_{1}(s')\,\rangle\biggr.\notag\\
	 &\quad+\biggl.\Bigl[d_{2}^{(+)}(t-s)+d_{2}^{(-)}(t-s)\Bigr]\Bigl[\dot{d}_{2}^{(+)}(t-s')-\dot{d}_{2}^{(-)}(t-s')\Bigr]\langle\,\xi_{2}(s)\xi_{2}(s')\,\rangle\biggr\}\notag\\
	&\quad+\text{homogeneous terms independent of stochastic forces}\,.
\end{align}
In the limit $t\to\infty$, the homogeneous terms vanish and we are left with
\begin{align}\label{E:bnxiryt}
	&\quad\;P_{21}(\infty)\\
	 &=i\,2\pi\sigma\gamma\int^{\infty}_{-\infty}\frac{d\omega}{2\pi}\;\omega\,\Bigl[\tilde{d}_{2}^{(+)}(\omega)-\tilde{d}_{2}^{(-)}(\omega)\Bigr]\Bigl[\tilde{d}_{2}^{(+)}(\omega)+\tilde{d}_{2}^{(-)}(\omega)\Bigr]^{*}\Bigl[\tilde{G}_{H}^{\beta_{2}}(\omega)-\tilde{G}_{H}^{\beta_{1}}(\omega)\Bigr]\,.\notag
\end{align}
Observe the similarity in form with \eqref{E:mar} except for the sign difference in the second square bracket.

Let us now compute the average power $P_{12}(t)=-m\sigma\langle\,\chi^{(1)}(t)\dot{\chi}^{(2)}(t)\,\rangle$ Subsystem  1 delivers to Subsystem 2 via their mutual interaction. In the same manners as we arrive at \eqref{E:bnxiryt}, we find that at late time $t\to\infty$, the power $P_{12}(\infty)$ is given by
\begin{align}
	 P_{12}(\infty)&=i\,2\pi\sigma\gamma\int^{\infty}_{-\infty}\frac{d\omega}{2\pi}\;\omega\,\Bigl[\tilde{d}_{2}^{(+)}(\omega)-\tilde{d}_{2}^{(-)}(\omega)\Bigr]\Bigl[\tilde{d}_{2}^{(+)}(\omega)+\tilde{d}_{2}^{(-)}(\omega)\Bigr]^{*}\notag\\
	 &\qquad\qquad\qquad\qquad\qquad\times\Bigl[\tilde{G}_{H}^{\beta_{2}}(\omega)-\tilde{G}_{H}^{\beta_{1}}(\omega)\Bigr]\notag\\
	&=-P_{21}(\infty)\,.
\end{align}
Note this is NOT the consequence of Newton's third law because $P_{12}$ is not the time rate of work ( = power ) done by the reaction force, namely, the force Subsystem 2 exerts on Subsystem 1.

\subsubsection{Energy Flow between $S_{2}$ and $B_{2}$}

Finally let us look at the energy flow between Subsystem 2 and Bath 2. First, the average power delivered by the stochastic force  $\xi_{2}$ from Bath 2 on Subsystem 2 is $P_{\xi_{2}}(t)=\langle\xi_{2}(t)\dot{\chi}^{(2)}(t)\rangle$. At late limit $t\to\infty$, we have
\begin{align}
	 P_{\xi_{2}}(\infty)&=4\pi\gamma\int^{\infty}_{-\infty}\frac{d\omega}{2\pi}\;-i\,\omega\,\Bigl[\tilde{d}_{2}^{(+)}(\omega)+\tilde{d}_{2}^{(-)}(\omega)\Bigr]\,\tilde{G}_{H}^{\beta_{2}}(\omega)\,.\label{E:gdswe}
\end{align}
Similarly the power delivered by the dissipation force $-2m\gamma\dot{\chi}^{(2)}$ to this subsystem is defined by $P_{\gamma_{2}}(t)=-2m\gamma\langle\,\dot{\chi}^{(2)\,2}(t)\,\rangle$, and it becomes
\begin{align}
	 P_{\gamma_{2}}(\infty)&=-4\pi\gamma^{2}\int^{\infty}_{-\infty}\frac{d\omega}{2\pi}\;\omega^{2}\biggl\{\Bigl[\tilde{d}_{2}^{(+)}(\omega)-\tilde{d}_{2}^{(-)}(\omega)\Bigr]\Bigl[\tilde{d}_{2}^{(+)}(\omega)-\tilde{d}_{2}^{(-)}(\omega)\Bigr]^{*}\,\tilde{G}_{H}^{\beta_{1}}(\omega)\biggr.\notag\\
	 &\qquad\qquad+\biggl.\Bigl[\tilde{d}_{2}^{(+)}(\omega)+\tilde{d}_{2}^{(-)}(\omega)\Bigr]\Bigl[\tilde{d}_{2}^{(+)}(\omega)+\tilde{d}_{2}^{(-)}(\omega)\Bigr]^{*}\,\tilde{G}_{H}^{\beta_{2}}(\omega)\biggr\}\,,
\end{align}
in the limit $t\to\infty$. Following the procedures that lead to \eqref{E:mar}, we find that after the subsystems are fully relaxed, the net power flows into Subsystem 2 from Bath 2 is
\begin{align}
	&\quad\;P_{\xi_{2}}(\infty)+P_{\gamma_{2}}(\infty)\label{E:msar}\\
	 &=4\pi\gamma^{2}\int^{\infty}_{-\infty}\frac{d\omega}{2\pi}\;\omega^{2}\Bigl[\tilde{d}_{2}^{(+)}(\omega)-\tilde{d}_{2}^{(-)}(\omega)\Bigr]\Bigl[\tilde{d}_{2}^{(+)}(\omega)-\tilde{d}_{2}^{(-)}(\omega)\Bigr]^{*}\Bigl[\tilde{G}_{H}^{\beta_{2}}(\omega)-\tilde{G}_{H}^{\beta_{1}}(\omega)\Bigr]\,.\notag
\end{align}
Compared to the energy flow from $B_{1}$ to $S_{1}$, namely, $P_{\xi_{1}}(\infty)+P_{\gamma_{1}}(\infty)$ derived in \eqref{E:mar} , this carries the opposite sign.  This has to be the case for a stationary state to be established.  It says that after the system $\mathfrak{S}= S_{1}+ S_{2}$ is fully relaxed, the energy which flows into $\mathfrak{S}$ from Bath 1 at temperature $\beta_{1}^{-1}$ is the same as that out of $\mathfrak{S}$ into Bath 2 at temperature $\beta_{2}^{-1}$.

One last task remains in demonstrating the existence of a NESS:  we must show that the magnitude of energy flow between the system S and either bath is also the same as the energy flow between the two subsystems, that is, $-P_{21}(\infty)=P_{\xi_{1}}(\infty)+P_{\gamma_{1}}(\infty)$. We now show that this is indeed so.

\subsection{Condition of Stationarity}

Owing to the facts that
\begin{align}
	 \tilde{d}_{2}^{(+)}(\omega)-\tilde{d}_{2}^{(-)}(\omega)&=-2\sigma\,\tilde{d}_{2}^{(+)}(\omega)\tilde{d}_{2}^{(-)}(\omega)\,,\label{E:dlfjv}\\
	 \tilde{d}_{2}^{(+)}(\omega)+\tilde{d}_{2}^{(-)}(\omega)&=2\bigl(\omega_{R}^{2}-\omega^{2}-i\,2\gamma\omega\bigr)\,\tilde{d}_{2}^{(+)}(\omega)\tilde{d}_{2}^{(-)}(\omega)\,,\label{E:dlfjw}
\end{align}
we can write $P_{21}(\infty)$ as
\begin{align}
	&\quad\;P_{21}(t)\notag\\
	 &=i\,2\pi\sigma\gamma\int^{\infty}_{-\infty}\frac{d\omega}{2\pi}\;\omega\,\Bigl[\tilde{d}_{2}^{(+)}(\omega)-\tilde{d}_{2}^{(-)}(\omega)\Bigr]\Bigl[\tilde{d}_{2}^{(+)}(\omega)+\tilde{d}_{2}^{(-)}(\omega)\Bigr]^{*}\Bigl[\tilde{G}_{H}^{\beta_{2}}(\omega)-\tilde{G}_{H}^{\beta_{1}}(\omega)\Bigr]\notag\\
	 &=16\pi\gamma^{2}\sigma^{2}\int^{\infty}_{-\infty}\frac{d\omega}{2\pi}\;\omega^{2}\left|\tilde{d}_{2}^{(+)}(\omega)\right|^{2}\left|\tilde{d}_{2}^{(-)}(\omega)\right|^{2}\Bigl[\tilde{G}_{H}^{\beta_{2}}(\omega)-\tilde{G}_{H}^{\beta_{1}}(\omega)\Bigr]\,,\label{E:sahq}
\end{align}
because the imaginary of the integrand is an odd function of $\omega$.

As for $P_{\xi_{1}}(\infty)+P_{\gamma_{1}}(\infty)$, we can also show that
\begin{align}
	&\quad\;P_{\xi_{1}}(\infty)+P_{\gamma_{1}}(\infty)\notag\\
	 &=-4\pi\gamma^{2}\int^{\infty}_{-\infty}\frac{d\omega}{2\pi}\;\omega^{2}\Bigl[\tilde{d}_{2}^{(+)}(\omega)-\tilde{d}_{2}^{(-)}(\omega)\Bigr]\Bigl[\tilde{d}_{2}^{(+)}(\omega)-\tilde{d}_{2}^{(-)}(\omega)\Bigr]^{*}\Bigl[\tilde{G}_{H}^{\beta_{2}}(\omega)-\tilde{G}_{H}^{\beta_{1}}(\omega)\Bigr]\notag\\
	 &=-16\pi\gamma^{2}\sigma^{2}\int^{\infty}_{-\infty}\frac{d\omega}{2\pi}\;\omega^{2}\left|\tilde{d}_{2}^{(+)}(\omega)\right|^{2}\left|\tilde{d}_{2}^{(-)}(\omega)\right|^{2}\Bigl[\tilde{G}_{H}^{\beta_{2}}(\omega)-\tilde{G}_{H}^{\beta_{1}}(\omega)\Bigr]\label{E:fdke}\,,
\end{align}
with $\omega_{\pm}^{2}=\omega_{R}^{2}\pm\sigma$. Eqs.~\eqref{E:sahq} and \eqref{E:fdke} indicate that indeed we have $-P_{21}(\infty)=P_{\xi_{1}}(\infty)+P_{\gamma_{1}}(\infty)$, or
\begin{equation}\label{E:fkwaz}
	P_{21}(\infty)+P_{\xi_{1}}(\infty)+P_{\gamma_{1}}(\infty)=0\,.
\end{equation}
This has an interesting consequence. From the Langevin equations \eqref{E:hgwieywia} and \eqref{E:igwieywia}, if we multiply them with $\dot{\chi}^{(1)}$ and $\dot{\chi}^{(2)}$ respectively and take their individual average, we arrive at
\begin{align}
	\frac{d}{dt}\,\langle\,E^{(1)}_{k}\,\rangle&=P_{21}+P_{\xi_{1}}+P_{\gamma_{1}}\,,\\
	\frac{d}{dt}\,\langle\,E^{(2)}_{k}\,\rangle&=P_{12}+P_{\xi_{2}}+P_{\gamma_{2}}\,,
\end{align}
where $E_{k}^{(i)}$ is the \textit{mechanical energy} of each subsystem,
\begin{equation}
	E_{k}^{(i)}=\frac{1}{2}\,m\dot{\chi}^{(i)\,2}+\frac{1}{2}\,m\omega_{R}^{2}\chi^{(i)\,2}\,.
\end{equation}
Eq.~\eqref{E:fkwaz} then says that the mechanical energy of each subsystem is conserved when the whole system reaches relaxation. In addition, the \textit{condition of stationarity}
\begin{equation}\label{E:dfns}
	P_{\xi_{1}}+P_{\gamma_{1}}=-P_{\xi_{2}}-P_{\gamma_{2}}
\end{equation}
implies the energy of the whole system will go to a fixed value at late time
\begin{equation}
	 \frac{d}{dt}\,\langle\,\left[\sum_{i=1,\,2}\frac{1}{2}\,m\dot{\chi}^{(i)\,2}+\frac{1}{2}\,m\omega_{R}^{2}\chi^{(i)\,2}\right]+m\sigma\,\chi^{(1)}\chi^{(2)}\,\rangle=0\,,\qquad\qquad\text{as $t\to\infty$}\,.
\end{equation}
Eq.~\eqref{E:dfns} also says that in the end we must have $P_{21}+P_{12}=0$. This is not obvious when compared with the corresponding closed systems. If there is no reservoir in contact with either subsystem, then although the total energy of the whole system (internal energy) is a constant value, the mechanical energy in each subsystem is not. The energy is transferred back and forth between subsystems via their mutual coupling $m\sigma\,\chi^{(1)}\chi^{(2)}$. Thus in the case of the closed systems, $P_{21}+P_{12}\neq0$ but oscillate with time. The key difference between an open and a closed system may lie in the fact that, as $t\to\infty$, the dynamics of an open system is determined largely by the reservoirs, at least from the viewpoint of the Langevin equation.

\subsection{A Mathematically More Concise Derivation}\label{S:beje}

In the above we sought a balance relation by displaying the energy flows between components explicitly. This has the advantage of seeing the physical processes in great detail and clarity. There is a mathematically more concise formulation of energy transport which we present here. We adopt a matrix notation for ease of generalization to the harmonic chain case treated in the following sections.

Since Subsystem 2 exerts a force $-m\sigma\,\chi^{(2)}$ on Subsystem 1, the average power delivered by Subsystem 2 to Subsystem 1 is
\begin{align}
	P_{21}(t)&=-m\sigma\,\langle\chi^{(2)}(t)\dot{\chi}^{(1)}(t)\rangle=-m\sigma\lim_{t'\to t}\frac{d}{d\tau'}\langle\chi^{(2)}(t)\chi^{(1)}(t')\rangle\notag\\
	 &=-m\sigma\left\{\left[\frac{\varsigma^{2}}{2}\,\mathbf{D}_{1}(t)\cdot\dot{\mathbf{D}}_{1}(t)+\frac{1}{2m^{2}\varsigma^{2}}\,\mathbf{D}_{2}(t)\cdot\dot{\mathbf{D}}_{2}(t)\right]_{21}\right.\notag\\
	 &\qquad+\left.\frac{e^{2}}{m^{2}}\int_{0}^{t}\!ds\,ds'\;\Bigl[\mathbf{D}_{2}(t-s)\cdot\mathbf{G}_{H}^{ab}(s-s')\cdot\dot{\mathbf{D}}_{2}^{b1}(t-s')\Bigr]_{21}\right\}\,,
\end{align}
where we have used \eqref{E:mnxeruea} and the properties that $\mathbf{D}_{i}$ are symmetric.
As noted before the expressions within the square brackets approach zero at late times, so in the limit $\tau\to\infty$, the power $P_{21}(t)$ becomes
\begin{align}
	 P_{21}(\infty)=-\frac{e^{2}\sigma}{m}\int_{-\infty}^{\infty}\!ds\,ds'\;\Bigl[\mathbf{D}_{2}(s)\cdot\mathbf{G}_{H}(s-s')\cdot\dot{\mathbf{D}}_{2}(s')\Bigr]_{21}\,.
\end{align}
Recall that since $\mathbf{D}_{2}^{ij}(s)=0$ for $s<0$, we can extend the lower limit of the integration to minus infinity. Expressing the integrand by the Fourier transform of each kernel function yields
\begin{equation}\label{E:reuskyte}
	 P_{21}(\infty)=-\frac{e^{2}\sigma}{m}\int_{-\infty}^{\infty}\!\frac{d\omega}{2\pi}\,\bigl(-i\,\omega\bigr)\Bigl[\widetilde{\mathbf{D}}_2^{*}(\omega)\cdot\widetilde{\mathbf{G}}_{H}(\omega)\cdot\widetilde{\mathbf{D}}_{2}(\omega)\Bigr]_{21}\,.
\end{equation}
It says that the average power delivered by Subsystem 2 $(S_{2})$ on Subsystem 1 $(S_{1})$ approaches a constant eventually.

Next we  examine the corresponding power transfer between the $S_{1}$ and its private bath $(B_{1})$. The average power delivered by the stochastic force (noise) from $B_{1}$ to $S_{1}$ is
\begin{align}\label{E:lernddf}
	 P_{\xi_{1}}(t)=\langle\,\xi^{(1)}(t)\dot{\chi}^{(1)}(t)\,\rangle&=\frac{1}{m}\int_{0}^{t}\!ds\;\dot{D}^{(2)}_{1a}(t-s)\langle\,\xi_{1}(t)\xi_{a}(s)\,\rangle\notag\\
	&=\frac{e^{2}}{m}\int_{0}^{t}\!ds\;\Bigl[\dot{\mathbf{D}}_{2}(t-s)\cdot\mathbf{G}_{H}(t-s)\Bigr]_{11}\,.
\end{align}
Hence at late times $t\to\infty$, the average power $P_{\xi_{1}}$ becomes
\begin{align}
	 P_{\xi_{1}}(\infty)&=\frac{e^{2}}{m}\int_{-\infty}^{\infty}\!\frac{d\omega}{2\pi}\;\bigl(i\,\omega\bigr)\Bigl[\widetilde{\mathbf{D}}^{*}_{2}(\omega)\cdot\tilde{\mathbf{G}}_{H}(\omega)\Bigr]_{11}\,.
\end{align}
Likewise, the average power delivered by the dissipative force in $S_{1}$ from the backaction of $B_{1}$  is described by
\begin{align}
	 P_{\gamma_{1}}(t)=-2m\gamma\,\langle\,\dot{\chi}^{(1)\,2}(t)\,\rangle&=-\frac{2e^{2}\gamma}{m}\int_{0}^{t}\!ds\,ds'\;\Bigl[\dot{\mathbf{D}}_{2}(s)\cdot\mathbf{G}_{H}(s-s')\cdot\dot{\mathbf{D}}_{2}(s')\Bigr]_{11}\,.
\end{align}
Again, we have ignored contributions which are exponentially small at late times. The value of $P_{\gamma_{1}}$ in the limit $t\to\infty$ is given by
\begin{align}
	 P_{\gamma_{1}}(\infty)&=-\frac{2e^{2}\gamma}{m}\int_{-\infty}^{\infty}\!\frac{d\omega}{2\pi}\;\omega^{2}\Bigl[\widetilde{\mathbf{D}}^{*}_{2}(\omega)\cdot\widetilde{\mathbf{G}}_{H}(\omega)\cdot\widetilde{\mathbf{D}}_{2}(\omega)\Bigr]_{11}\,.
\end{align}
Therefore the net energy transfer at late times between $B_{1}$ and $S_{1}$ is
\begin{align}
	P_{S_{1}}&=P_{\xi_{1}}(\infty)+P_{\gamma_{1}}(\infty)\notag\\
	 &=\frac{e^{2}}{m}\int_{-\infty}^{\infty}\!\frac{d\omega}{2\pi}\;\bigl(i\,\omega\bigr)\,\widetilde{\mathbf{D}}^{1a\,*}_{2}(\omega)\Bigl[\widetilde{\mathbf{G}}^{1a}_{H}(\omega)+i\,2\gamma\omega\,\widetilde{\mathbf{D}}^{1b}_{2}(\omega)\widetilde{\mathbf{G}}^{ab}_{H}(\omega)\Bigr]\notag\\
	 &=\frac{e^{2}}{m}\int_{-\infty}^{\infty}\!\frac{d\omega}{2\pi}\;\bigl(i\,\omega\bigr)\widetilde{\mathbf{D}}^{1a\,*}_{2}\Bigl[\mathbf{I}_{1b}+i\,2\gamma\omega\,\widetilde{\mathbf{D}}^{1b}_{2}(\omega)\Bigr]\widetilde{\mathbf{G}}_{H}^{ab}(\omega)\notag\\
	 &=\frac{e^{2}}{m}\int_{-\infty}^{\infty}\!\frac{d\omega}{2\pi}\;\bigl(i\,\omega\bigr)\left\{\Bigl[\mathbf{I}+i\,2\gamma\omega\,\widetilde{\mathbf{D}}_{2}(\omega)\Bigr]\cdot\widetilde{\mathbf{G}}_{H}(\omega)\cdot\widetilde{\mathbf{D}}_{2}^{\dagger}(\omega)\right\}_{11}\,,\label{E:dkekssa}
\end{align}
where we have used the symmetric property of $\mathbf{G}_{H}$. We next note that the Fourier transform $\widetilde{\mathbf{D}}_{2}(\omega)$ satisfies
\begin{align}
	 \widetilde{\mathbf{D}}_{2}^{-1}(\omega)&=\pmb{\Omega}^{2}-\omega^{2}\mathbf{I}-i\,2\gamma\omega\,\mathbf{I}\,,&&\Rightarrow&&\Bigl[\pmb{\Omega}^{2}-\omega^{2}\mathbf{I}-i\,2\gamma\omega\,\mathbf{I}\Bigr]\cdot\widetilde{\mathbf{D}}_{2}(\omega)=\mathbf{I}\,,\notag\\
	& &&\Rightarrow &&\Bigl[\pmb{\Omega}^{2}-\omega^{2}\mathbf{I}\Bigr]\cdot\widetilde{\mathbf{D}}_{2}(\omega)=\mathbf{I}+i\,2\gamma\omega\,\widetilde{\mathbf{D}}_{2}(\omega)\,,
\end{align}
with $\mathbf{I}$ being a $2\times2$ identity matrix. Putting this result back into \eqref{E:dkekssa}, we arrive at
\begin{align}\label{E:erksjaq}
	 P_{S_{1}}&=\frac{e^{2}}{m}\int_{-\infty}^{\infty}\!\frac{d\omega}{2\pi}\;\bigl(i\,\omega\bigr)\left\{\Bigl[\pmb{\Omega}^{2}-\omega^{2}\mathbf{I}\Bigr]\cdot\widetilde{\mathbf{D}}_{2}(\omega)\cdot\widetilde{\mathbf{G}}_{H}(\omega)\cdot\widetilde{\mathbf{D}}_{2}^{\dagger}(\omega)\right\}_{11}\,.
\end{align}
From the definition of $\pmb{\Omega}^{2}$, we see that $\bigl[\pmb{\Omega}^{2}-\omega^{2}\mathbf{I}\bigr]_{ab}$ is in fact $\sigma\,\bigl(\delta_{a1}\delta_{b2}+\delta_{a2}\delta_{b1}\bigr)$, so that \eqref{E:erksjaq} becomes
\begin{equation}
	 P_{S_{1}}=-\frac{e^{2}\sigma}{m}\int_{-\infty}^{\infty}\!\frac{d\omega}{2\pi}\;\bigl(-i\,\omega\bigr)\,\Bigl[\widetilde{\mathbf{D}}_{2}^{*}(\omega)\cdot\widetilde{\mathbf{G}}_{H}(\omega)\cdot\widetilde{\mathbf{D}}_{2}(\omega)\Bigr]_{12}\,,\label{E:hgeywowa}
\end{equation}
where we have used the symmetry property of $\widetilde{\mathbf{D}}_{2}$ and $\widetilde{\mathbf{G}}_{H}$. Compared to \eqref{E:reuskyte}, we see that $P_{S_{1}}$ is equal to $P_{12}(\infty)$ at late time. It means that if ${P}_{S_{1}}>0$, then the energy flow from the bath 1 to the subsystem 1 is equal to the energy flow from the subsystem 1 to subsystem 2.

In addition from \eqref{E:reuskyte}, we can show that at late time ${P}_{12}(\infty)=-{P}_{21}(\infty)$ as follows. Since by construction ${P}_{21}(\infty)$ is a real physical quantity, if we take the complex conjugate of ${P}_{21}(\infty)$ we should return to the very same ${P}_{21}(\infty)$, that is
\begin{align}\label{E:uryebd}
	 {P}_{21}(\infty)={P}_{21}^{*}(\infty)&=-\frac{e^{2}\sigma}{m}\int_{-\infty}^{\infty}\!\frac{d\omega}{2\pi}\,\bigl(i\,\omega\bigr)\widetilde{\mathbf{D}}_{2}^{2a}(\omega)\widetilde{\mathbf{G}}_H^{ab}(\omega)\widetilde{\mathbf{D}}_{2}^{b1\,*}(\omega)\\
	 &=\frac{e^{2}\sigma}{m}\int_{-\infty}^{\infty}\!\frac{d\omega}{2\pi}\,\bigl(-i\,\omega\bigr)\widetilde{\mathbf{D}}_{2}^{1b\,*}(\omega)\widetilde{\mathbf{G}}_H^{ab}(\omega)\widetilde{\mathbf{D}}_2^{a2}(\omega)=-{P}_{12}(\infty)\,.\notag
\end{align}
Thus we also establish that ${P}_{S_{1}}=-{P}_{21}(\infty)$.

To make connection with \eqref{E:sahq}, we observe that from \eqref{E:zdkjres}, we can relate the elements of the fundamental solution matrices with the corresponding fundamental solutions of the normal modes by
\begin{align*}
	 \widetilde{\mathbf{D}}_{2}^{11}(\omega)&=\widetilde{\mathbf{D}}_{2}^{22}(\omega)=\frac{1}{2}\Bigl[\widetilde{d}_{2}^{(+)}(\omega)+\widetilde{d}_{2}^{(-)}(\omega)\Bigr]=\bigl(\omega_{R}^{2}-\omega^{2}-i\,2\gamma\omega\bigr)\,\tilde{d}_{2}^{(+)}(\omega)\tilde{d}_{2}^{(-)}(\omega)\,\,,\\
	 \widetilde{\mathbf{D}}_{2}^{12}(\omega)&=\widetilde{\mathbf{D}}_{2}^{21}(\omega)=\frac{1}{2}\Bigl[\widetilde{d}_{2}^{(+)}(\omega)-\widetilde{d}_{2}^{(-)}(\omega)\Bigr]=-\sigma\,\tilde{d}_{2}^{(+)}(\omega)\tilde{d}_{2}^{(-)}(\omega)\,,
\end{align*}
from \eqref{E:dlfjw}--\eqref{E:dlfjv}. This enable us to write \eqref{E:uryebd} as
\begin{align}
	 P_{21}(\infty)&=16\pi\gamma^{2}\sigma^{2}\int^{\infty}_{-\infty}\frac{d\omega}{2\pi}\;\omega^{2}\left|\tilde{d}_{2}^{(+)}(\omega)\right|^{2}\left|\tilde{d}_{2}^{(-)}(\omega)\right|^{2}\Bigl[\tilde{G}_{H}^{\beta_{2}}(\omega)-\tilde{G}_{H}^{\beta_{1}}(\omega)\Bigr]\,.
\end{align}
Thus we recover \eqref{E:sahq}.

\subsection{Steady State Energy Flow at High and Low Temperatures}

We may define the \textit{steady energy flow} $J$ by
\begin{align}\label{E:deukfhd}
	J&\equiv P_{21}=P_{\xi_{2}}(\infty)+P_{\gamma_{2}}(\infty)\notag\\
	 &=16\pi\gamma^{2}\sigma^{2}\int^{\infty}_{-\infty}\frac{d\omega}{2\pi}\;\omega^{2}\left|\tilde{d}_{2}^{(+)}(\omega)\right|^{2}\left|\tilde{d}_{2}^{(-)}(\omega)\right|^{2}\Bigl[\tilde{G}_{H}^{\beta_{2}}(\omega)-\tilde{G}_{H}^{\beta_{1}}(\omega)\Bigr]\,,
\end{align}
with
\begin{equation}
	\tilde{G}_{H}^{\beta}(\omega)=\frac{\omega}{4\pi}\,\coth\frac{\beta\omega}{2}\,.
\end{equation}
In the high temperature limit $\beta_{i}\to0$, we have $\tilde{G}_{H}^{\beta_{i}}(\omega)\approx1/(2\pi\beta_{i})$ so that the steady energy current becomes, when $\beta_{i}\omega\ll1 $
\begin{align}
\qquad\qquad J&\simeq8\gamma^{2}\sigma^{2}\bigl(\beta_{2}^{-1}-\beta_{1}^{-1}\bigr)\int^{\infty}_{-\infty}\frac{d\omega}{2\pi}\;\omega^{2}\left|\tilde{d}_{2}^{(+)}(\omega)\right|^{2}\left|\tilde{d}_{2}^{(-)}(\omega)\right|^{2}\propto(T_{2}-T_{1})\,.\label{E:rkdgrs}
\end{align}
The integral in \eqref{E:rkdgrs} can be exactly carried out, and it is given by\begin{align}
	 \int^{\infty}_{-\infty}\frac{d\omega}{2\pi}\;\omega^{2}\left|\tilde{d}_{2}^{(+)}(\omega)\right|^{2}\left|\tilde{d}_{2}^{(-)}(\omega)\right|^{2}&=\frac{1}{8\gamma}\frac{1}{\sigma^{2}+4\gamma^{2}\omega_{R}^{2}}\,,
\end{align}
with $\omega_{\pm}^{2}=\omega_{R}^{2}\pm\sigma$. Therefore the steady energy current in the high temperature limit is given by
\begin{equation}\label{E:cbvkds}
	J=\frac{\gamma\sigma^{2}}{\sigma^{2}+4\gamma^{2}\omega_{R}^{2}}\,\Delta T=\begin{cases}
		\gamma\,\Delta T\,,&\gamma\omega_{R}\ll\sigma\,,\\
		\dfrac{\sigma^{2}}{4\gamma\omega_{R}^{2}}\,\Delta T\,,&\gamma\omega_{R}\gg\sigma\,,
	\end{cases}
\end{equation}
where $\Delta T=T_{2}-T_{1}$, for different relative coupling strengths between the subsystems and the reservoirs. When $\gamma\to0$, that is, when the coupling between the subsystems and their baths is turned off, there is no energy flow. Likewise, if there is no coupling between the subsystems $\sigma\to0$, the energy flow also terminates, as also expected.

\subsection{Heat Conductance}

\begin{figure}
\centering
    \scalebox{0.45}{\includegraphics{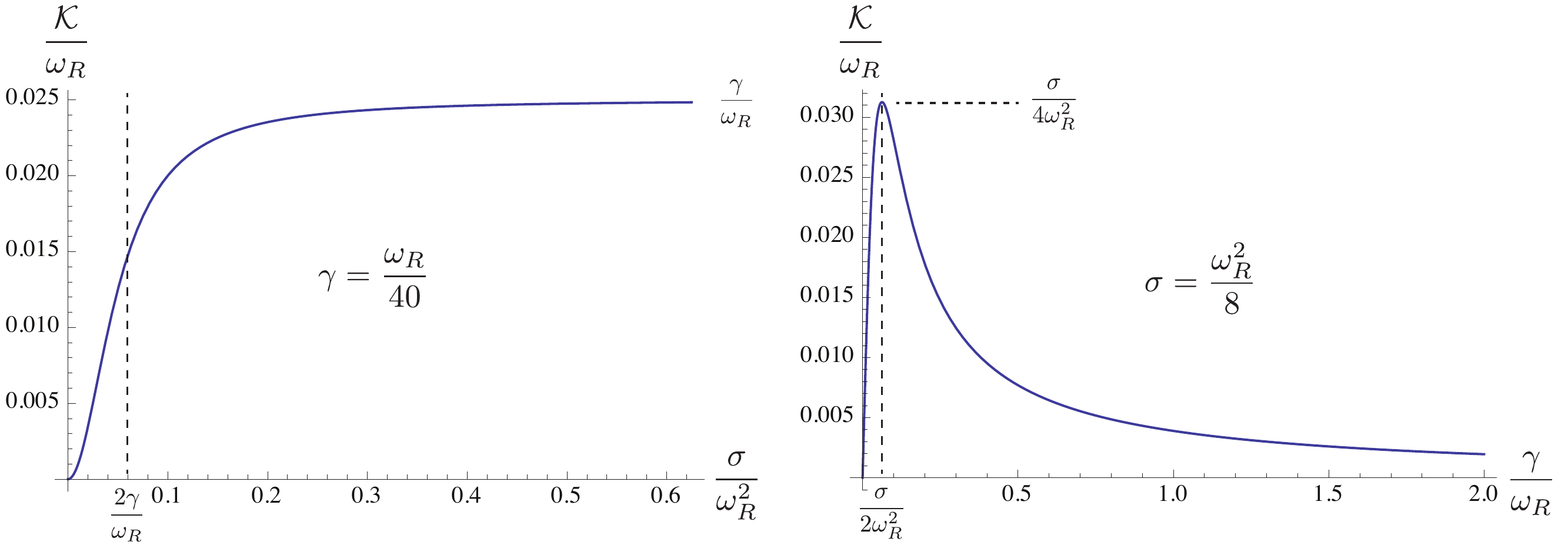}}
    \caption{variation of the conductance $\mathcal{K}$ with respect to the coupling constants $\sigma$ or $\gamma$ in the high temperature limit.}\label{Fi:conductance}
\end{figure}

We define the thermal conductance $\mathcal{K}$ by the ratio of the steady current over the temperature difference between the reservoirs,
\begin{equation}\label{E:bdhredw}
	\mathcal{K}=\lim_{\Delta T\to0}\frac{J}{\Delta T}\,.
\end{equation}
Thus we find that in the high temperature limit $\beta\omega_{R}\gg1$, the conductance
\begin{equation}\label{E:dfmsdrrer}
	\mathcal{K}=\frac{\gamma\,\sigma^{2}}{\sigma^{2}+4\gamma^{2}\omega_{R}^{2}}\,,
\end{equation}
becomes independent of temperature but only depends on the parameters $\sigma$, $\gamma$ and $\omega_{R}$. From Fig.~\ref{Fi:conductance}, we see that the conductance monotonically increases with the inter-oscillator coupling $\sigma$, and gradually approaches the value $\gamma$ as long as the constraint $\sigma\leq\omega_{R}^{2}$ is still satisfied. On the other hand, when we fix the inter-oscillator coupling, the conductance rises up to a maximum value $\sigma/4\omega_{R}$ at $\gamma=\sigma/2\omega_{R}$, and then gradually decreases to zero as the system-environment coupling $\gamma$ increases.
From the expression of $\bigl|\tilde{d}_{2}^{(\pm)}(\omega)\bigr|^{2}$ we note that it traces out a Breit-Wigner resonance curve with respect to $\omega$. The resonance feature is well-defined only when $\gamma$ is sufficiently small, that is, $\gamma\ll\omega_{R}$. The peak is located at about $\omega=(\omega_{R}^{2}\pm\sigma)^{1/2}$ and the width of the peak is about $2\gamma$. Therefore for a fixed value of $\Omega$, the inter-oscillator coupling constant $\sigma$ determines the location of the resonance peak, while the system-environment coupling constant $\gamma$ determines the width of the resonance. The integrand \eqref{E:deukfhd} contains a product of $\bigl|\tilde{d}_{2}^{(+)}(\omega)\bigr|^{2}\bigl|\tilde{d}_{2}^{(-)}(\omega)\bigr|^{2}$, which indicates that there are two resonance peaks at $(\omega_{R}^{2}\pm\sigma)^{1/2}$ respectively. Hence the distance between these two peaks is
\begin{equation*}
	 \bigl(\omega_{R}^{2}+\sigma\bigr)^{\frac{1}{2}}-\bigl(\omega_{R}^{2}-\sigma\bigr)^{\frac{1}{2}}\simeq\frac{\sigma}{\omega_{R}}\,.
\end{equation*}
When the separation of peaks is much greater than the width, it has two distinct, well-defined peaks. If the separation becomes smaller than the characteristic width of each peak, $\sigma<\gamma\omega_{R}$, then the two peaks gradually fuse into one peak. This change of the dominant scale is reflected in the behavior of the conductance $\mathcal{K}$ in the respective regimes,
\begin{equation}\label{E:derjsfw}
	\mathcal{K}=\begin{cases}
		\gamma\,,&\gamma\omega_{R}\ll\sigma\,,\\
		\dfrac{\sigma^{2}}{4\gamma\omega_{R}^{2}}\,,&\gamma\omega_{R}\gg\sigma\,,
	\end{cases}
\end{equation}
as can be seen from \eqref{E:cbvkds}.

In the low temperature limit $\beta\omega_{\pm}\gg1$, we may write the Bose-Einstein distribution factor in \eqref{E:deukfhd} as
\begin{equation*}
	 \tilde{G}_{H}^{\beta_{2}}(\omega)-\tilde{G}_{H}^{\beta_{1}}(\omega)=\frac{\omega}{2\pi}\sum_{n=1}^{\infty}\left[e^{-n\beta_{2}\omega}-e^{-n\beta_{1}\omega}\right]\,,
\end{equation*}
and the steady state current becomes
\begin{align}\label{E:nmeukfhd}
	 J&=\frac{8\pi\gamma^{2}\sigma^{2}}{\pi}\sum_{n=1}^{\infty}\int^{\infty}_{0}\!d\omega\;\omega^{3}\left|\tilde{d}_{2}^{(+)}(\omega)\right|^{2}\left|\tilde{d}_{2}^{(-)}(\omega)\right|^{2}\Bigl[e^{-n\beta_{2}\omega}-e^{-n\beta_{1}\omega}\Bigr]\notag\\
	 &=\frac{48\pi\gamma^{2}\sigma^{2}}{\pi\omega_{+}^{4}\omega_{-}^{4}}\left[\frac{1}{\beta_{2}^{4}}-\frac{1}{\beta_{1}^{4}}\right]\sum_{n=1}^{\infty}\frac{1}{n^{4}}+\cdots\,,\notag\\
	 &=\frac{8\pi^{3}}{15}\frac{\gamma^{2}\sigma^{2}}{\bigl(\omega_{R}^{4}-\sigma^{2}\bigr)^{2}}\left[\frac{1}{\beta_{2}^{4}}-\frac{1}{\beta_{1}^{4}}\right]\,.
\end{align}
The summation over $n$ gives $\zeta(4)=\pi^{4}/90$. This familiar number comes from the higher order expansions of the $\coth z$ function in the limit $z\to\infty$. If we write the steady current \eqref{E:nmeukfhd} in terms of temperature, we obtain
\begin{align}\label{E:dfberhe}
	 J=\frac{8\pi^{3}}{15}\frac{\gamma^{2}\sigma^{2}}{\bigl(\omega_{R}^{4}-\sigma^{2}\bigr)^{2}}\bigl(T_{2}^{4}-T_{1}^{4}\bigr)&=\frac{8\pi^{3}}{15}\frac{\gamma^{2}\sigma^{2}}{\bigl(\omega_{R}^{4}-\sigma^{2}\bigr)^{2}}\bigl(4\mathfrak{T}^{3}\Delta T+\mathfrak{T}\,\Delta T^{3}\bigr)\,,
\end{align}
where $\Delta T=T_{2}-T_{1}$ and $\mathfrak{T}=(T_{2}+T_{1})/2$. We see that the temperature dependence of the steady current is different from the high temperature limit. In the low temperature limit it is proportional to $T_{2}^{4}-T_{1}^{4}$. However, for fixed $\mathfrak{T}$, the current turns out more or less linearly proportional to the temperature difference between the reservoirs except for the case $\Delta T\simeq\mathfrak{T}$, which is equivalent to $T_{1}<3T_{2}$, where the contributions of the $\Delta T^{3}$ terms appreciable.
\begin{figure}
\centering
    \scalebox{0.45}{\includegraphics{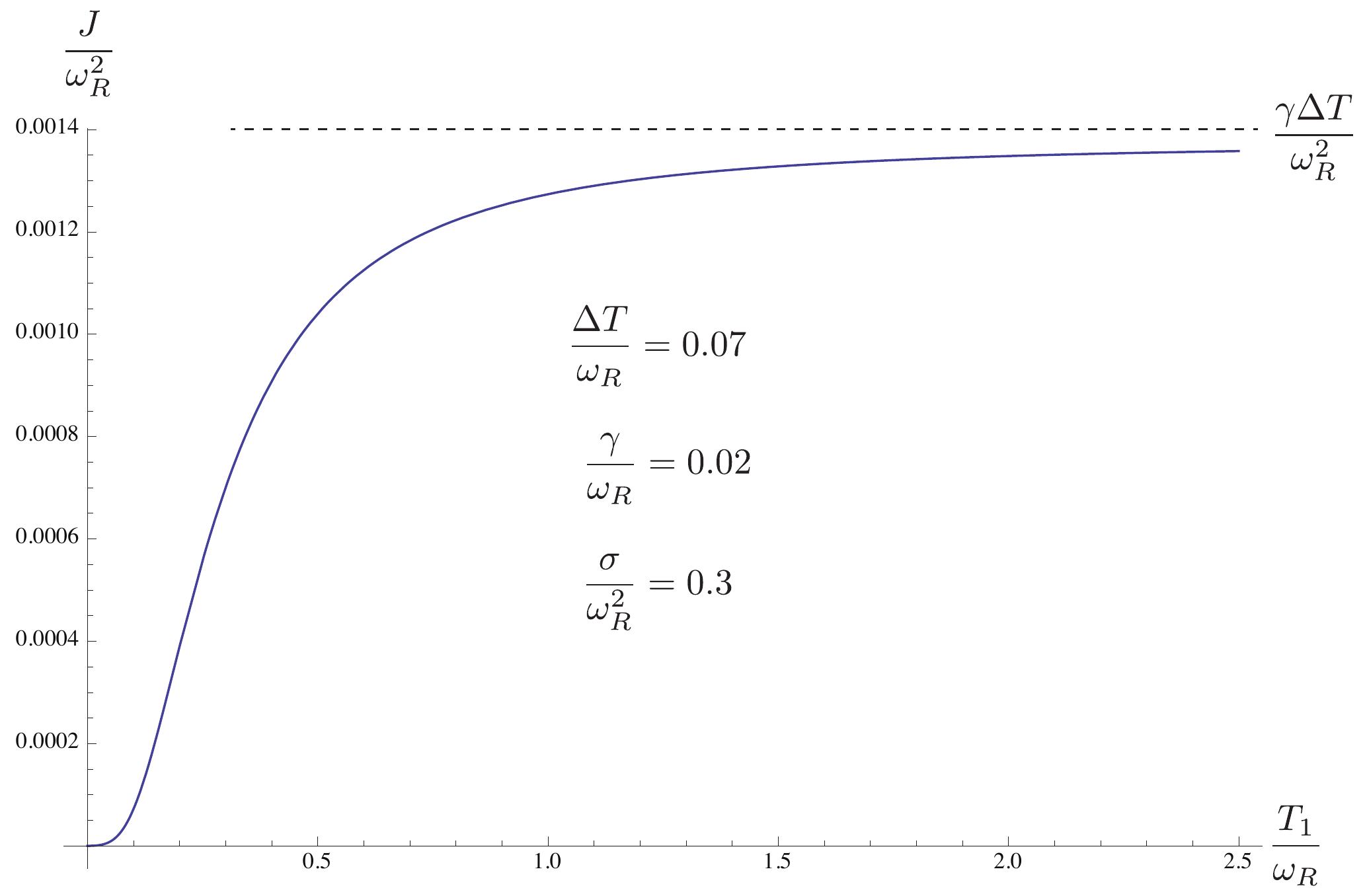}}
    \caption{variation of the steady state current $J$ with respect to the temperature $T_{1}$ of the Bath 1 . The temperature difference $\Delta T$ between two reservoirs is fixed.}\label{Fi:currents}
\end{figure}
In the regime $\Delta T\ll\mathfrak{T}$, the steady state current is
\begin{equation}
	 J\simeq\frac{32\pi^{3}}{15}\frac{\gamma^{2}\sigma^{2}}{\bigl(\omega_{R}^{4}-\sigma^{2}\bigr)^{2}}\,\mathfrak{T}^{3}\Delta T\,,
\end{equation}
and then we may use the definition of the conductance \eqref{E:bdhredw} to find:
\begin{equation}\label{E:mnhruewa}
	 \mathcal{K}=\frac{32\pi^{3}}{15}\frac{\gamma^{2}\sigma^{2}}{\bigl(\omega_{R}^{4}-\sigma^{2}\bigr)^{2}}\,\mathfrak{T}^{3}\,.
\end{equation}
Apparently in the low temperature limit, the conductance depends on $\mathfrak{T}$, which is the mean temperature of the two reservoirs. In Fig.~\ref{Fi:currents}, we plot the steady state current as a function of the temperature $T_{1}$ of Bath 1, with the temperature difference $\Delta T$ fixed, according to \eqref{E:deukfhd}. We see in the high temperature limit, the current approaches  a constant, independent of $T_{2}$, $T_{1}$, as long as $\Delta T$ is fixed, which is consistent with \eqref{E:cbvkds}.

\section{Harmonic Chain}
\label{HarCh}

We now extend the previous results to a one-dimensional chain of $n$ harmonic oscillators. The oscillators at both ends, labelled as  $O_{1}$ and $O_{n}$, are attached to their own pivate baths of respective temperatures $T_1 > T_n$. The remaining $n-2$ oscillators called collectively ${\bf k} = \{2, 3, ..., n-1\}$ are insulated from these two baths, and only interact with their nearest neighbors bilinearly with coupling strength $\sigma$.

In analogy with the case of two oscillators in the previous sections, here the column matrix $\pmb{\chi}$ has $n$ entries, so does the row matrix $\pmb{\chi}^{T}=(\chi^{(1)},\chi^{(2)},\cdots,\chi^{(n)})$. The matrices $\pmb{\Omega}^{2}$, $\mathbf{I'}$ and $\mathbf{G}$ are now spanned to $n\times n$ matrices,
\begin{align}
	\pmb{\Omega}^{2}&=\begin{pmatrix}
						\omega^{2}	&\sigma	&0 &0 &\cdots &0\\
						\sigma	&\omega^{2}	&\sigma &0 &\cdots &0\\
						0	&		&\ddots & & &\vdots\\
						\vdots	& & &\ddots&&0\\
						0	&\cdots &0 &\sigma&\omega^{2}&\sigma\\
						0		&\cdots	&0	&0&\sigma&\omega^{2}
					\end{pmatrix}\,,
					\qquad\mathbf{I'}=\begin{pmatrix}
						1	&0	&0 &0 &\cdots &0\\
						0	&0	&0 &0 &\cdots &0\\
						\vdots	&		&\ddots & & &\vdots\\
						\vdots	& & &\ddots&&\vdots\\
						0	&\cdots &0 &0&0&0\\
						0		&\cdots	&0	&0&0&1
					\end{pmatrix}\,,\notag
\intertext{and}
	\mathbf{G}&(s,s')=\begin{pmatrix}\label{E:dkbgjte}
						G_{\beta_{1}}(s,s')	&0	&0 &0 &\cdots &0\\
						0	&0	&0 &0 &\cdots &0\\
						0	&		&\ddots & & &\vdots\\
						\vdots	& & &\ddots&&0\\
						0	&\cdots &0 &0&0&0\\
						0		&\cdots	&0	&0&0&G_{\beta_{n}}(s,s')
					\end{pmatrix}\,.
\end{align}
In addition, we have the same stochastic effective action as in \eqref{E:hunxms} except that now the stochastic force is a column vector $\pmb{\xi}^{T}=(\xi^{(1)},\,0,\,\cdots,\,0,\,\xi^{(n)})$ and its moments satisfy the Gaussian statistics
\begin{equation}
	 \langle\pmb{\xi}(s)\rangle=0\,,\qquad\qquad\qquad\langle\pmb{\xi}(s)\cdot\pmb{\xi}^{T}(s')\rangle=\mathbf{G}_{H}(s,s')\,.
\end{equation}
Note that although the matrix $\mathbf{G}$ in \eqref{E:dkbgjte} is not invertible, it does not prevent us from writing down the stochastic effective action. In fact we can do it by components and then write them back into the tensor notation. Note the stochastic average is defined only with respect to the first and the final components of the (vectorial) stochastic force $\pmb{\xi}$.

Taking the variation of the stochastic effective action with respect to $\mathbf{q}$ and letting $\mathbf{q}=0$, we arrive at the Langevin equation,
\begin{align}
	 m\,\ddot{\pmb{\chi}}(s)+2m\gamma\,\mathbf{I'}\cdot\dot{\pmb{\chi}}(s)+m\,\pmb{\Omega}_{R}^{2}\cdot\pmb{\chi}(s)&=\pmb{\xi}(s)\,.\label{E:jnkdnfks}
\end{align}
where $\pmb{\Omega}_{R}^{2}$ is the same as $\pmb{\Omega}^{2}$ in the structure except that we replace $\omega^{2}$ by $\omega_{R}^{2}$ due to renormalization.

\subsection{Existence of a Steady Current}

The derivation of the energy currents between the components of the total system is similar to those presented in Sec.~\ref{S:beje}. Thus the net energy flow from Bath 1 ($B_{1}$) to Oscillator 1 ($O_{1}$) at late times is  given by
\begin{align}
	J_{1}&=P_{\xi_{1}}(\infty)+P_{\gamma_{1}}(\infty)\notag\\
		 &=8\gamma^{2}\int^{\infty}_{-\infty}\!d\omega\;\omega^{2}\Bigl[\widetilde{\mathbf{D}}(\omega)\cdot\mathbf{I'}\cdot\widetilde{\mathbf{D}}^{*}(\omega)\cdot\widetilde{\mathbf{G}}_{H}^{T}(\omega)-\widetilde{\mathbf{D}}(\omega)\cdot\widetilde{\mathbf{G}}_{H}(\omega)\cdot\widetilde{\mathbf{D}}^{\dagger}(\omega)\Bigr]_{11}\notag\\
	 &=8\gamma^{2}\int^{\infty}_{-\infty}\!d\omega\;\omega^{2}\bigl|\widetilde{D}^{1n}(\omega)\bigr|{}^{2}\Bigl[\widetilde{G}_{H}^{11}(\omega)-\widetilde{G}_{H}^{nn}(\omega)\Bigr]\,.\label{E:tvrsmd}
\end{align}
Here we note that the prefactor $\bigl|\widetilde{D}^{1n}(\omega)\bigr|{}^{2}$ in the integrand of \eqref{E:tvrsmd} has dependence on the length of the chain $n$. We will take a closer look at its behavior later.

To demonstrate the existence of a steady state for the present configuration, we have to show that the steady current between the neighboring oscillators is the same as $J_{1}$. Let the late-time energy current flow in the intermediate oscillators from $O_{k}$ to $O_{k+1}$ be $J_{k,k+1}$ with $k=2,3,\ldots,n-1$. From \eqref{E:reuskyte}, we know it is given by
\begin{align}\label{E:hmcnvw}
	 J_{k,k+1}&=-i\,\frac{e^{2}\sigma}{m}\int_{-\infty}^{\infty}\!\frac{d\omega}{2\pi}\;\omega\Bigl[\widetilde{\mathbf{D}}(\omega)\cdot\widetilde{\mathbf{G}}_{H}(\omega)\cdot\widetilde{\mathbf{D}}^{\dagger}(\omega)\Bigr]_{k,k+1}\\
	 &=-i\,4\gamma\sigma\int_{-\infty}^{\infty}\!d\omega\;\omega\,\Bigl[\widetilde{D}^{k,1}(\omega)\,\widetilde{D}^{k+1,1\,*}(\omega)\,\widetilde{G}_{H}^{11}(\omega)+\widetilde{D}^{k,n}(\omega)\,\widetilde{D}^{k+1,n\,*}(\omega)\,\widetilde{G}_{H}^{nn}(\omega)\Bigr]\,.\notag
\end{align}
Eq.~\eqref{E:hmcnvw} does not have any reference to  time, so the energy current from $O_{k}$ to $O_{k+1}$ is also a time-independent constant.

To show the equality between \eqref{E:tvrsmd} and \eqref{E:hmcnvw}, we will relate $\widetilde{D}^{k,1}(\omega)\,\widetilde{D}^{k+1,1\,*}(\omega)$ or $\widetilde{D}^{k,n}(\omega)\,\widetilde{D}^{k+1,n\,*}(\omega)$ to $\bigl|\widetilde{D}^{1n}(\omega)\bigr|{}^{2}$ so that we can factor out the noise kernel $\widetilde{G}_{H}(\omega)$ in \eqref{E:hmcnvw}. These relations are provided in Appendix A.  Using the results in \eqref{E:yeibsw} and \eqref{E:lkejo}
\begin{align}
	 \widetilde{D}^{k,1}\,\widetilde{D}^{k+1,1\,*}&=+\frac{i\,c}{\sigma}\,\bigl|\widetilde{D}^{1n}\bigr|{}^{2}+\cdots\,,\\
	 \widetilde{D}^{n-k,1\,*}\,\widetilde{D}^{n-k+1,1}&=-\frac{i\,c}{\sigma}\,\bigl|\widetilde{D}^{1n}\bigr|{}^{2}+\cdots\,,
\end{align}
where $\dots$ denotes terms which will have vanishing contributions to the integral \eqref{E:hmcnvw}, we can rewrite \eqref{E:hmcnvw} as
\begin{align}
	 J_{k,k+1}&=8\gamma^{2}\int_{-\infty}^{\infty}\!d\omega\;\omega^{2}\bigl|\widetilde{D}^{1n}(\omega)\bigr|{}^{2}\Bigl[\widetilde{G}_{H}^{11}(\omega)-\widetilde{G}_{H}^{nn}(\omega)\Bigr]\,,\label{E:zvrsmd}
\end{align}
recalling that $c=2\gamma\omega$. We immediately see that for any neighboring oscillators $O_{k}$ and $O_{k+1}$ located along the chain, the energy current $J_{k,k+1}$ between them is exactly the same as the current $J_{1}$ transported from Bath 1 to Oscillator $O_{1}$ in \eqref{E:tvrsmd}.

As a final touch, we compute the energy current from Bath $B_{n}$, located at the opposite end of the chain, to Oscillator $O_{n}$. From \eqref{E:tvrsmd}, we find this current is given by
\begin{align}
	 J_{n}&=8\gamma^{2}\int^{\infty}_{-\infty}\!d\omega\;\omega^{2}\Bigl[\widetilde{\mathbf{D}}(\omega)\cdot\mathbf{I'}\cdot\widetilde{\mathbf{D}}^{*}(\omega)\cdot\widetilde{\mathbf{G}}_{H}^{T}(\omega)-\widetilde{\mathbf{D}}(\omega)\cdot\widetilde{\mathbf{G}}_{H}(\omega)\cdot\widetilde{\mathbf{D}}^{\dagger}(\omega)\Bigr]_{nn}\notag\\
	 &=-8\gamma^{2}\int^{\infty}_{-\infty}\!d\omega\;\omega^{2}\bigl|\widetilde{D}^{1n}(\omega)\bigr|{}^{2}\Bigl[\widetilde{G}_{H}^{11}(\omega)-\widetilde{G}_{H}^{nn}(\omega)\Bigr]\,.\label{E:uvrsmd}
\end{align}
It has the same magnitude as $J_{1}$ in \eqref{E:tvrsmd} and $J_{k,k+1}$ in \eqref{E:zvrsmd}, but opposite in sign, which says that the current flows from Oscillator $O_{n}$ to Bath $B_{n}$, as is also expected.

In summary, given a quantum harmonic oscillator chain, where each oscillator interacts with its nearest neighbors via bilinear coupling, if the two end-oscillators of the chain are placed in contact with two thermal baths of different temperatures, while the oscillators in between are kept insulated from those baths,  we have explicitly shown that after a time when all the oscillators have fully relaxed, the energy flow along the chain becomes independent of time and the currents between the neighboring oscillators are the same in both magnitude and direction
\begin{equation}\label{E:utrsmd}
	 J_{NESS}=8\gamma^{2}\int^{\infty}_{-\infty}\!d\omega\;\omega^{2}\bigl|\widetilde{D}^{1n}(\omega)\bigr|{}^{2}\Bigl[\widetilde{G}_{H}^{11}(\omega)-\widetilde{G}_{H}^{nn}(\omega)\Bigr]\,.
\end{equation}
This implies that a NESS exists for a quantum harmonic oscillator chain and a steady current flows from the high temperature front along the chain to the low temperature end. There is no buildup or localization of energy at any site along the chain.  We emphasize that in the transient phase before the constituent oscillators come to full relaxation, additional contributions from the homogeneous solutions of the oscillators' modes  render the current between  neighboring oscillators unequal, but in the course of the order of the relaxation time, the energy along the chain is re-distributed to a final constant value while the whole system settles down to a NESS.

\subsection{Scaling Behavior of the NESS Current}
We have shown that after the motion of the constituents of the chain reaches relaxation, a steady thermal energy current exists flowing along the harmonic chain across from the hot  thermal reservoir to the cold one. However we have not addressed the scaling behavior of the steady current with the length of the chain. To shed some light on this problem, we first analyze the prefactor $\bigl|\widetilde{D}^{1n}(\omega)\bigr|^{2}$, which is proportional to the transmission coefficient in the Landauer formula.

From \eqref{E:uidkfhw}, we have
\begin{equation}\label{E:mcnvdh1}
	\bigl|\widetilde{D}^{1n}(\omega)\bigr|^{2}=\frac{\sigma^{2n-2}}{\lvert\theta_{n}\rvert^{2}}\,,
\end{equation}
where
\begin{align}
	\lvert\theta_{n}\rvert^{2}&=f_{2}^{2}\,c^{4}+2f_{1}\,c^{2}+f_{0}^{2}+2\sigma^{2(n-1)}c^{2}>0\,.\label{E:icmer}
\end{align}
Once  $f_{0}$, $f_{1}$, $f_{2}$ is found we can derive the analytic expression of $\lvert\theta_{n}\rvert^{2}$. Indeed we have shown this in the Appendix, where the general expression of $f_{k}$ is given by
\begin{equation}\label{E:mcnvdh2}
	f_{k}=\frac{\mu_{1}^{n-k+1}-\mu_{2}^{n-k+1}}{\mu_{1}-\mu_{2}}\,,
\end{equation}
with the roots of the characteristic equation given by
\begin{align}
	\mu_{1}&=\frac{a+\sqrt{a^{2}-4\sigma^{2}}}{2}\,,&\mu_{2}&=\frac{a-\sqrt{a^{2}-4\sigma^{2}}}{2}\,,
\end{align}
we immediately see that when $a^{2}<4\sigma^{2}$, that is, when $\omega$ lies in the interval $\sqrt{\Omega^{2}-2\sigma}<\omega<\sqrt{\Omega^{2}+2\sigma}$, two roots $\mu_{1}$, $\mu_{2}$ are complex-conjugated. It in turn implies that $\theta_{n}$, as well as $\bigl|\widetilde{D}^{1n}(\omega)\bigr|^{2}$,  will be highly oscillatory with $\omega$ in this interval. Analytically $\bigl|\theta_{n}\bigr|^{2}$ is described by
\begin{align}\label{E:dersw}
	 \lvert\theta_{n}\rvert^{2}&=\frac{\sigma^{2n}}{\sin^{2}\psi}\left[\sin^{2}(n+1)\psi+2\frac{c^{2}}{\sigma^{2}}\Bigl(1+\sin^{2}n\psi\Bigr)+\frac{c^{4}}{\sigma^{4}}\sin^{2}(n-1)\psi\right]\,,
\end{align}
with $\psi$ being
\begin{align}
	\psi&=\tan^{-1}\frac{\sqrt{4\sigma^{2}-a^{2}}}{a}\,.
\end{align}
\begin{figure}
\centering
    \scalebox{0.45}{\includegraphics{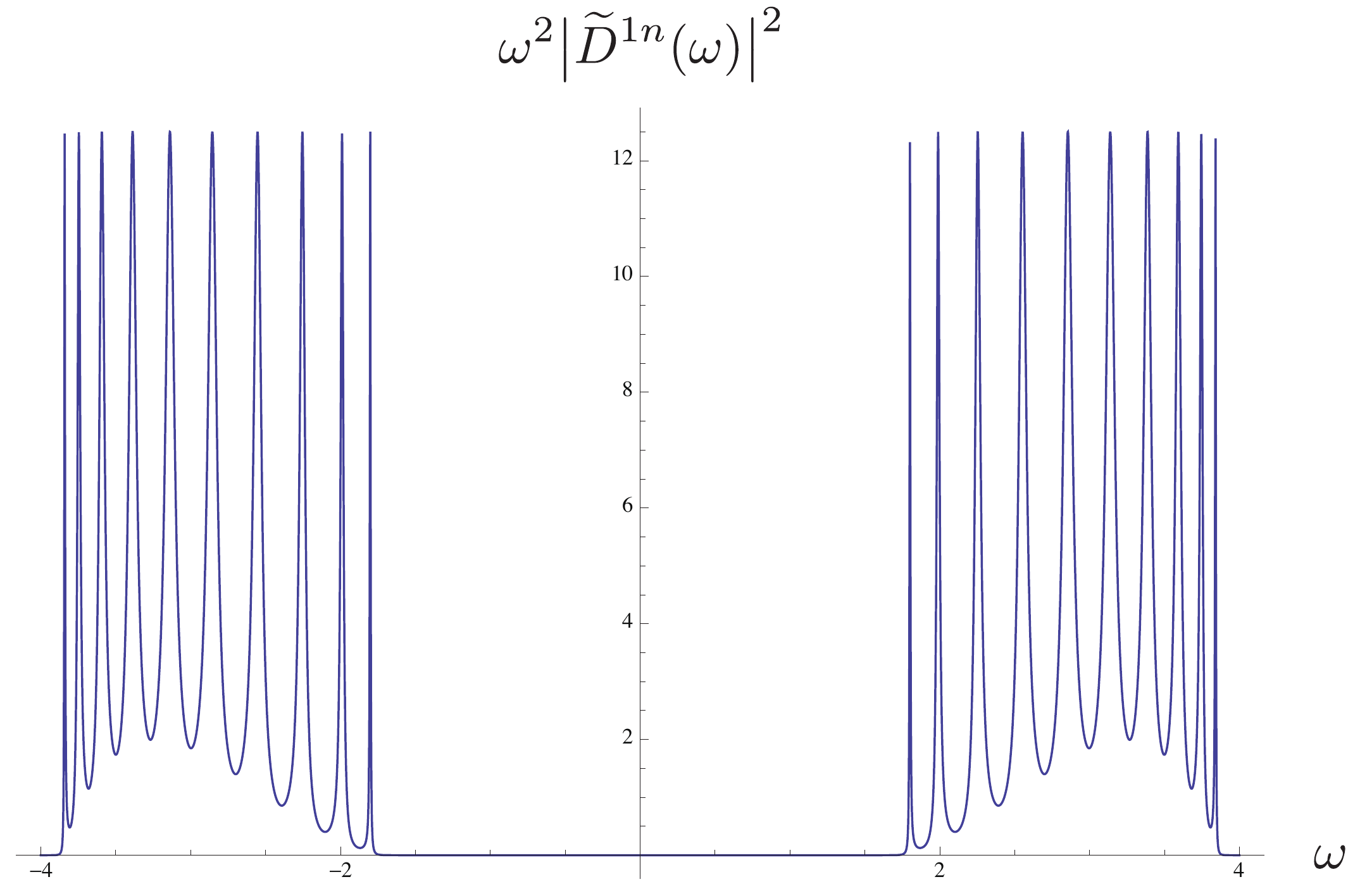}}
    \caption{The generic structure of $\omega^{2}\bigl|\widetilde{D}^{1n}(\omega)\bigr|^{2}$. We choose $n=10$ for example.}\label{Fi:dhjfw}
\end{figure}
We also argue in the Appendix that outside the interval $\sqrt{\Omega^{2}-2\sigma}<\omega<\sqrt{\Omega^{2}+2\sigma}$, the transmission coefficient falls to a vanishingly small value very rapidly. Therefore, generically speaking $\omega^{2}\bigl|\widetilde{D}^{1n}(\omega)\bigr|^{2}$ will have a comb-like structure within the band $\sqrt{\Omega^{2}-2\sigma}<\omega<\sqrt{\Omega^{2}+2\sigma}$, as is shown in Fig~\ref{Fi:dhjfw}. The lower envelope of the comb structure is traced by $\sin^{2}\psi$, while the upper envelope is determined by the subleading term in \eqref{E:dersw} because the maxima occur approximately at the locations where $\sin^{2}(n+1)\psi$ vanishes. In addition, recall that $c=2\gamma\omega$, so the upper envelope of the comb structure is almost constant. The number of the spikes of the comb structure is equal to the number of the oscillators in the harmonic chain. However, since the bandwidth of the interval is independent of $n$, the width of each spike will scale as $n^{-1}$. This implies, as is argued in the Appendix, that the contribution from each spike to the integral over $\omega$ in \eqref{E:utrsmd} also scales as $n^{-1}$. The argument supplied in the Appendix improves with growing $n$, so we see for sufficiently large $n$, the scaling behavior of the contribution of each spike to the steady current will be nicely counteracted by the increasing number of spikes with the band  $\sqrt{\Omega^{2}-2\sigma}<\omega<\sqrt{\Omega^{2}+2\sigma}$. Therefore it implies that the NESS energy current, in the case of the harmonic chain, is independent of the length of the chain.
\begin{figure}
\centering
    \scalebox{0.55}{\includegraphics{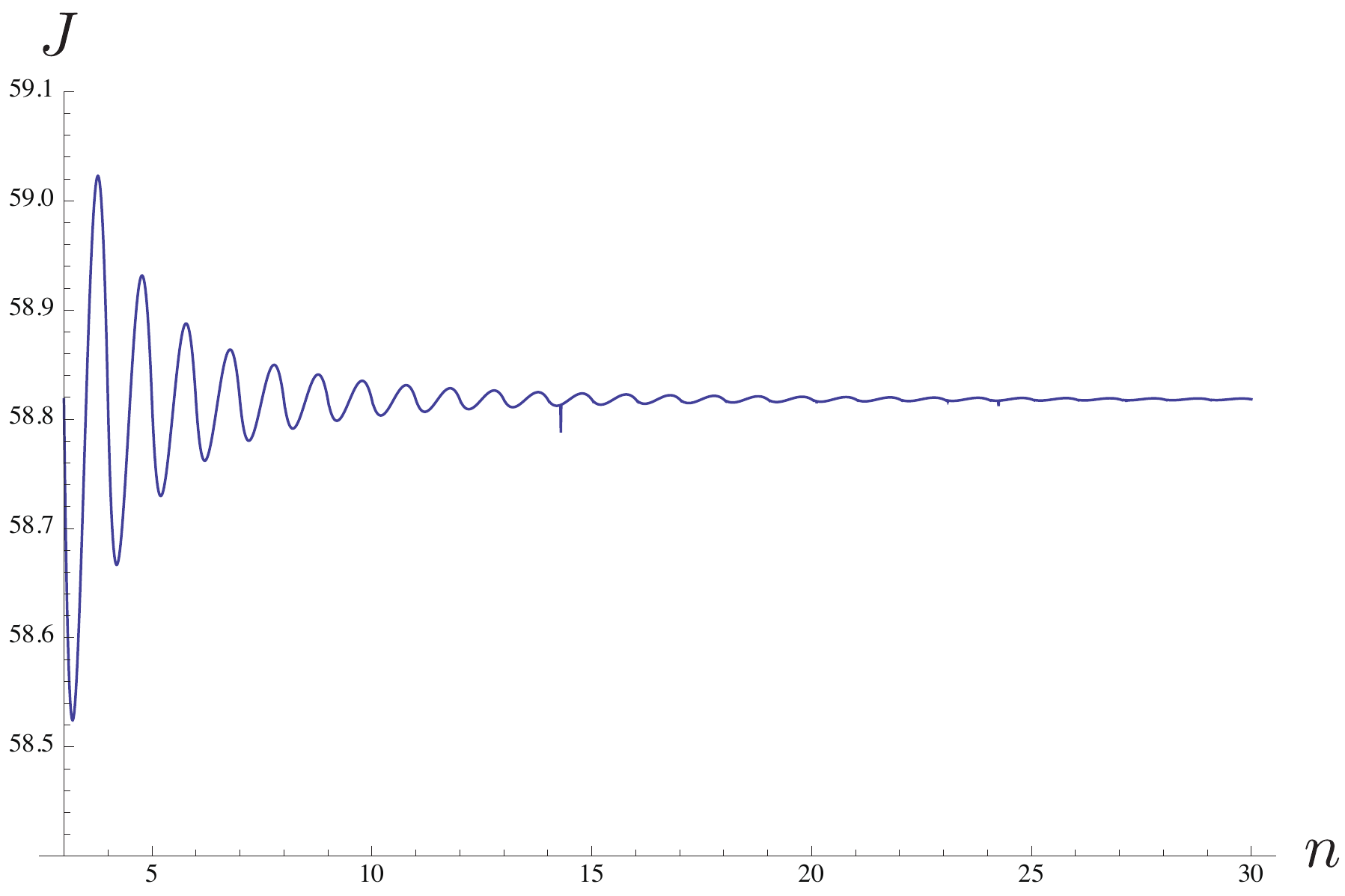}}
    \caption{The scaling behavior of the NESS energy current $J$ for a particular set of parameters. The number $n$ ranges from 3 to 30.}\label{Fi:scaling}
\end{figure}
We show in Fig.~\ref{Fi:scaling} an example of the scaling behavior of the NESS energy current along the chain. The length of the chain ranges from $n=3$ to $n=30$. We see that the value quickly converges to an almost $n$-independent constant.

While this qualitative behavior for a perfect quantum harmonic chain (with no defect, impurity or nonlinearity) may  be well-known or can be reasoned out, we have hereby provided an explicit quantitative proof of it.

\section{Concluding Remarks}

The setup of an open system interacting with two heat baths serves as the basis for a wide range of investigations in physics, chemistry and biology. The existence of a nonequilibrium steady state in such a system is an issue of fundamental importance because, to name just one, it is the pre-condition for nonequilibrium thermodynamics, which serves as a powerful springboard for investigations in many areas of sciences and engineering. The existence and uniqueness of NESS have been studied for classical systems decades ago with rigorous mathematical proofs. We want to do the same now for quantum open systems, starting with the simpler case of continuous variable harmonic systems (in contradistinction to discrete variables such as spin chains, a subject which has seen many flowering results).

The key findings of this investigation have been enumerated in the Introduction so there is no need to repeat them here. A few general concluding remarks would suffice.

The broader value of this work as we see it is twofold: 1) A demonstration of the existence of a NESS for the system of interest. Rather than constructing mathematical proofs we provide the full dynamics of the system and derive explicit expressions for the energy flow in each component leading to a proof that an energy flow balance relation exists. 2) Presenting a toolbox whereby one can derive the stochastic equations and calculate the average values of physical variables in open quantum systems -- this involves both taking the expectation values of quantum operators of the system and the distributional averages of stochastic variables originating from the environment. The functional method we adopt here has the advantage that it is compact and powerful, and it can easily accommodate perturbative techniques and diagrammatic methods developed in quantum field theory to deal with weakly nonlinear open quantum systems, as we will show in a sequel paper.  The somewhat laborious and expository construction presented in this paper is necessary to build up a platform for systematic investigations of nonequilibrium open quantum systems, some important physical issues therein will be discussed in future communications.

To expand the second point somewhat, our approach is characterized by two essential features;  a) we use a \textit{microphysics model }of generic nature, namely here, a chain of harmonic oscillators interacting with two baths described by two scalar fields at different temperatures. b) this allows us to derive everything from \textit{first principles}, e.g., starting with an action principle describing the interaction of all the microscopic constituents and components in the model. This way of doing things has the advantage that one knows the physics which goes into all the approximations made, in clearly marked stages. For Gaussian systems, namely, bilinear coupling between harmonic oscillators and with baths, one can solve this problem exactly, providing the fully nonequilibrium evolution of the open system with the influences of it environment (the two heat baths here) accounted for in a self-consistent manner.

Self-consistency is an absolutely essential requirement which underlies the celebrated relation of Onsager, for example (one may refer to the balance relations we obtained here as the quantum Onsager relations) and realization of the symmetries in the open systems which were used for the mathematical proofs of NESS. This consistency condition is not so well appreciated in the open quantum system literature, but we see it as crucial in the exploitation of the symmetry principles mentioned above as well as in treating physical processes when memory effects (in non-Markovian processes) and when the effects of backaction are important. This includes situations when one wants to a) treat strongly correlated systems or systems subjected to colored noises  b) design feedback control of quantum systems  c) engineer an environment with sensitive interface with the open quantum system, to name a few.\\

\noindent{\bf Acknowledgment}   This work began in the summer of 2013 when both authors visited Fudan University's Center for Theoretical Physics at the invitation of Prof. Y. S. Wu.  Earlier that year BLH visited the group of Prof. Baowen Li at the National University of Singapore.  Thanks are due to them for their warm hospitality.  The leitmotiv to understand nonequilibrium energy transport began when BLH attended a seminar by Prof. Bambi Hu at Zhejiang University in 2010. He also thanks Prof. Dhar for useful discussions.  JTH's research is supported by the National Science Council, R.O.C.

\section{Appendix}

In this Appendix, we will derive the relations between $\widetilde{D}^{k,1}(\omega)\,\widetilde{D}^{k+1,1\,*}(\omega)$ or $\widetilde{D}^{k,n}(\omega)\,\widetilde{D}^{k+1,n\,*}(\omega)$ to $\bigl|\widetilde{D}^{1n}(\omega)\bigr|{}^{2}$. They are used in Sec.~\ref{HarCh} to set up the equality of the energy current between the neighboring sites along the chain.

To this goal, we first establish some useful relations between the elements of the fundamental solution matrix $\widetilde{\mathbf{D}}$. From the definition of the fundamental solution,
\begin{equation}
	 \Bigl[-\omega^{2}\mathbf{I}-i\,2\gamma\omega\,\mathbf{I'}+\pmb{\Omega}^{2}\Bigr]\cdot\widetilde{\mathbf{D}}(\omega)=\mathbf{I}\,,
\end{equation}
we see that the matrix
\begin{equation}\label{E:jkdfexa}
	 \widetilde{\mathbf{D}}^{-1}(\omega)=\Bigl[-\omega^{2}\mathbf{I}-i\,2\gamma\omega\,\mathbf{I'}+\pmb{\Omega}^{2}\Bigr]
\end{equation}
is symmetric with respect to the diagonal and the anti-diagonal, its inverse, the fundamental solution matrix $\widetilde{\mathbf{D}}$, also has these properties, that is,
\begin{align}
	\widetilde{D}^{j,k}&=\widetilde{D}^{k,j}\,,&\widetilde{D}^{j,k}&=\widetilde{D}^{n+1-k,n+1-j}\,.
\end{align}
In particular, it implies
\begin{align}
	\widetilde{D}^{j,1}&=\widetilde{D}^{1,j}=\widetilde{D}^{n+1-j,n}\,,
\end{align}
and \eqref{E:hmcnvw} becomes
\begin{align}\label{E:qmcnvw}
	 J_{j,j+1}&=-i\,4\gamma\sigma\int_{-\infty}^{\infty}\!d\omega\;\omega\,\Bigl[\widetilde{D}^{j,1}(\omega)\,\widetilde{D}^{j+1,1\,*}(\omega)\,\widetilde{G}_{H}^{11}(\omega)\Bigr.\notag\\
	 &\qquad\qquad\qquad\qquad\qquad+\Bigl.\widetilde{D}^{n-j,1\,*}(\omega)\,\widetilde{D}^{n-j+1,1}(\omega)\,\widetilde{G}_{H}^{nn}(\omega)\Bigr]\,.
\end{align}
Now the problem reduces to identifying a relation between $\widetilde{D}^{j,1}(\omega)\,\widetilde{D}^{j+1,1\,*}(\omega)$ or $\widetilde{D}^{n-j,1\,*}(\omega)\,\widetilde{D}^{n-j+1,1}(\omega)$ and $\bigl|\widetilde{D}^{1n}(\omega)\bigr|{}^{2}$.

To find the explicit expressions for the elements of the fundamental matrix $\widetilde{\mathbf{D}}$, we can invert \eqref{E:jkdfexa}. Since $\widetilde{\mathbf{D}}^{-1}$ forms a tridiagonal matrix, its inverse can be given by the recursion relations
\begin{align}\label{E:gndkees}
	\widetilde{D}^{j,1}&=\bigl(-1\bigr)^{j+1}\sigma^{j-1}\,\frac{\Upsilon_{j+1}}{\theta_{n}}\,,&j&>1\,,
\end{align}
where $\Upsilon_{n}=a_{n}$, $\Upsilon_{n+1}=1$ and
\begin{align}\label{E:nbfjes}
	\Upsilon_{j+1}&=a_{j+1}\,\Upsilon_{j+2}-\sigma^{2}\,\Upsilon_{j+2}\,,&\Upsilon_{j}&\in\mathbb{C}\,,
\end{align}
with $j=1,2,\ldots,n-1$ and $\Upsilon_{j}\in\mathbb{C}$. The quantity $\theta_{n}$ is the determinant of the inverse of the fundamental matrix, that is, $\theta_{n}=\det\widetilde{\mathbf{D}}^{-1}$, and satisfies the recursion relation
\begin{equation}
	\theta_{k}=a_{k}\,\theta_{k-1}-\sigma^{2}\,\theta_{k-2}\,,
\end{equation}
with $\theta_{0}=1$, $\theta_{-1}=0$ and $\theta_{k}\in\mathbb{C}$.

We introduce two shorthand notations $a'$ an $a$ by assigning $a'\equiv a_{1}=a_{n}=-\omega^{2}-i\,2\gamma\omega+\omega_{R}^{2}$, and $a\equiv a_{2}=\cdots=a_{n-1}=-\omega^{2}+\omega_{R}^{2}$. We note that $a'$ is a complex number and its imaginary part is an odd function of $\omega$. It then proves useful to express $\Upsilon_{j+1}$ explicitly in terms of $a'$. The motivation behind this lies in the fact that inside the square brackets of \eqref{E:qmcnvw}, only terms that are odd with respect to $\omega$ can have nontrivial contributions to the current. On the other hand, the only source that may contribute to the odd power in $\omega$ is the imaginary part of $a'$.

From the recursion relation \eqref{E:nbfjes}, we see that in general the variable $\Upsilon_{j}$ can be expanded by $f_{j}$
\begin{align}\label{E:pods}
	\Upsilon_{j}&=f_{j}\,a'-f_{j+1}\,\sigma^{2}\,,
\end{align}
where $f_{j}$ is an $(n-j)$--order polynomial of $a$ with $f_{n}=1$, $f_{n+1}=0$, and it satisfies the recursion relation,
\begin{align}\label{E:oewmda}
	f_{j}&=a\,f_{j+1}-\sigma^{2}\,f_{j+2}\,,&f_{j}&\in\mathbb{R}
\end{align}
Here we write down the first a few entries in the sequence $\{f_{j}\}$,
\begin{align}
	f_{n+1}&=0\,,&f_{n}&=1\,,\notag\\
	f_{n-1}&=a\,,&f_{n-2}&=a^{2}-\sigma^{2}\,, \notag\\
	f_{n-3}&=a^{3}-2a\sigma^{2}\,,&f_{n-4}&=a^{4}-3a^{2}\sigma^{2}+\sigma^{4}\,,\notag\\
	f_{n-5}&=a^{5}-4a^{3}\sigma^{2}+3a\sigma^{4}\,,&f_{n-6}&=a^{6}-5a^{4}\sigma^{2}+6a^{2}\sigma^{4}-\sigma^{4}\,.
\end{align}
Although $f_{j}$ follows a similar recursion relation to \eqref{E:nbfjes}, introduction of $f_{j}$ makes it easier to identify the imaginary part of $\Upsilon_{j}$. Therefore once we find the general solution of $f_{j}$ via the recursion relation \eqref{E:oewmda}, we will have $\Upsilon_{j}$, which is useful to construct the general expression for the elements of the fundamental solution matrix $\widetilde{\mathbf{D}}$.

The recursion relation \eqref{E:oewmda} can be solved if we substitute
\begin{equation}
	f_{k}\propto \mu^{-k}
\end{equation}
into \eqref{E:oewmda}, we find that $\mu$ will satisfy a characteristic equation
\begin{equation}\label{E:vneruw}
	\mu^{2}-a\,\mu+\sigma^{2}=0\,.
\end{equation}
The two solutions, labeled by $\mu_{1}$ and $\mu_{2}$,
\begin{align}
	\mu_{1}&=\frac{a+\sqrt{a^{2}-4\sigma^{2}}}{2}\,,&\mu_{2}&=\frac{a-\sqrt{a^{2}-4\sigma^{2}}}{2}\,,
\end{align}
are assumed distinct, and then the general solution for $f_{k}$ is given by
\begin{equation}
	f_{k}=p_{1}\,\mu_{1}^{-k}+p_{2}\,\mu_{2}^{-k}\,.
\end{equation}
We can use the conditions $f_{n}=1$, $f_{n+1}=0$ to fix the unknown coefficients $p_{1}$ and $p_{2}$,
\begin{align}
	f_{n}&=p_{1}\,\mu_{1}^{-n}+p_{2}\,\mu_{2}^{-n}=1\,,\\
	f_{n+1}&=p_{1}\,\mu_{1}^{-n-1}+p_{2}\,\mu_{2}^{-n-1}=0\,,
\end{align}
so they are
\begin{align}
	p_{1}&=\frac{\mu_{1}^{n+1}}{\mu_{1}-\mu_{2}}\,,&p_{2}&=-\frac{\mu_{2}^{n+1}}{\mu_{1}-\mu_{2}}\,.
\end{align}
Thus the general solution of $f_{k}$ takes the form
\begin{equation}\label{E:bvkdeor}
	f_{k}=\frac{\mu_{1}^{n-k+1}-\mu_{2}^{n-k+1}}{\mu_{1}-\mu_{2}}=\sum_{m=0}^{n-k}\mu_{1}^{n-k-m}\mu_{2}^{m}\,.
\end{equation}
With the help of these results, we can set up relations between $\widetilde{D}^{k,1}(\omega)\,\widetilde{D}^{k+1,1\,*}(\omega)$ or $\widetilde{D}^{n-k,1\,*}(\omega)\,\widetilde{D}^{n-k+1,1}(\omega)$ and $\bigl|\widetilde{D}^{1n}(\omega)\bigr|{}^{2}$.

To gain some insight, we first write down $\left|\widetilde{D}^{1n}(\omega)\right|{}^{2}$ by \eqref{E:gndkees}. Since
\begin{equation}
	 \widetilde{D}^{1n}=\bigl(-1\bigr)^{n+1}\sigma^{n-1}\,\frac{\Upsilon_{n+1}}{\theta_{n}}=\bigl(-1\bigr)^{n+1}\,\frac{\sigma^{n-1}}{\theta_{n}}\,,
\end{equation}
we find
\begin{equation}\label{E:uidkfhw}
	\bigl|\widetilde{D}^{1n}\bigr|{}^{2}=\frac{\sigma^{2n-2}}{\lvert\theta_{n}\rvert^{2}}\,.
\end{equation}
Next, $\widetilde{D}^{s,1}\,\widetilde{D}^{s+1,1\,*}$ can be given by
\begin{align}
	 \widetilde{D}^{s,1}\,\widetilde{D}^{s+1,1\,*}&=\left[\bigl(-1\bigr)^{s+1}\sigma^{s-1}\,\frac{\Upsilon_{s+1}}{\theta_{n}}\right]\left[\bigl(-1\bigr)^{s+2}\sigma^{s}\,\frac{\Upsilon^{*}_{s+2}}{\theta^{*}_{n}}\right]\notag\\
	&=-\sigma^{2s-1}\frac{\Upsilon_{s+1}\Upsilon^{*}_{s+2}}{\lvert\theta_{n}\rvert^{2}}\,.\label{E:ioertskw}
\end{align}
We can expand the product $\Upsilon_{s+1}^{\vphantom{*}}\Upsilon^{*}_{s+2}$ by \eqref{E:pods}, and get
\begin{align}
	 \Upsilon_{s+1}^{\vphantom{*}}\Upsilon^{*}_{s+2}&=\Bigl[f_{s+1}\,a'-f_{s+2}\,\sigma^{2}\Bigr]\Bigl[f_{s+2}\,a'^{*}-f_{s+3}\,\sigma^{2}\Bigr]\notag\\
	&=-\Bigl(f_{s+1}f_{s+3}\,a'+f_{s+2}^{2}\,a'^{*}\Bigr)\sigma^{2}+\cdots\,,\label{E:gnmcroe}
\end{align}
where $\dots$ are terms that will not contribute to the integral \eqref{E:qmcnvw}. Now recall that the imaginary part of $a'$ is odd with respect to $\omega$, so we write $a'$ explicitly as $a'=a-i\,c$, where $a=-\omega^{2}+\omega_{R}^{2}$ has been defined before while $c$ is equal to $2\gamma\omega$. In so doing, we are able to condense \eqref{E:gnmcroe} further to highlight its dependence on the imaginary part of $a'$, that is, $c$,
\begin{align}
	 \Upsilon_{s+1}^{\vphantom{*}}\Upsilon^{*}_{s+2}&=i\,c\Bigl(f_{s+1}f_{s+3}-f_{s+2}^{2}\Bigr)\sigma^{2}+\cdots\,,\label{E:hnmcroe}
\end{align}
The expression in the parentheses can be evaluated by \eqref{E:bvkdeor}, and we find
\begin{align}
	f_{s+1}f_{s+3}-f_{s+2}^{2}=-\sigma^{2(n-s-2)}\,,
\end{align}
where we have used the fact that $\mu_{1}\mu_{2}=\sigma^{2}$ in \eqref{E:vneruw} at the final step. Thus eq.~\eqref{E:hnmcroe} becomes
\begin{align}
	\Upsilon_{s+1}^{\vphantom{*}}\Upsilon^{*}_{s+2}&=-i\,c\,\sigma^{2(n-s-1)}+\cdots\,.\label{E:dloe}
\end{align}
We put it back to \eqref{E:ioertskw} and arrive at
\begin{align}\label{E:yeibsw}
	 \widetilde{D}^{s,1}\,\widetilde{D}^{s+1,1\,*}&=+i\,c\,\frac{\sigma^{2n-3}}{\lvert\theta_{n}\rvert^{2}}+\cdots=+\frac{i\,c}{\sigma}\,\bigl|\widetilde{D}^{1n}\bigr|{}^{2}+\cdots\,,
\end{align}
where we have compared the result with \eqref{E:uidkfhw}. Again the dots represent terms that do not contribute to the integral \eqref{E:qmcnvw}.

Next we proceed to evaluate $\widetilde{D}^{n-s,1\,*}\,\widetilde{D}^{n-s+1,1}$. By \eqref{E:gndkees}, we obtain
\begin{align}
	 \widetilde{D}^{n-s,1\,*}\,\widetilde{D}^{n-s+1,1}&=\left[\bigl(-1\bigr)^{n-s+1}\sigma^{n-s-1}\,\frac{\Upsilon_{n-s+1}^{*}}{\theta_{n}^{*}}\right]\left[\bigl(-1\bigr)^{n-s+2}\sigma^{n-s}\,\frac{\Upsilon_{n-s+2}}{\theta_{n}}\right]\notag\\
	&=-\sigma^{2(n-s)-1}\,\frac{\Upsilon_{n-s+1}^{*}\Upsilon_{n-s+2}}{\lvert\theta_{n}\rvert^{2}}\,.
\end{align}
The factor $\Upsilon_{n-s+1}^{*}\Upsilon_{n-s+2}$ is then further expanded by $f_{j}$ as shown in \eqref{E:pods}, and we identify the imaginary part of $a'$,
\begin{align}
	 \Upsilon_{n-s+1}^{*}\Upsilon_{n-s+2}&=\Bigl[f_{n-s+1}\,a'^{*}-f_{n-s+2}\,\sigma^{2}\Bigr]\Bigl[f_{n-s+2}\,a'-f_{n-s+3}\,\sigma^{2}\Bigr]\notag\\
	&=-\Bigl(f_{n-s+2}^{2}\,a'+f_{n-s+1}f_{n-s+3}\,a'^{*}\Bigr)\sigma^{2}+\cdots\notag\\
	&=-i\,c\,\Bigl(f_{n-s+1}f_{n-s+3}-f_{n-s+2}^{2}\Bigr)\sigma^{2}+\cdots\notag\\
	&=i\,c\,\sigma^{2s-2}+\cdots\,,
\end{align}
where we have used \eqref{E:bvkdeor} and the fact $a'=a-i\,c$. This implies
\begin{align}\label{E:lkejo}
	 \widetilde{D}^{n-s,1\,*}\,\widetilde{D}^{n-s+1,1}&=-i\,c\,\frac{\sigma^{2n-3}}{\lvert\theta_{n}\rvert^{2}}+\cdots=-\frac{i\,c}{\sigma}\,\bigl|\widetilde{D}^{1n}\bigr|{}^{2}+\cdots\,,
\end{align}
where $\dots$ will have vanishing contributions to the integral \eqref{E:qmcnvw}. Eqs.~\eqref{E:yeibsw} and \eqref{E:lkejo} are the sought-after relations between $\widetilde{D}^{s,1}\,\widetilde{D}^{s+1,1\,*}$ or $\widetilde{D}^{n-s,1\,*}\,\widetilde{D}^{n-s+1,1}$ and $\bigl|\widetilde{D}^{1n}\bigr|{}^{2}$.

Next we turn to the scaling behavior of $\bigl|\widetilde{D}^{1n}\bigr|{}^{2}$ with $n$. From \eqref{E:uidkfhw} and following the procedures that lead to the general expression of $\Upsilon_{j}$, we find
\begin{align}
	\lvert\theta_{n}\rvert^{2}&=f_{2}^{2}\,c^{4}+2f_{1}\,c^{2}+f_{0}^{2}+2\sigma^{2(n-1)}c^{2}>0\,,\label{E:jcmer}
\end{align}
where we have used \eqref{E:bvkdeor} for the case $k=n$
\begin{equation}
	f_{1}^{2}-f_{0}\,f_{2}=\sigma^{2(n-1)}\,,
\end{equation}
to simplify $\lvert\theta_{n}\rvert^{2}$.

To draw further information about $\lvert\theta_{n}\rvert^{2}$, we would like to discuss the generic behavior of $f_{k}$ with respect to $\omega$. We first note that when $a^{2}-4\sigma^{2}<0$, the two solutions $\mu_{1}$, $\mu_{2}$ of the characteristic equation \eqref{E:vneruw} become complex-conjugated. If we write them in terms of polar coordinate, then we have
\begin{align}
	 \mu_{1}&=\sigma\,e^{i\,\psi}\,,&\mu_{2}&=\sigma\,e^{-i\,\psi}\,,&\psi&=\tan^{-1}\frac{\sqrt{4\sigma^{2}-a^{2}}}{a}\,,
\end{align}
with $0\leq\psi\leq\pi$. Recall that $a=-\omega^{2}+\Omega^{2}$ and $c=2\omega\gamma$. In terms of the frequency the condition $a^{2}-4\sigma^{2}<0$ corresponds to the frequency band
\begin{align}
	\sqrt{\Omega^{2}-2\sigma}&<\omega<\sqrt{\Omega^{2}+2\sigma}\,,
\end{align}
within which the parameter $a$ monotonically decreases from $+2\sigma$ to $-2\sigma$ and $\psi$ steadily grows from 0 to $\pi$ as $\omega$ increases. Hence $f_{k}$ can be written as
\begin{align}\label{E:bfghbr}
	f_{k}&=\sigma^{n-k}\,\frac{\sin(n-k+1)\psi}{\sin\psi}\,,
\end{align}
which is heavily oscillating in $\omega$. If we substitute \eqref{E:bfghbr} into \eqref{E:jcmer}, we find that $\lvert\theta_{n}\rvert^{2}$ becomes
\begin{align}
	 \lvert\theta_{n}\rvert^{2}&=\sigma^{2n}\,\frac{\sin^{2}(n+1)\psi}{\sin^{2}\psi}+2\sigma^{2(n-1)}c^{2}\,\frac{\sin^{2}n\psi}{\sin^{2}\psi}+\sigma^{2(n-2)}\,\frac{\sin^{2}(n-1)\psi}{\sin^{2}\psi}+2\sigma^{2(n-1)}c^{2}\notag\\
	 &=\frac{\sigma^{2n}}{\sin^{2}\psi}\left[\sin^{2}(n+1)\psi+2\frac{c^{2}}{\sigma^{2}}\Bigl(1+\sin^{2}n\psi\Bigr)+\frac{c^{4}}{\sigma^{4}}\sin^{2}(n-1)\psi\right]\,.
\end{align}
It can be greatly simplified when the coupling with the environment is weak such that $\gamma\Omega\ll\sigma$. In this case $\lvert\theta_{n}\rvert^{2}$ reduces to
\begin{equation}
	\lvert\theta_{n}\rvert^{2}\sim\frac{\sigma^{2n}}{\sin^{2}\psi}\Bigl[\sin^{2}(n+1)\psi+\varepsilon\Bigr]\,.
\end{equation}
Here $\varepsilon$ is a small positive number as a reminder that the expression in the squared brackets is not supposed to totally vanish so that when we put $\lvert\theta_{n}\rvert^{2}$ back to $\bigl|\widetilde{D}^{1n}(\omega)\bigr|^{2}$, it will not introduce artefact poles,
\begin{equation}\label{E:fbmjwiu}
	 \bigl|\widetilde{D}^{1n}(\omega)\bigr|^{2}=\frac{1}{\sigma^{2}}\frac{\sin^{2}\psi}{\sin^{2}(n+1)\psi+\varepsilon}\,.
\end{equation}
Owing to the factor $\sin^{2}(n+1)\psi$ in the denominator, $\bigl|\widetilde{D}^{1n}(\omega)\bigr|^{2}$ will have maxima at $\psi=k\pi/(n+1)$ for $k=1,2,\ldots,n$. As for $k=0$, or $n+1$, since the numerator $\sin^{2}\psi$ cancels with the denominator $\sin^{2}(n+1)\psi$, there is no maximum of $\bigl|\widetilde{D}^{1n}(\omega)\bigr|^{2}$ at these two locations. It implies that there are $n$ peaks distributed evenly\footnote{in particular when $\sigma\ll\Omega^{2}$ because in that limit,
\begin{equation*}
	\omega=\sqrt{\Omega^{2}-2\sigma\cos\psi}\sim\Omega-\dfrac{\sigma}{\Omega}\,\cos\psi\,.
\end{equation*}
For the neighboring maxima, the separation between them is given by
\begin{equation*}
	-\frac{\sigma}{\Omega}\Bigl[\cos(k+1)\Delta-\cos k\Delta\Bigr]=-\frac{\sigma}{\Omega}\Bigl[\cos k\Delta\,\cos\Delta-\sin k\Delta\,\sin\Delta-\cos k\Delta\Bigr]\simeq\left(\frac{\sigma}{\Omega}\,\Delta\right)\sin k\Delta+\mathcal{O}(\Delta^{2})\,,
\end{equation*}
where $\Delta=\pi/n\ll1$. Thus the separation is independent of $k$ in the neighborhood $k\Delta\ll1$ and $\pi-k\Delta\ll1$.} within the band $\sqrt{\Omega^{2}-2\sigma}<\omega<\sqrt{\Omega^{2}+2\sigma}$. As $n$ increases, the peaks become narrower with the width of the order $\pi/(n+1)$.

On the other hand, outside the band $\sqrt{\Omega^{2}-2\sigma}<\omega<\sqrt{\Omega^{2}+2\sigma}$, because we have $a^{2}-4\sigma^{2}>0$, both roots of the characteristic equations are real with $\mu_{1}>\mu_{2}>0$ and
\begin{align}
	\lvert\mu_{1}\rvert&>\lvert\mu_{2}\rvert\,,&a&>2\sigma\,,\\
	\lvert\mu_{2}\rvert&>\lvert\mu_{1}\rvert\,,&a&<-2\sigma\,.
\end{align}
Hence we note that when $a>2\sigma$, $\mu_{1}^{k}$ rapidly dominates over $\mu_{2}^{k}$ as $k$ increases, but when $a<-2\sigma$, $\mu_{2}^{k}$ quickly outgrows $\mu_{1}^{k}$ for large enough $k$. Thus we have
\begin{equation}\label{E:cvkdeor}
	f_{k}\sim\begin{cases}
					\dfrac{\mu_{1}^{n-k+1}}{\mu_{1}-\mu_{2}}\,,&a>+2\sigma\,,\\
					\\
					\dfrac{\mu_{2}^{n-k+1}}{\mu_{1}-\mu_{2}}\,,&a<-2\sigma\,,
			\end{cases}
\end{equation}
for $n-k\gg1$. If we further assume $\gamma\Omega\ll\sigma$, then $\left|\theta_{n}\right|^{2}$ will be approximately given by
\begin{align}
	 \left|\theta_{n}\right|^{2}&\sim\frac{\mu_{i}^{2n+2}}{a^{2}-4\sigma^{2}}\left\{1+\frac{2c^{2}}{\mu_{i}^{2}}\left[1+\left(\frac{\sigma}{\mu_{i}}\right)^{2n}\right]+\cdots\right\}\,,
\end{align}
and then
\begin{equation}
	 \bigl|\widetilde{D}^{1n}(\omega)\bigr|^{2}\simeq\frac{a^{2}-4\sigma^{2}}{\mu_{i}^{4}}\left(\frac{\sigma}{\mu_{i}}\right)^{2(n-1)}\left\{1+\frac{2c^{2}}{\mu_{i}^{2}}\left[1+\left(\frac{\sigma}{\mu_{i}}\right)^{2n}\right]+\cdots\right\}^{-1}\,,
\end{equation}
with $i=1$ for $a>2\sigma$ but $i=2$ for $a<-2\sigma$. Now we note that although the strong inter-oscillator coupling is allowed, the coupling strength $\sigma$ is required smaller than $\Omega^{2}/2$ to avoid instability of the system. We also observe that since $a=-\omega^{2}+\Omega^{2}$, we have $a^{2}-4\sigma^{2}>0$ outside the band $\omega<\sqrt{\Omega^{2}-2\sigma}<0<\omega<\sqrt{\Omega^{2}+2\sigma}$. It implies
\begin{align}
	\mu_{1}-\sigma&=\frac{a+\sqrt{a^{2}-4\sigma^{2}}}{2}-\sigma>0\,,&0&<\frac{\sigma}{\mu_{1}}<1\,,&a&>2\sigma\,,\\
	-\mu_{2}-\sigma&=\frac{-a+\sqrt{a^{2}-4\sigma^{2}}}{2}-\sigma>0\,,&-1&<\frac{\sigma}{\mu_{2}}<0\,,&a&<-2\sigma\,.
\end{align}
We then may conclude that the factor
\begin{align}
	\left(\frac{\sigma}{\mu_{i}}\right)^{2(n-1)}\ll1\,,
\end{align}
drops to zero very fast for sufficiently large $n$, if $0<\omega<\sqrt{\Omega^{2}-2\sigma}$ or $\omega>\sqrt{\Omega^{2}+2\sigma}$. Thus, $\bigl|\widetilde{D}^{1n}(\omega)\bigr|^{2}$ will monotonically and rapidly falls to a relatively small value outside the band $\sqrt{\Omega^{2}-2\sigma}<\omega<\sqrt{\Omega^{2}+2\sigma}$.

At this point we may draw some conclusions about $\bigl|\widetilde{D}^{1n}(\omega)\bigr|^{2}$. In the frequency space, $\bigl|\widetilde{D}^{1n}(\omega)\bigr|^{2}$ falls monotonically and rapidly to a relatively small value outside the band $\sqrt{\Omega^{2}-2\sigma}<\omega<\sqrt{\Omega^{2}+2\sigma}$; on the other hand, within the frequency band, it possesses comb-like structure. The number of spikes grows with the length of the chain, but the width of the spike, on the contrary, becomes narrower and narrower, inversely proportional to the length of the chain.

Next how the behavior of the transmission coefficient helps to understand the dependence of the steady current on the length of the chain? A simpler question to ask is how the contribution from each spike in \eqref{E:fbmjwiu} will scale with $n$ within the frequency band? We first make an observation for the integral
\begin{equation}
	 I_{n}^{(k)}=\int_{\frac{k-1/2}{n+1}\pi}^{\frac{k+1/2}{n+1}\pi}\,d\psi\;\frac{\sin^{2}\psi}{\sin^{2}(n+1)\psi+\epsilon}
\end{equation}
for the $k^{th}$ spike among $n$ spikes confined within the interval $0<\psi<\pi$. The parameter $\epsilon$ is a very small positive number. Change of the variable $\psi$ to $\varpi=(n+1)\psi$ gives
\begin{equation}\label{E:bkdfeuw}
	 I_{n}^{(k)}=\frac{1}{n+1}\int_{(k-\frac{1}{2})\pi}^{(k+\frac{1}{2})\pi}\,d\varpi\;\frac{\sin^{2}\frac{\varpi}{n+1}}{\sin^{2}\varpi+\epsilon}\,.
\end{equation}
For large $n$, the numerator of the integrand is slowly varying compared with the denominator, so it can be pulled out of the integral and is evaluated for $\varpi=k\pi$,
\begin{align}\label{E:hbsene}
	 I_{n}^{(k)}&\simeq\frac{1}{n+1}\sin^{2}\frac{k\,\pi}{n+1}\int_{(k-\frac{1}{2})\pi}^{(k+\frac{1}{2})\pi}\,d\varpi\;\frac{1}{\sin^{2}\varpi+\epsilon}\notag\\
	 &=\frac{1}{n+1}\left[\sin^{2}\frac{k\,\pi}{n+1}\int_{-\frac{\pi}{2}}^{\frac{\pi}{2}}\,d\varpi\;\frac{1}{\sin^{2}\varpi+\epsilon}\right]\,.
\end{align}
The approximation improves for a larger value of $n$ because the denominator of \eqref{E:bkdfeuw} becomes more slowly varying with $\omega$. The contribution from the squared brackets is approximately the same for the spike at about the same locations within the interval $0<\psi<\pi$. Thus $I_{n}^{(k)}$ will scale with $n^{-1}$ for sufficiently large $n$. For example, let us pick one spike, say, at $k=n_{0}/5$ for $n=n_{0}$. Now suppose we rescale $n$ from $n=n_{0}$ to $n=3n_{0}$, and then we see that the three spikes centered at about $k=3_{0}/5$ will have about the same height but only about one third of width, for sufficiently large $n_{0}$. Therefore each spike in the $n=3n_{0}$ case will contribute one third as much as that in the $n=n_{0}$ case to the integral in $I_{n}^{(k)}$, as we can see from \eqref{E:hbsene}. When we consider all the spikes with the range $0<\psi<\pi$, we expect
\begin{equation}
	I_{n}=\sum_{k=1}^{n}I_{n}^{(k)}
\end{equation}
should independent of $n$ for sufficiently large $n$.

Following this argument, we see the steady current $J$ becomes
\begin{align}
	 J&=8\gamma^{2}\int^{\infty}_{-\infty}\!d\omega\;\omega^{2}\bigl|\widetilde{D}^{1n}(\omega)\bigr|^{2}\Bigl[\widetilde{G}_{H}^{11}(\omega)-\widetilde{G}_{H}^{nn}(\omega)\Bigr]\notag\\
	 &\simeq\frac{16\gamma^{2}}{\sigma^{2}}\int^{\sqrt{\Omega^{2}+2\sigma}}_{\sqrt{\Omega^{2}-2\sigma}}\!d\omega\;\frac{\sin^{2}\psi}{\sin^{2}(n+1)\psi+\varepsilon}\left\{\omega^{2}\Bigl[\widetilde{G}_{H}^{11}(\omega)-\widetilde{G}_{H}^{nn}(\omega)\Bigr]\right\}\notag\\
	 &=\frac{16\gamma^{2}}{\sigma^{}}\int^{\pi}_{0}\!d\psi\;\frac{\sin^{3}\psi}{\sin^{2}(n+1)\psi+\varepsilon}\,h(\omega)\,,
\end{align}
with $\omega=\sqrt{\Omega^{2}-2\sigma\cos\psi}$, and the function $h(\omega)$ being given by
\begin{equation}
	h(\omega)=\omega^{}\Bigl[\widetilde{G}_{H}^{11}(\omega)-\widetilde{G}_{H}^{nn}(\omega)\Bigr]\,.
\end{equation}
Since only the denominator $\sin^{2}(n+1)\psi+\varepsilon$ is vert rapidly oscillating with $\psi$ for large $n$, we can use the previous arguments to support that the steady current does not scale with $n$ for sufficiently large $n$, that is
\begin{equation}
	J\simeq\mathcal{O}(n^{0})\,,
\end{equation}
when $n>N_{0}$ for some large positive integer $N_{0}$.


\end{document}